\documentclass[letterpaper]{article} 

\usepackage[usenames,dvipsnames]{color}
\usepackage[utf8]{inputenc}
\usepackage{bm}
\usepackage[title]{appendix}
\usepackage{graphicx,fancyhdr,lscape,dsfont,amsmath,amssymb} 
\usepackage{color, amsthm} 
\usepackage{longtable}
\usepackage{psfrag}
\usepackage{tabularx}
\usepackage{boldline,multirow}
\usepackage{stmaryrd}
\usepackage[sl]{caption}
\usepackage{xcolor}
\usepackage{subcaption}
\usepackage{appendix}
\usepackage{soul}
\usepackage{cancel}
\usepackage[noend]{algpseudocode}
\usepackage{etoolbox}
\usepackage{algorithm2e}
\usepackage{mathtools}
\usepackage{sidecap}
\usepackage{enumitem}

\newcommand{\rev}{}
\newcommand{\revv}{}

\sidecaptionvpos{figure}{c}

\allowdisplaybreaks

\begin{document}

\title{Adaptation and validation of FFT methods for homogenization of lattice based materials}

\author{S. Lucarini$^{1}$, L. Cobian$^{1,2}$, A. Voitus$^{1}$ and J. Segurado$^{1,2}$
\\
\begin{small}
$^{1}$ Fundaci\'on IMDEA Materiales, C/ Eric Kandel 2, 
28906, Getafe, Madrid, Spain
\end{small}
 \\
\begin{small}
$^{2}$ Universidad Politécnica de Madrid, Department of Materials Science, 
\end{small}\\
\begin{small}
E.T.S.I. Caminos, C/ Profesor Aranguren 3, 28040, Madrid, Spain
\end{small}
}

\maketitle

\begin{abstract}
	
An FFT framework which preserves a good numerical performance in the case of domains with large regions of empty space is proposed and analyzed for its application to lattice based materials. \revv{Two spectral solvers specially suited to resolve problems} containing phases with zero stiffness are considered (1) a Galerkin approach combined with the MINRES linear solver and a discrete differentiation rule and  (2) a modification of a displacement FFT solver which penalizes the indetermination of strains in the empty regions, leading to a fully determined equation. The solvers are combined with several approaches to smooth out the lattice surface, based on modifying the actual stiffness of the voxels not fully embedded in the lattice or empty space. The accuracy of the resulting approaches is assessed for an octet-lattice by comparison with FEM solutions for different relative densities and discretization levels. It is shown that the \rev{adapted} Galerkin approach combined with a \rev{Voigt} surface smoothening was the best FFT framework considering accuracy, numerical efficiency and $h$-convergence. \rev{With respect to} numerical efficiency it was observed that FFT becomes competitive \rev{compared to} FEM for cells with relative densities above $\approx$7\%. Finally, to show the real potential of the approaches presented, the FFT frameworks are used to simulate the behavior of a printed lattice by using direct 3D tomographic data as input. The approaches proposed include explicitly in the simulation the actual surface roughness and internal porosity \revv{resulting from} the fabrication process. The \rev{simulations allowed to quantify} the reduction of the lattice stiffness \rev{as well as to} resolve the stress localization of $\approx$ 50\% near large pores.

\end{abstract}

\section{Introduction}\label{sec:intro}

The latest improvements in additive manufacturing techniques (AM) have made possible the fabrication of micro- and nano-architected metamaterials for mechanical applications with tailored stiffness, strength, toughness and energy absorption capacity \cite{YU2018114}. In most cases, the topology of these architected materials at the lower length scale is based on the periodic repetition of a unit cell \rev{made up} of bars or shells forming a lattice. These lattice \revv{materials} can reach high strength-to-weight ratios and other specific properties as well as achieving non-standard elastic responses such as negative compressibility or zero Poisson's ratio \cite{YU2018114,Deshpande2001,Xu2016}. In addition, unit cells of lattice metamaterials can be designed using unstable structural elements \rev{to achieve} energy dissipation with fully reversible deformation or programmable/switchable properties \cite{FINDEISEN2017151}.

In order to design an optimal lattice for a specific property is fundamental to perform accurate and computationally efficient simulations of the effective response and the local fields developed within the microstructure. Although discrete mechanical models are a good approach \rev{for capturing} the overall response \cite{lenstringant_2020}, full-field homogenization approaches allow obtaining a more complete, and often more accurate, result. Under this approach, a micromechanical problem \rev{is solved} on a representative volume element (RVE) which \rev{contains a unit cell or a collection of unit cells in which the lattice geometry is explicitly represented}. In particular, the Finite Element Method (FEM) is the most common approach to carry out these simulations since the designed \emph{ideal} microstructure can be accurately reproduced with adaptative meshes \cite{McMeeking2017}. However, the real microstructures arising from an AM process can differ from the  \emph{ideal} ones designed in-silico, and usually present geometrical deviations from the target geometry, porosity and surface roughness \cite{AMANI2018395}. Although these \rev{features} can be neglected for the lattice design phase, a reliable simulation of the material deformation in the non linear regime \rev{which considers} its irreversible response or fracture should take these differences into account. FEM models of realistic geometries including surface roughness or porosity are very difficult to generate and mesh. Moreover, the real geometrical data \rev{is usually obtained} from tomographic images \cite{Lhuisser2016,AMANI2018395}, and transforming this data into a FEM model is a very complex process that results in a very large number of elements. 

In this context, the use of spectral methods, based on the Fast Fourier Transform (FFT) algorithm, can be an interesting alternative. \rev{FFT methods do not require meshing because} the calculations are carried out on a regular grid \rev{and the phase belonging to each point of the grid} can be obtained directly from digital images or tomographic data. In addition, FFT approaches are very efficient, \rev{with a computational cost which grows} as $n\log (n)$, \rev{improving} the FEM computational efficiency in the homogenization of bulk heterogeneous materials by orders of magnitude \cite{Prakash_2009,LUCARINI2019_CM}. There are also other tangential benefits \rev{coming from the} use of spectral approaches for studying lattice \revv{materials} \rev{like the possibility of studying brittle fracture or ductile damage using phase-field fracture or gradient damage approaches, methods that show a very efficient performance in FFT \cite{MA2020,ERNESTI2020112793,Magri2020b}}.

The basic ideas of FFT homogenization were proposed by Moulinec and Suquet \cite{Moulinec1994}. Since then, different approaches have been proposed to improve the convergence rate of the method. Some approaches are derived from the original method, based on the use of Green's functions for a reference medium and the solution of the Lippmann-Schwinger equation \cite{Moulinec1994,Moulinec1998,Eyre1999,Kabohm2005,Zeman2010,Kabel2014,Schneider2019,Wicht2021}. \rev{Alternative approaches} have also been proposed, based on either using a Galerkin approximation of the equilibrium \rev{using trigonometric polynomials to discretize test and trial functions}, \cite{Vondrejc2014,Zeman2017,DeGeus2017,Lucarini2019a} or on solving the strong form of equilibrium using displacements as unknown \cite{Lucarini2019b}. Despite the clear potential benefits of FFT solvers for lattice materials, two main limitations arise that have prevented their extensive use in this field. First, FFT approaches present a convergence rate and accuracy strongly dependent on the contrast between the phases represented in the domain. In the present case, the contrast is infinite \rev{because} a large amount of the RVE voxels are empty space, with zero stiffness. Second, \rev{although} the voxelized representation is very convenient for using digital microstructure data\rev{, its use for representing} smooth geometries might result in poor results near the boundaries. In addition, since FFT and inverse FFT transformations \rev{should include all the points of the RVE} including the empty space, the efficiency of the method \rev{with respect to} FEM will depend on the relative volume fraction, and it will not be competitive for very small \rev{relative} densities.

\rev{Regarding} the first limitation, several studies can be found which \rev{aim at overcoming} the problems with the high phase stiffness contrast in FFT solvers and try to extend their use to study materials with voids. Michel et al. \cite{Michel2001} developed an Augmented Lagrangian formulation to solve a non linear problem including non-compatible fields. \rev{Although this formulation allows introducing zero stiffness phases, it might require a very large number of iterations to fulfil both stress equilibrium and strain compatibility, as noticed in \cite{MouSilva2014}}. 
Brisard and Dormieux \cite{Brisard2010} developed a variational formulation based on the energy principle of Hashin and Shtrikman applied to a porous media. Their approach allows to accurately predict the overall response of porous materials but it involves the pre-computation of a consistent Green operator which is computationally very expensive. More recently, a method for solving the conductivity problem in the presence of voids has been developed by To and Bonnet \cite{To2020}. This approach is focused on solving the equilibrium only in the bulk phases including a flux term at the inter-phase between bulk and void phases. This method is suitable for scalar fields, but cannot be directly extended to the vector and tensor fields that arise in the mechanical problem since the flux term in the internal boundaries \rev{does} not restrict tensor components parallel to the interphase. Another recent method proposed by Schneider \cite{Schneider2020} consists in searching solutions in a subspace of solutions on which the homogenization problem is nondegenerate for the resolution of a material with pores.  \rev{In parallel to these techniques specifically suited for porous materials, an efficient and simple alternative to improve the convergence rate under very large phase contrast is the use of methods to reduce the numerical oscillations that may occur due to Gibb's phenomenon or aliasing effects. A first possibility is filtering the high frequencies, as proposed for example in \cite{KABOHM200690,GELEBART2015,Shanthraj2015}. A second possibility is replacing the continuum differential operators in the formulation of the partial differential equations in the real space by some finite difference differentiation rule. This idea was first introduced in \cite{Mueller98} who incorporated the finite difference definitions of the derivatives in the real space using the FFT algorithm and the definition of Fourier derivation using modified frequencies. The finite difference stencil used in the real space defines the particular form of the modified frequencies to be used. Among the different discrete differential approaches, the so-called rotated scheme \cite{Willot2015} shows a significant reduction of noise and improves the convergence \cite{LUCARINI2019_CM,DJAKA2020136}. \revv{Other two finite differentiation schemes which are not based on the use of modified frequencies were also proposed in \cite{Schneider2016} and \cite{Schneider2017} showing a clear improvement of the spurious oscillations.} A third approach to reduce the noise in the solution was proposed by Eloh et. al, \cite{Eloh2019} who instead of using the DFT as the discrete counterpart of the continuous Fourier transform, considered the continuous Fourier transform of a piecewise constant operator in the real space to derive \emph{consistent periodized discrete Green operators.} 

\rev{The second disadvantage of using FFT homogenization for lattice materials arises when, instead of using direct images or tomographic data to construct the model, the objective is representing smoothly the lattice material boundaries for general geometries defined analytically or through CAD models. In these cases, the voxelized representation can be inaccurate for small number of voxels, not achieving the actual relative density of the lattice. This can lead to local inaccurate results due to the combination of a high phase contrast and a non-smooth interface. This issue has been treated by smoothening the sharp inter-phase using \emph{composite voxels}  \cite{GELEBART2015,Kabel2015}. These approaches are based on defining the material response for the voxels crossed by an internal interphase as the homogenization of the two phases contained.}

Due to these limitations, only a few previous attempts of using FFT to study lattice \rev{materials} can be found on the literature. \rev{In all the cases the motivation was to describe the actual lattice geometry, including imperfections, using data obtained from tomographic images. In \cite{Suard2015}, the augmented Lagrangian approach \cite{Michel2001} was used only as a preprocessing step, using tomographic images of a single strut as data, in order to determine an equivalent diameter of the struts. Then, the simulation of full lattices was done in FEM using ideal geometries with equivalent radius. In \cite{Lhuisser2016} and \cite{Chen2019}, FFT was used to determine directly the response of lattice \rev{materials}. In both cases, the studies were limited to the linear elastic regime and the FFT approaches used did not consider an infinite compliant phase for the empty space, but used a material with small but finite stiffness to achieve convergence instead.  None of these works included an assessment of the accuracy of the FFT approach in order to determine the effect of the artificial stiffness used to represent the empty areas. }

\rev{The present work presents a systematic and critical assessment of the accuracy and efficiency of FFT approaches for predicting the mechanical response of lattice based materials,  \revv{both in the linear and the nonlinear regime, in order to establish an optimal framework for the homogenization of this type of materials.} After a preliminary study, two linear FFT approaches are selected as potential candidates, \revv{including a novel algorithm for RVEs including pores.} In parallel, different geometrical approaches to represent smoothly the lattice geometries are combined with each approach. The accuracy and efficiency of the different combinations of FFT solvers and geometrical approaches are compared against FEM to model both the linear and non linear responses of an ideal octet lattice. Finally, both frameworks are applied to study a real octet lattice cell, including fabrication defects, obtained with 3D-tomography.

\section{FFT homogenization for RVEs with empty areas}

\rev{As previously discussed}, a clear limitation for the use of FFT techniques for homogenization is the deficient convergence rate in the case of microstructures with a large contrast in the phase properties. This problem becomes critical in the case of materials with voids where one of the phases is infinitely compliant. In this case, the problem of the low convergence rate (or no convergence in many cases) is superposed to the singularity of the problem: the solution is not unique since any compatible strain field in the empty phase is admissible. The problem has been circumvented on many occasions by setting a very compliant elastic behavior for the void phase, but this artificial stiffness might have an effect on the cell response, especially for large volume fractions of empty voxels. 

The proper adaptation of FFT homogenization algorithms to account for actual zero stiffness has been studied almost from the first developments of this numerical technique. The augmented Lagrangian approach \cite{Michel2001} is a modification of the original basic scheme \rev{to improve convergence for RVEs containing phases with very low or zero stiffness}. The algorithm is based on the combination of two strain fields,  one of them forced to be compatible, and two stress fields, one of them forced to be in equilibrium. Then, the solution is obtained by the iterative minimization of a Lagrangian. Although the method is in theory able to resolve cases with infinite phase contrast, \rev{the convergence rate becomes really small when controlling a residual which enforces both stress equilibrium and strain compatibility in addition to the macroscopic constraints \cite{MouSilva2014}}. Other acceleration methods also based on using additional fields have been developed to account for infinite phase contrast as \cite{monchiet2012}. Nevertheless,  these methods present a similar convergence rate when small tolerances are imposed for both equilibrium and compatibility \cite{MouSilva2014}. \rev{Another potential method specifically developed for considering voids is the variational approach by Brisard et.al.  \cite{Brisard2010}. Although the idea presented is very smart, the consistent differential operator derived is very complex to compute and the authors themselves do not use it finally,  but an approximate version based on discrete derivatives.} 

As mentioned in the introduction, an alternative (or additional) way of improving the convergence rate under large phase property contrast is using methods that reduce the numerical oscillations. \revv{In particular, the standard continuous differentiation scheme can be replaced by finite difference schemes through different approaches as using staggered grids \cite{Schneider2016} or modifying the Fourier derivative definition with modified frequencies which corresponds to different finite difference stencils \cite{Mueller98,Willot2015}.} These alternative differentiation schemes can be combined with different FFT solvers, improving the convergence of the original ones.

\rev{In this paper the Galerkin FFT approach \cite{Vondrejc2014,Zeman2017} combined with the use of mixed control \cite{Lucarini2019a} and the rotated finite difference scheme \cite{Willot2015} has been chosen as a first candidate for the lattice \rev{materia}l homogenization. This combination has been made after a preliminary study and is based on its very fast convergence rate but also on its relative simplicity and the ability of the method to be efficiently extended for non linear cases. This approach does not break the underdetermination of the solution, but allows to converge in the presence of regions with zero stiffness to an equilibrated stress and compatible strain with relatively low noise.} Note also that any other Krylov based approach with the appropriate reference medium and combined with the same discretization scheme would eventually lead to similar results.

As second candidate, a new method based on a modification of the DBFFT approach \cite{Lucarini2019b}  is proposed. In this method, the standard equilibrium is augmented with additional conditions for the void regions and interfaces \rev{to break the underdetermination leading to a non-singular discrete system of equations. In the next subsections, both methods will be presented including their extension for non linear response.}

\subsection{Galerkin-FFT with discrete differences and mixed loading control}\label{sec:gal}

\rev{The FFT-Galerkin method} was initially developed by Vondr{\v e}jc et al. \cite{Vondrejc2014} to homogenize the elastic behavior of \rev{fully dense heterogeneous materials. This approach presents a very fast convergence rate for limited phase contrast, but is not able by itself to converge in the presence of infinitely compliant phases. In order to extend this scheme for RVEs with empty regions, the original method is adapted by changing the iterative linear solver, introducing an alternative differentiation scheme \cite{Willot2015} and using mixed macroscopic control \cite{Lucarini2019b}. Both the original method and these modifications will be presented below.}

Following \cite{Vondrejc2014}, starting from the weak form of the equilibrium in small strains for a given heterogeneous periodic domain $\Omega$ the next equation can be derived
\begin{equation}\label{eq:gal}
    \mathcal{F}^{-1} \left\{ \widehat{\mathbb{G}} \left( \boldsymbol{\xi} \right) : \mathcal{F} \left\{ \boldsymbol{\sigma} \left( \mathbf{x} \right) \right\} \right\} =\mathbf{0} \text{ ,}
    \end{equation}
where $\mathbf{x}$ represents the spatial position, $\boldsymbol{\xi}$ the spatial frequency vector, $\boldsymbol{\sigma}$ is the Cauchy stress ---determined by the local constitutive equations--- and $\widehat{\mathbb{G}}$ is the Fourier transform of a linear operator which projects any arbitrary tensor field into its compatible part. The Fourier transform and the inverse Fourier transform are represented in eq. \eqref{eq:gal} by $\mathcal{F}$ and $\mathcal{F}^{-1}$ respectively.

In equation (\ref{eq:gal}) the domain is \revv{discretized in} a regular grid in which each voxel center $\mathbf{x}=(x_1,x_2,x_3)$ is given by $$x_p=\frac{L_p}{2 N_p} + \frac{L_p}{N_p} n_p, \text{with}\  n_p=0,\dots, N_p - 1 ; \ p=1,2,3 \text{ ,} $$
where $L_p$ and $N_p$ stand for the length of the cell edge and number of voxels in direction $p$. The discrete form of the equation \eqref{eq:gal} is a linear system of \rev{algebraic equations} in which the unknown is the value of the strain $\boldsymbol{\epsilon}$ at the center of each voxel. The frequency vector $\boldsymbol{\xi}$ is given by 
\begin{equation}\label{eq:freq}
    \xi_p=i   q_p N_p/L_p \quad \text{with} \quad q_p=2 \pi \frac{n_p-N_p/2}{N_p} \in [-\pi,\pi] \text{ ,}
\end{equation}
where $i=\sqrt{-1}$ is the imaginary unit. The Fourier transforms correspond to the direct and inverse Discrete Fourier Transform that are carried out using the FFT algorithm. \rev{The ability of the FFT algorithm to reduce the computational cost of the DFT transforms from $\mathcal{O}(n^2)$ to $\mathcal{O}(n\log n)$ is the main reason behind the high performance of spectral solvers.}

The macroscopic state is provided as a combination of macroscopic strain components $kl$, $\overline{\boldsymbol{\varepsilon}}=\overline{\varepsilon}_{kl} (\mathbf{e}_k\otimes  \mathbf{e}_l)^{sym}$ or macroscopic stress  components $KL$ \rev{$\overline{\boldsymbol{\sigma}}=\overline{\sigma}_{KL} (\mathbf{e}_K\otimes  \mathbf{e}_L)^{sym}$} with $kl \cap KL = \emptyset$. For these general loading conditions \cite{Lucarini2019a} the expression of the projector operator in the Fourier space follows
\begin{equation}\label{eq:galop}
\widehat{G}_{ijkl}(\boldsymbol{\xi}) = \left\{ 
\begin{array}{lr}
I^s_{ijKL} &   \text{if }\boldsymbol{\xi}=\mathbf{0} \text{ for components  $KL$ }  \\ 
0_{ijkl} &  \text{if }\boldsymbol{\xi}=\mathbf{0} \text{ for components $kl$} \\
0_{ijkl} & \text{for Nyquist frequencies} \\
\left[ I^s_{ijkl} \xi_j \xi_l \right]^{-1}\xi_j \xi_l   & \text{for } \boldsymbol{\xi}\ne \mathbf{0}
\end{array} 
\right.  \text{ ,}
\end{equation}
where $\mathbb{I}^s$ is the fourth order symmetric \rev{identity tensor and $\widehat{\mathbb{G}}$ accounts for \rev{major} and minor symmetries.}

In the case of linear elastic phases eq. (\ref{eq:gal}) yields  
\begin{equation}\label{eq:gallin}
    \mathcal{F}^{-1} \left\{ \widehat{\mathbb{G}} \left( \boldsymbol{\xi} \right) : \mathcal{F} \left\{ \mathbb{C} \left( \mathbf{x} \right) : \boldsymbol{\varepsilon} \left( \mathbf{x} \right) \right\} \right\} = - \mathcal{F}^{-1} \left\{ \widehat{\mathbb{G}} \left( \boldsymbol{\xi} \right) : \mathcal{F} \left\{ \mathbb{C} \left( \mathbf{x} \right) : \overline{\boldsymbol{\varepsilon}} \left( \mathbf{x} \right) - \overline{\boldsymbol{\sigma}} \right\} \right\} \text{ ,}
\end{equation}
where $\mathbb{C}$ is the local fourth order stiffness tensor. Equation \eqref{eq:gallin} represents a linear system of equations in which the left-hand side is a symmetric \rev{semidefinite positive} linear operator acting on a discrete strain field
\begin{equation}
 \mathcal{A}(\bullet)= \mathcal{F}^{-1} \left\{ \widehat{\mathbb{G}} \left( \boldsymbol{\xi} \right) : \mathcal{F} \left\{ \mathbb{C} \left( \mathbf{x} \right) : \bullet \right\} \right\} \label{eq:linearOp}
 \end{equation}
and the right-hand side is the independent term $\mathbf{b}$
$$\mathbf{b} = - \mathcal{F}^{-1} \left\{ \widehat{\mathbb{G}} \left( \boldsymbol{\xi} \right) : \mathcal{F} \left\{ \mathbb{C} \left( \mathbf{x} \right) : \overline{\boldsymbol{\varepsilon}} \left( \mathbf{x} \right) - \overline{\boldsymbol{\sigma}} \right\} \right\}.$$
The system $ \mathcal{A}(\boldsymbol{\varepsilon})=\mathbf{b}$ can be solved efficiently using the \rev{Conjugate Gradient method} (CG) method for domains with a relatively low stiffness contrast. The residual of this system of equations is defined as the \rev{$L_2$ norm} of the difference between the linear operator applied on the candidate solution and the right-hand side over the norm of the right-hand side,
\begin{equation}\label{eq:galres}
r_{lin}=\frac{\left\| \mathcal{A}(\boldsymbol{\varepsilon} ) - \mathbf{b}  \right\|_{L_2}}{\left\| \mathbf{b} \right\|_{L_2}}
\end{equation}
reaching the solution when $r_{lin}$ is below a tolerance. \rev{Note that eq. \eqref{eq:gallin} is undetermined \revv{independently of} the phase properties because any incompatible field $\boldsymbol{\varepsilon}^{INC}$ added to the solution still fulfils eq. \eqref{eq:gallin},
 $$\mathcal{A}(\boldsymbol{\varepsilon} +\boldsymbol{\varepsilon}^{INC}) = \mathcal{A}(\boldsymbol{\varepsilon})=\mathbf{b}.$$
 Nevertheless, the solution of eq. \eqref{eq:gallin} is unique in the subspace of compatible strain fields, so the CG is able to handle the underdetermination and recovers the unique compatible solution.}

In the case of lattices, there are empty regions in the RVE, which do not transfer stresses and that should be properly accounted for with zero stiffness. \revv{The solution of the problem in the full RVE is singular even in the subspace of compatible strain fields, because any compatible strain field 
which is zero outside of the empty region can be added to the solution providing the same equilibrated stress.} This singularity is transferred to the numerical method increasing the underdetermination of the system also to compatible fields. As a result, it is observed that the CG method is not able to reach convergence and the \rev{Minimal Residual Method (MINRES)}, an alternative Krylov subspace solver able to handle efficiently singular systems, is used as linear solver to ovecome this limitation.} 

Finally, in order to improve both smoothness of the solution and convergence rate, an alternative discretization scheme is used, the rotated forward finite difference rule \cite{Willot2015}. This discretization is introduced through modified frequencies in the Fourier derivation leading to an alternative projection operator (eq. \ref{eq:galop}). The modified frequencies correspond to
\begin{equation}\label{eq:freqrot}
    \xi'_p=  i 2 \frac{N_p}{L_p} \tan{ \left( q_p/2 \right) }
    \prod_{p = 1}^{d} \frac{1}{2} \left( 1 + e^{ i q_p}\right)   \quad \text{with} \quad q_p=2 \pi \frac{n_p-N_p/2}{N_p} \in [-\pi,\pi] \text{ ,}
\end{equation}
where $d=1,2$ or 3 is the space dimension. 
\revv{
It is important to remark that standard Galerkin approach does not converge using standard Fourier discretization and a conjugate gradient solver. On the contrary, the combination of the alternative differentiation scheme and the use of MINRES allows the Galerkin approach to reach a solution in a relatively small number of iterations, as it will be shown in the numerical results.} 
\subsubsection{Non linear extension}\label{sec:galerkinnl}

In order to take into account material non linearities on the Galerkin FFT formulation \rev{the macroscopic strain/stress history is applied} as function of the time (or pseudo-time for rate independent materials) in several time increments. The non linear equilibrium at each increment is solved iteratively using the Newton-Raphson method, as proposed in \cite{Zeman2017}.  If the solution at time $t$ and the macroscopic stress and strain components applied at $t+\Delta t$, $\overline{\boldsymbol{\varepsilon}}^{t+\Delta t}, \overline{\boldsymbol{\sigma}}^{t+\Delta t}$ are known, the non linear equation at time $t+\Delta t$ is linearized at each iteration $i$ around the strain field at previous iteration $\boldsymbol{\varepsilon}^{i-1}$.  Let  $\delta\boldsymbol{\varepsilon}\left( \mathbf{x} \right)$ be the strain field correction to be obtained at iteration $i$, then the linearized stress corresponds to 
\begin{equation}\label{eq:Slin}
    \boldsymbol{\sigma}^i \left( \mathbf{x} \right) = \boldsymbol{\sigma}^{i-1} \left( \mathbf{x} \right)+ \left. \frac{\partial \boldsymbol{\sigma} \left( \mathbf{x} \right)}{\partial \boldsymbol{\varepsilon}}\right|_{\epsilon = \epsilon^{i-1}} : \delta\boldsymbol{\varepsilon} \left( \mathbf{x} \right)=
 \boldsymbol{\sigma}^{i-1} \left( \mathbf{x} \right)+ \mathbb{C}^{i-1} \left( \mathbf{x} \right) : \delta\boldsymbol{\varepsilon} \left( \mathbf{x} \right) \text{ ,}
\end{equation}
where $\mathbb{C}^{i-1}$ is the material consistent tangent evaluated using the solution of the previous  iteration $i-1$. The equilibrium equation linearized at $\boldsymbol{\varepsilon}^{i-1}$ reads
\begin{equation}\label{eq:galnonlin}
    \mathcal{F}^{-1} \left\{ \widehat{\mathbb{G}} \left( \boldsymbol{\xi} \right) : \mathcal{F} \left\{ \mathbb{C}^{i-1} \left( \mathbf{x} \right) : \delta \boldsymbol{\varepsilon} \left( \mathbf{x} \right) \right\} \right\} =
    - \mathcal{F}^{-1} \left\{ \widehat{\mathbb{G}} \left( \boldsymbol{\xi} \right) : \mathcal{F} \left\{ \boldsymbol{\sigma} \left( \boldsymbol{\varepsilon}^{i-1} \left( \mathbf{x} \right) \right)  - \overline{\boldsymbol{\sigma}}^{t+\Delta t} \right\} \right\} \text{ ,}
\end{equation}
\rev{where the solution at the previous time step enters in the first iteration} as $\boldsymbol{\varepsilon}^{0}\left( \mathbf{x} \right) = \boldsymbol{\varepsilon}^{t} \left( \mathbf{x} \right)+ \overline{\boldsymbol{\varepsilon}}^{t+\Delta t}-\overline{\boldsymbol{\varepsilon}}^{t}$. 

\rev{The left-hand side of equation \eqref{eq:galnonlin} corresponds to a linear operator $\mathcal{A}_i$, acting on the correction strain field  $\delta\boldsymbol{\varepsilon}$, that is equivalent to the one defined in \rev{eq. \eqref{eq:linearOp}} using the tangent stiffness at iteration $i-1$ instead of the elastic one}. Similarly, the right-hand side forms an independent vector $\mathbf{b}_i$. The solution of the non linear problem \rev{at} each time increment is \rev{obtained} solving the linear equation \eqref{eq:galnonlin} for each Newton iteration, and adding the successive \rev{solution corrections} until the convergence is reached. 

Special care has to be \rev{paid} to the definition of the residuals in the non linear case. As in the original approach \cite{Zeman2017}, two residuals are \rev{proposed} for the non linear solver, but they are redefined in order to avoid \rev{oversolving}.  
The linear residual controls the accuracy of the solution in the linear system resulting from each Newton iteration. \rev{This linear system $\mathcal{A}_i(\delta\boldsymbol{\varepsilon})=\mathbf{b}_i$ is solved iteratively up to a given tolerance and the non linear algorithm becomes an inexact damped Newton method \cite{Wicht2020}.}   \rev{The standard non-linear approach for Galerkin FFT establish the convergence criterion on the relative residual,} defined as $$\frac{\left\|\mathcal{A}_i\left(\mathbf{x}\right)-\mathbf{b}_i\right\|_{L_2}}{\left\|\mathbf{b}_i\right\|_{L_2}} \text{ ,}$$ \rev{where the norm used is the $L_2$ norm}. However in this case, since the right-hand-side $\mathbf{b}_i$ changes at each Newton iteration and should converge towards $\mathbf{0}$, the relative residual becomes too restrictive near \rev{the solution} and results in additional meaningless linear iterations. To avoid this problem, the norm in the first iteration $\left\|\mathbf{b}_0\right\|$ is used here to normalize the residual for the rest of the Newton iterations. The \rev{convergence criterion} for the linear solver can be then rewritten as
\begin{equation}\label{eq:galrlin}
 r_{lin}=\frac{\left\| \mathcal{A}_i(\delta \boldsymbol{\varepsilon})-\mathbf{b}_i\right\|_{L_2}}{\left\|\mathbf{b}_0\right\|_{L_2}}.
\end{equation}
Note that if the increment is linear, this expression corresponds to eq. \eqref{eq:galres}. Moreover, since $\mathbf{b}_i$ is zero for an equilibrated stress field, the ratio $\|\mathbf{b}_i\|/\left\|\mathbf{b}_0\right\|$ is a relative measure of the internal equilibrium \rev{and as a result} the number of iterations required for the linear solver decreases with the number of Newton iterations. The second residual is the Newton residual for the non linear equation. It is defined as the ratio between the infinity norm of the last deformation gradient correction and the infinity norm of the change in the total strain field within each time increment 
\begin{equation}\label{eq:galrnewton}
    r_{newton}=\frac{\left\|\delta \boldsymbol{\varepsilon}\right\|_{\infty}}{\left\|\boldsymbol{\varepsilon}^{i}-\boldsymbol{\varepsilon}^{t}\right\|_{\infty}}.
\end{equation}
The solution is accepted only when both of the \rev{residuals (eqs. \eqref{eq:galrlin} and \eqref{eq:galrnewton}) are below their respective tolerances.  \revv{Note that although the choice of the Newton forcing term has not been optimized (as proposed in \cite{Wicht2020}), it is observed that when a consistent tangent is used the number of Newton iterations per strain increment was quite small, always less than five}.}
\subsection{Modified displacement based FFT for infinite contrast (MoDBFFT)}

A modification of the displacement based FFT approach presented in \cite{Lucarini2019b}, called from now on MoDBFFT for brevity, is proposed here for simulating lattice materials. \rev{The objective is to derive a method that presents accurate results and a good convergence rate for infinite phase contrast maintaining standard continuum discretization and derivation in the Fourier space, without the need of using alternative discretization approaches and modified frequencies.} 

The starting point of the method is the strong formulation of the conservation of linear momentum on a periodic domain which can be divided into two sub-domains (see Figure \ref{sfig:scheme1}) representing the two phases: $\Omega_m$, the matrix, and $\Omega_v$, the void. The interphase between the two phases is $\Gamma$.

\begin{figure}[h]\centering
\begin{subfigure}{.49\linewidth}\centering
\includegraphics[width=0.9\linewidth]{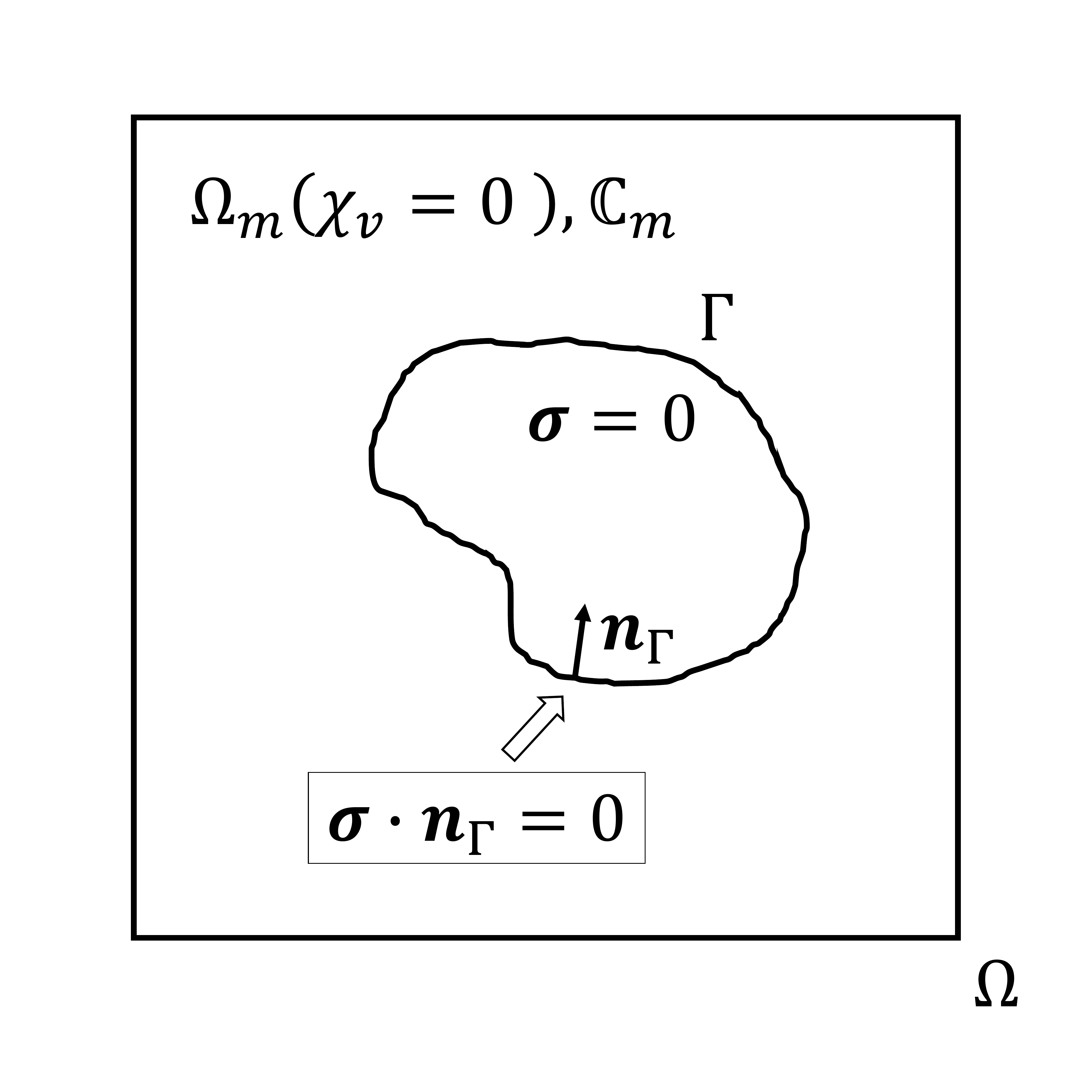}
\caption{Two domain scheme}\label{sfig:scheme1}
\end{subfigure}
\begin{subfigure}{.49\linewidth}\centering
\includegraphics[width=0.9\linewidth]{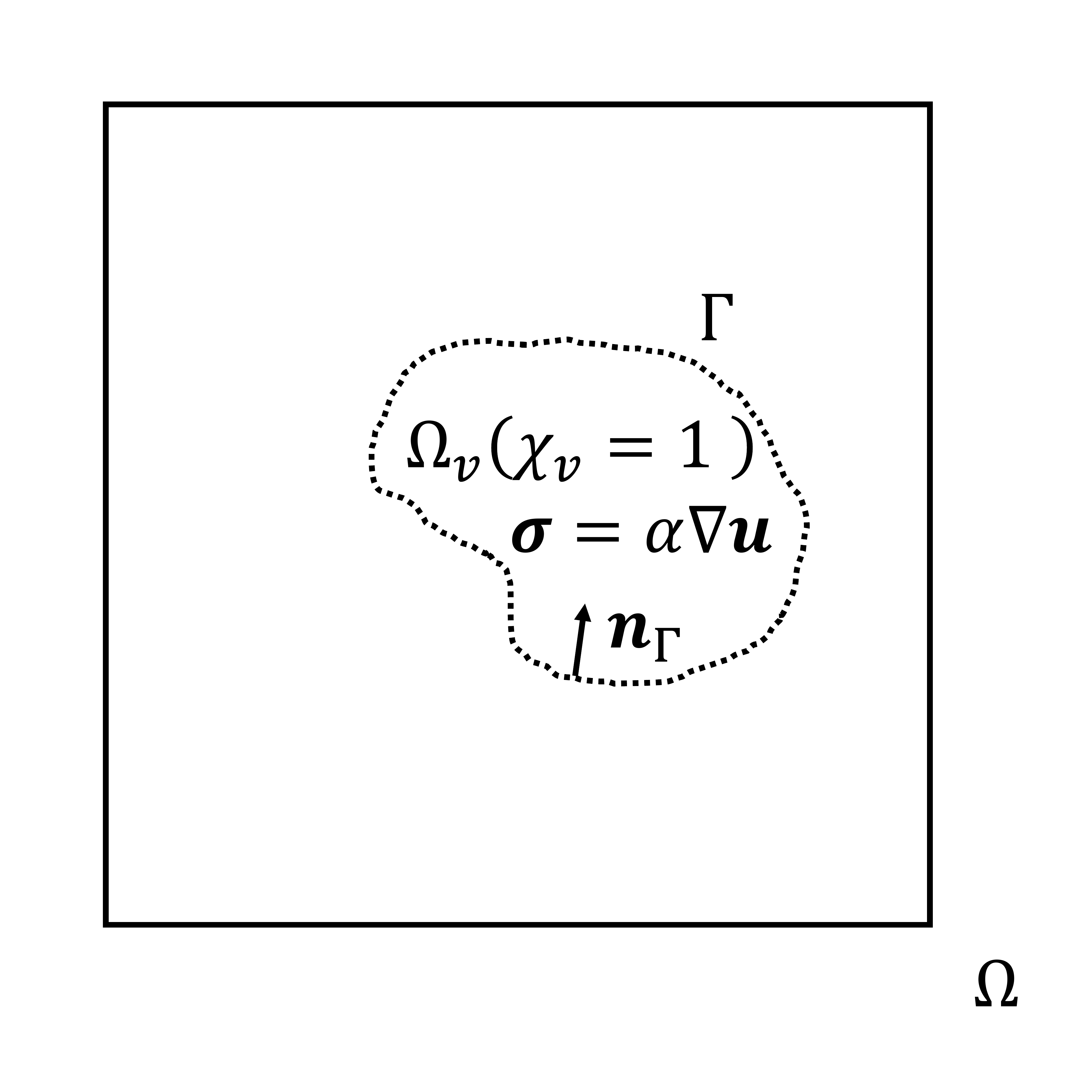}
\caption{MoDBFFT scheme}\label{sfig:scheme2}
\end{subfigure}
\caption{Schematic representation of the domains for the standard FFT and the present approach}\label{fig:schemes}
\end{figure}

\rev{The real boundary value problem is defined only in the domain $\Omega_m$ and corresponds to finding the displacement field $\mathbf{u} \in \Omega_m$ such that}
\begin{equation}\label{eq:strongr}
\left\{ \begin{array}{c}
\nabla \cdot \boldsymbol{\sigma}\left( \nabla^s \mathbf{u} \left( \mathbf{x} \right) \right) = \mathbf{0} \quad \text{in} \quad \Omega_m \\
\boldsymbol{\sigma}(\mathbf{x}) \cdot \mathbf{n}_\Gamma = \mathbf{0} \ \text{in} \ \mathbf{x} \in \Gamma \\
\boldsymbol{\sigma} \ \text{and} \ \boldsymbol{\varepsilon} \ \text{periodic}\ \text{in} \ \Omega 
\end{array}
\right. .
\end{equation}
Due to the periodicity in $\Omega$, the weak formulation of this problem with free Neumann boundary conditions on $\Gamma$ is simply
\begin{equation}
\int_{\Omega_m} \boldsymbol{\sigma}(\mathbf{u} ):\nabla^s \delta \mathbf{u} \ \mathrm{d} \Omega_m = 0\text{ ,}
\label{eq:weak_orig}
\end{equation}
where  \rev{$ \mathbf{u}$ and $ \delta \mathbf{u}$ are the trial and virtual displacement fields respectively.} 

\rev{FFT methods need to resolve the fields in the full cell, including the points in $\Omega_v$ where no material exists and fields are \revv{not defined}. On the contrary, the balance equation \eqref{eq:weak_orig} is not defined \revv{in the entire unit cell} $\Omega$ so any compatible strain field is acceptable for the stress equilibrium in the region $\Omega_v$  and the system is underdetermined.}

In the new approach, the standard formulation of the two-phase domain is modified (see Figure \ref{sfig:scheme2}) in order to extend the weak formulation to every point of the domain $\Omega$ , including the free Neumann conditions. To this aim, first, an artificial elastic energy density is defined in $\Omega_v$ which aims to prevent the indeterminacy of the displacement in that region. This energy density depends on the square of the displacement gradient is weighted by a numerical parameter $\alpha$, \revv{with stress dimensions} and which represents an artificial stiffness in the interior of $\Omega_v$. The first variation of the total energy in $\Omega_v$ leads to a new term in the weak formulation which is defined only in that region and corresponds to
\begin{equation}
\int_{\Omega_v} \alpha \nabla^s \mathbf{u}: \nabla^s \delta \mathbf{u} \ \mathrm{d} \Omega_v  \text{ ,}
\label{eq:weakalphar}
\end{equation}
where $\alpha$ is assumed to be small. Note that when $\alpha = 0$, the original underdetermined formulation is recovered. 

To extend the weak formulation of the original problem (eq. \ref{eq:weak_orig}) in $\Omega_v$ to the full domain $\Omega$,  the indicator function of the voided region $\chi_v(\mathbf{x})$ which defines the microstructure, is introduced (eq. \ref{eq:indicator})
\begin{align}\label{eq:indicator}
    \chi_v  =
    \begin{cases}
      1 & \forall \mathbf{x}\in \Omega_v\\
      0 & \forall \mathbf{x}\in \Omega_m\\
    \end{cases} .
\end{align}
The eqs. \eqref{eq:weak_orig} and \eqref{eq:weakalphar}, defined in $\Omega_m$ and $\Omega_v$ respectively, are then premultiplied by their corresponding indicator functions to be defined in a unique domain $\Omega$, leading to
\rev{
\begin{eqnarray}
\label{eq:weak1}
\int_\Omega (1-\chi_v)  \boldsymbol{\sigma}(\mathbf{u}):\nabla^s \delta \mathbf{u}  + \chi_v \  \alpha  \ \nabla^s \mathbf{u}:\nabla^s \delta \mathbf{u} \ \mathrm{d} \Omega  = 0.
\end{eqnarray}}

Using the divergence theorem and removing boundary terms because of the periodicity, the corresponding strong formulation of eq. \eqref{eq:weak1} follows
$$
\nabla \cdot \left( 1-\chi_v \left( \mathbf{x} \right) \right) \boldsymbol{\sigma}\left( \nabla^s \mathbf{u} \left( \mathbf{x} \right) \right) + \nabla \cdot \chi_v \left( \mathbf{x} \right) \alpha\left( \nabla^s \mathbf{u} \left( \mathbf{x} \right) \right) = 0
$$
and if the chain rule is applied to the previous equation and terms are regrouped, the result is
\begin{equation}\label{eq:split}
 \left( 1-\chi_v \right) \nabla \cdot \boldsymbol{\sigma}+\chi_v  \nabla \cdot \alpha\left( \nabla^s \mathbf{u} \left( \mathbf{x} \right) \right) +\nabla \chi_v \cdot ( \alpha\left( \nabla^s \mathbf{u} \left( \mathbf{x} \right) \right) - \boldsymbol{\sigma} )
 = 0.
\end{equation}

Second, the stress free condition at the interphase $\Gamma$ of the boundary value problem, defined in the second equation of eq. (\ref{eq:strongr}) should be imposed in equation \eqref{eq:split}. To this aim, the boundary condition on $\Gamma$  is diffused over a thin volume which is defined by the surface $\Gamma$ with an infinitesimal thickness \cite{To2020}.  This translation can be done using the surface delta function $\delta(\mathbf{x})_\Gamma $, defined as \cite{Lange2012}
\begin{equation}
\delta(\mathbf{x})_\Gamma =   \mathbf{n}_\Gamma  \cdot \nabla \chi_v (\mathbf{x})\label{eq:delta} \text{ ,}
\end{equation}
where the direction of $ \mathbf{n}_\Gamma$ is represented in Fig. \ref{sfig:scheme2}. The gradient of the indicator function vanishes everywhere except near the surface $\Gamma$, where it points in the normal direction \cite{Lange2012}. Therefore, multiplying the previous equation by the normal vector leads to
\begin{equation}
  \mathbf{n}_\Gamma  \delta(\mathbf{x})_\Gamma = \nabla \chi_v (\mathbf{x}).\label{eq:delta2}
\end{equation} 


\rev{Using the surface delta (eqs. \ref{eq:delta} and \ref{eq:delta2}), the stress free condition can be expressed as a volume integral as}
\begin{equation}\label{eq:stressfree}
\int_\Gamma \boldsymbol{\sigma}(\mathbf{x}) \cdot \mathbf{n}_\Gamma \mathrm{d}\Gamma = 
\int_\Omega \boldsymbol{\sigma}(\mathbf{x}) \cdot \mathbf{n}_\Gamma  \delta(\mathbf{x})_\Gamma  \mathrm{d}\Omega =
\int_\Omega \boldsymbol{\sigma}(\mathbf{x}) \cdot \nabla \chi_v \mathrm{d}\Omega .
\end{equation}

In order to apply the stress free condition, eq. \eqref{eq:stressfree} has to be incorporated to eq. \eqref{eq:split}. Since for a non-vanishing $\alpha$ and continuous displacement field in the surface $\Gamma$ it is fulfilled
$$
 \boldsymbol{\sigma} \cdot \mathbf{n} = \alpha \left( \nabla^s \mathbf{u} \left( \mathbf{x} \right) \right) \cdot \mathbf{n} \ \ \text{in} \  \Gamma$$
then, the volume counterparts of any of these terms are also identical and any of them can be set to zero in eq. \eqref{eq:split}. 
 $$\boldsymbol{\sigma}(\mathbf{x}) \cdot \nabla \chi_v  =\alpha \left( \nabla^s \mathbf{u} \left( \mathbf{x} \right) \right) \cdot  \nabla \chi_v \ \text{in} \ \Omega.$$
Choosing for simplicity $\alpha \left( \nabla^s \mathbf{u} \left( \mathbf{x} \right) \right)  \cdot  \nabla \chi_v =0$ leads to the final expression of the strong formulation

\begin{equation}\label{eq:strongnew}
    \nabla \cdot \left( 1-\chi_v \left( \mathbf{x} \right) \right) \boldsymbol{\sigma}\left( \nabla^s \mathbf{u} \left( \mathbf{x} \right) \right)  + \chi_v \left( \mathbf{x} \right) \nabla \cdot \alpha \nabla^s \mathbf{u} \left( \mathbf{x} \right)  = \mathbf{0} 
\end{equation}
together with periodic boundary conditions in $\nabla^s \mathbf{u}$.

\rev{In order to impose periodicity conditions in the strain field, the displacement field is split into two contributions as
\begin{equation}
\mathbf{u}(\mathbf{x})=\widetilde{\mathbf{u}}(\mathbf{x})+\overline{\boldsymbol{\varepsilon}}\cdot\mathbf{x} \text{ ,}
\end{equation}
where $\widetilde{\mathbf{u}}(\mathbf{x})$, the fluctuation of the displacement field which is periodic and has zero average, becomes the new unknown of the problem}. If this new formulation is particularized for a linear elastic matrix using as input a general mixed macroscopic state as a macroscopic strain $\overline{\boldsymbol{\varepsilon}}$ or macroscopic stress $\overline{\boldsymbol{\sigma}}$, the final equation to solve the fluctuations in the displacement reads
\begin{equation}\label{eq:modbfft}
    \nabla \cdot \left[ \left( 1-\chi_v \left( \mathbf{x} \right) \right) \mathbb{C} \left( \mathbf{x} \right) : \left( \nabla^s \widetilde{\mathbf{u}} \left( \mathbf{x} \right) + \overline{\boldsymbol{\varepsilon}}_{\overline{\boldsymbol{\sigma}}} \right) \right] + \chi_v \left( \mathbf{x} \right) \nabla \cdot \left( \alpha \nabla^s \widetilde{\mathbf{u}} \left( \mathbf{x} \right) \right) =\\
    - \nabla \cdot \left( \mathbb{C} \left( \mathbf{x} \right) : \overline{\boldsymbol{\varepsilon}} - \overline{\boldsymbol{\sigma}} \right) \text{ ,}
\end{equation}
where the field $\widetilde{\mathbf{u}}$ is the unknown, that has to be solved together with the $IJ$ components of the overall strain tensor,  $\overline{\boldsymbol{\varepsilon}}_{\overline{\boldsymbol{\sigma}}}$, which are conjugate of the applied macroscopic stress. For this last unknown, an extra equation is used linking the $IJ$ components of $\overline{\boldsymbol{\varepsilon}}_{\overline{\boldsymbol{\sigma}}}$ with the corresponding components of \revv{ the macroscopic stress,
\begin{equation}\label{eq:modbfft2}
\frac{1}{V_\Omega} \int_{\Omega} \left[ \mathbb{C} \left( \mathbf{x} \right) : \left( \nabla^s \widetilde{\mathbf{u}} \left( \mathbf{x} \right) + \overline{\boldsymbol{\varepsilon}}_{\overline{\boldsymbol{\sigma}}} \right) \right]_{IJ} \mathrm{d} \Omega = - \frac{1}{V_\Omega} \int_{\Omega} \left[ \mathbb{C} \left( \mathbf{x} \right) : \overline{\boldsymbol{\varepsilon}} \right]_{IJ} \mathrm{d} \Omega + \overline{\boldsymbol{\sigma}}_{IJ}.
\end{equation}
}

As usual in FFT methods, the differential operators can be defined by their Fourier space counterparts using spatial frequencies as
\begin{gather}\label{eq:diffops}
    \nabla^s (\bullet) = \mathcal{F}^{-1} \left\{ \mathcal{F} \left\{\bullet \right\} \otimes^s \boldsymbol{\xi} \right\}= \mathcal{F}^{-1} \left\{ \frac{1}{2}\left( \mathcal{F} \left\{\bullet \right\} \otimes \boldsymbol{\xi} + \boldsymbol{\xi} \otimes \mathcal{F} \left\{\bullet \right\} \right) \right\}\\
    \nabla \cdot (\bullet) = \mathcal{F}^{-1} \left\{   \mathcal{F}\left\{\bullet \right\} \cdot \boldsymbol{\xi} \right\}
\end{gather}

To solve \eqref{eq:modbfft} in the Fourier space, the spatial domain is discretized in a standard regular grid (section \ref{sec:gal}). The result is a linear system of equations that now is fully determined. \rev{ Nevertheless, to solve the system iteratively, the use of a preconditioner is unavoidable for a competitive convergence rate}. The linear operator \rev{$\mathbb{M} \left( \mathbf{x} \right) (\bullet)$}, proposed for the DBFFT approach \cite{Lucarini2019b}, is also used here as preconditioner \rev{  
\begin{equation}\label{eq:prec}
    \mathbb{M} (\bullet)= \left[ \nabla \cdot \overline{\mathbb{C}} \ \nabla \right]^{-1} (\bullet) \text{ ,}
\end{equation}}
where $\overline{\mathbb{C}}$ is the volume averaged stiffness tensor,
\begin{equation}\label{eq:avstiff}
    \overline{\mathbb{C}}=\frac{1}{V_\Omega} \int_{\Omega} \mathbb{C}\left(\mathbf{x}\right) \mathrm{d}\Omega
\end{equation}
and $V_\Omega$ represents the volume of the entire domain $\Omega$.

\revv{The resulting equilibrium equation written in Fourier space yields 
\begin{equation}\label{eq:modbfftf}
\mathcal{F}\left\{ \left( 1-\chi_v\right) \mathbb{C} : \left( \mathcal{F}^{-1}\left\{\widehat{\widetilde{\mathbf{u}}} \otimes^s \boldsymbol{\xi}\right\} + \overline{\boldsymbol{\varepsilon}}_{\overline{\boldsymbol{\sigma}}} \right) \right\} \cdot \boldsymbol{\xi} +  \mathcal{F}\left\{\chi_v \mathcal{F}^{-1}\left\{\alpha \widehat{\widetilde{\mathbf{u}}} \otimes^s \boldsymbol{\xi} \cdot \boldsymbol{\xi} \right\} \right\} = -  \mathcal{F}\left\{ \mathbb{C} : \overline{\boldsymbol{\varepsilon}} - \overline{\boldsymbol{\sigma}} \right\} \cdot \boldsymbol{\xi} \text{ ,}
\end{equation}
and the Fourier transform of the equation to impose the macroscopic stress components  $IJ$ (eq.\eqref{eq:modbfft2}) corresponds to
\begin{equation}\label{eq:averstrain_F}
\mathcal{F}\left\{ \left( 1-\chi_v\right) \mathbb{C} : \left( \mathcal{F}^{-1}\left\{\widehat{\widetilde{\mathbf{u}}} \otimes^s \boldsymbol{\xi}\right\} + \overline{\boldsymbol{\varepsilon}}_{\overline{\boldsymbol{\sigma}}} \right) \right\} \left( \mathbf{0} \right) = 
\mathcal{F}\left\{ \mathbb{C} : \overline{\boldsymbol{\varepsilon}} - \overline{\boldsymbol{\sigma}} \right\} \left( \mathbf{0} \right) \text{ .}
\end{equation}}
\revv{The equations (\ref{eq:modbfftf},\ref{eq:averstrain_F}) are linear and can be solved iteratively using a Krylov solver. To improve the convergence, a preconditioner $\widehat{\mathbb{M}}$ is used for eq. \eqref{eq:modbfftf}, which is the Fourier transform of the preconditioner defined in eq. \eqref{eq:prec}, and is given by
\begin{equation}\label{eq:precf}
    \widehat{\mathbb{M}}  (\ast)=
    \left[ \boldsymbol{\xi} \cdot \overline{\mathbb{C}} \cdot \boldsymbol{\xi} \right]^{-1} \cdot \ast \text{ ,}
\end{equation}
where $\ast$ represents a complex valued vector defined in the Fourier space for all non-zero frequencies. For preconditioning eq. \eqref{eq:averstrain_F}, the inverse of the volume averaged stiffness tensor (eq. \ref{eq:avstiff}) is directly used.}
\revv{Contrary to the Galerkin approach using standard Fourier differentiation, the linear system of equations defined in eq. \eqref{eq:modbfftf} is non-singular, and therefore the Conjugate Gradient method is able to converge efficiently and provide the solution of the system.} This is a potential benefit of this approach \rev{with respect to} the modified Galerkin which relies on the use of the more \revv{memory demanding} MINRES solver.

For the implementation, the problem unknowns are joined forming a vector composed of the fluctuating displacement field and the components of the macroscopic strain where the stress is imposed (eq. \ref{eq:averstrain_F}), $\left\{ \begin{array}{c}\widehat{\widetilde{\mathbf{u}} } \\ \overline{\boldsymbol{\varepsilon}}_{\overline{\boldsymbol{\sigma}}} \end{array}\right\}$. The left-hand side of \revv{eqs. \eqref{eq:modbfftf} and \eqref{eq:averstrain_F}} can be expressed as a linear \revv{operator that acts over the composed vector } following
\revv{\begin{equation}\label{eq:modbfftfop}
\widehat{\mathcal{A}}^\alpha\left( \left\{ \begin{array}{c}\widehat{\widetilde{\mathbf{u}} } \\ \overline{\boldsymbol{\varepsilon}}_{\overline{\boldsymbol{\sigma}}} \end{array}\right\} \right) = \left\{ 
\begin{matrix} 
\mathcal{F}\left\{ \left( 1-\chi_v\right)\mathbb{C} : \left(  \mathcal{F}^{-1}\left\{\widehat{\widetilde{\mathbf{u}} } \otimes^s \boldsymbol{\xi}\right\} + \overline{\boldsymbol{\varepsilon}}_{\overline{\boldsymbol{\sigma}}} \right) \right\} \cdot \boldsymbol{\xi} +
\mathcal{F}\left\{\chi_v \mathcal{F}^{-1}\left\{\alpha \widehat{\widetilde{\mathbf{u}} } \otimes^s \boldsymbol{\xi} \cdot \boldsymbol{\xi} \right\} \right\}\\ 
\mathcal{F}\left\{ \left( 1-\chi_v\right) \mathbb{C} : \left( \mathcal{F}^{-1}\left\{\widehat{\widetilde{\mathbf{u}} } \otimes^s \boldsymbol{\xi}\right\} + \overline{\boldsymbol{\varepsilon}}_{\overline{\boldsymbol{\sigma}}} \right) \right\}  \left( \mathbf{0} \right)
\end{matrix} \right\}
\end{equation}}
and the right-hand side can be written as a vector \revv{reading as
\begin{equation}\label{eq:modbfftfrhs}
\widehat{\mathbf{b}}^\alpha=\left\{
\begin{matrix} 
- \mathcal{F}\left\{ \mathbb{C} : \overline{\boldsymbol{\varepsilon}} - \overline{\boldsymbol{\sigma}} \right\} \cdot \boldsymbol{\xi} \\
\mathcal{F}\left\{ \mathbb{C} : \overline{\boldsymbol{\varepsilon}} - \overline{\boldsymbol{\sigma}} \right\} \left( \mathbf{0} \right)
\end{matrix} \right\}
\end{equation} }
It should be remarked that the linear operator (eq. \ref{eq:modbfftfop}) has a significantly higher computational cost (around 1.5x) compared to the linear operator of the Galerkin method (eq. \ref{eq:galop}) \rev{since it requires performing the additional Fourier transforms of a vector field in the extra term}. The equilibrium is reached when a linear residual defined as
\revv{
\begin{equation}\label{eq:modbfftresres}
r_{lin}=\frac{\left\| \widehat{\mathcal{A}}^\alpha\left(\left\{ \begin{array}{c}\widehat{\widetilde{\mathbf{u}} } \\ \overline{\boldsymbol{\varepsilon}}_{\overline{\boldsymbol{\sigma}}} \end{array}\right\} \right) - \widehat{\mathbf{b}}^\alpha  \right\|_{L_2}}{\left\| \widehat{\mathbf{b}}^\alpha \right\|_{L_2}}
\end{equation}}
is lower than a given tolerance.

\subsubsection{Non linear extension}\label{sec:modbfftnl}

In the case of non linear materials, a linearization of equation \eqref{eq:strongnew} is done similarly to the Galerkin method (Section \ref{sec:galerkinnl}). \rev{The stresses and strains are linearized following eq. \eqref{eq:Slin}}. The non linear problem is divided into time increments \rev{and an iterative Newton method is used at each time increment to solve the problem. The linearization at each Newton iteration leads to a system of equations
\begin{multline}\label{eq:modbfftfnonlin}
 \mathcal{F}\left\{ \left( 1-\chi_v\right) \mathbb{C}^{i-1} : \left( \mathcal{F}^{-1}\left\{\widehat{\delta\widetilde{\mathbf{u}}} \otimes \boldsymbol{\xi}\right\} + \delta\overline{\boldsymbol{\varepsilon}}_{\overline{\boldsymbol{\sigma}}} \right) \right\} \cdot \boldsymbol{\xi} +
 \mathcal{F}\left\{\chi_v \mathcal{F}^{-1}\left\{\alpha \widehat{\delta\widetilde{\mathbf{u}}} \otimes \boldsymbol{\xi} \cdot \boldsymbol{\xi} \right\} \right\}\\
 = - \mathcal{F}\left\{ \left( 1-\chi_v\right) \boldsymbol{\sigma} \left( \mathcal{F}^{-1}\left\{\widehat{\widetilde{\mathbf{u}}}^{i-1} \otimes \boldsymbol{\xi}\right\} + \overline{\boldsymbol{\varepsilon}}_{\overline{\boldsymbol{\sigma}}}^{i-1} + \overline{\boldsymbol{\varepsilon}}^{t+\Delta t} \right) - \overline{\boldsymbol{\sigma}}^{t+\Delta t} \right\} \cdot \boldsymbol{\xi}\\
 - \mathcal{F}\left\{ \chi_v \mathcal{F}^{-1}\left\{\alpha \widehat{\widetilde{\mathbf{u}}}^{i-1} \otimes \boldsymbol{\xi} \cdot \boldsymbol{\xi} \right\} \right\}
\end{multline}
in which the displacement correction and the average strain correction, $\widehat{\delta\widetilde{\mathbf{u}}}$ and $ \delta\overline{\boldsymbol{\varepsilon}}_{\overline{\boldsymbol{\sigma}}}$, are the unknowns and the solution for iteration $i$ is updated as} 
$$\widetilde{\mathbf{u}}^i=\widetilde{\mathbf{u}}^{i-1}+\delta\widetilde{\mathbf{u}}; \quad \overline{\boldsymbol{\varepsilon}}_{\overline{\boldsymbol{\sigma}}}^i=\overline{\boldsymbol{\varepsilon}}_{\overline{\boldsymbol{\sigma}}}^{i-1}+\delta\overline{\boldsymbol{\varepsilon}}_{\overline{\boldsymbol{\sigma}}}.$$

In equation \eqref{eq:modbfftfnonlin} the macroscopic prescribed strain and stress fields enter in the definition of the first iteration as
 $\widetilde{\mathbf{u}}^{0}=\widetilde{\mathbf{u}}^{t}$ and $\overline{\boldsymbol{\varepsilon}}_{\overline{\boldsymbol{\sigma}}}^{0}=\overline{\boldsymbol{\varepsilon}}_{\overline{\boldsymbol{\sigma}}}^{t}$.   The linear equation can be translated into a linear operator applied to the unknown \revv{$\widehat{\mathcal{A}}^\alpha_i\left( \left\{ \begin{array}{c} \delta\widehat{\widetilde{\mathbf{u}}} \\ \delta\overline{\boldsymbol{\varepsilon}}_{\overline{\boldsymbol{\sigma}}} \end{array}\right\} \right)$} and an independent right-hand side vector $\widehat{\mathbf{b}}^\alpha_i$, both defined in the $i$-th iteration. \revv{At each Newton iteration, the preconditioner given by eq. \eqref{eq:prec} is recomputed using the average tangent stiffness.}

Analogous to the Galerkin scheme, two residuals are used to solve the non linear problem. The linear solver residual is defined as
\revv{
\begin{equation}\label{eq:modbfftresresnl}
r_{lin}=\frac{\left\| \widehat{\mathcal{A}}_i^\alpha\left(\left\{ \begin{array}{c} \delta\widehat{\widetilde{\mathbf{u}}} \\ \delta\overline{\boldsymbol{\varepsilon}}_{\overline{\boldsymbol{\sigma}}} \end{array}\right\}\right) - \widehat{\mathbf{b}}_i^\alpha  \right\|_{L_2}}{\left\| \widehat{\mathbf{b}}_0^\alpha \right\|_{L_2}}
\end{equation}}
which normalizes the absolute error in the linear problem by the norm of the right-hand side vector of the first Newton iteration. The second residual is the Newton residual, which is formulated in strains and is identical to the one used in the Galerkin FFT in eq. \eqref{eq:galrnewton}.

\section{Geometrical adaptation}\label{sec:geoadapt}

FFT methods rely on a regular discretization of a \rev{hexahedral} domain in which the cell of the lattice material is embedded. This voxelized representation allows a simple generation of the microstructure and the direct use of image/tomographic data but presents two disadvantages when considering an ideal cell, especially for coarse discretizations. First, the final density represented can \rev{differ slightly} from the designed one and, second, the voxelized representation of the struts might impact the overall behavior of the lattice. 

In order to alleviate these problems, \rev{the properties assigned} to the voxels near the strut surface can be adapted to better capture the smoothness of the surfaces. In this section, different approaches for determining the properties of the voxels near the interfaces are presented. \rev{The geometry definition of the cross section of a circular strut is represented schematically in Fig. \ref{fig:geomsmooth}.}

\subsection{Plain Voxelized representation}

The plain voxelized (PV) geometry approach is the most simple representation: it assigns lattice material or empty space to a voxel based on whether the center of that voxel is inside or outside of the lattice struts. This method generates sharp boundaries for the struts and the resulting relative densities can slightly differ from the designed one in the case of coarse discretizations.

\subsection{Phase-field smoothening}\label{sec:pfs}

The phase-field smoothening (PFS) method consists in creating a smooth property transition, from the lattice material properties to zero, across the lattice interfaces. The property profile is dictated by the minimization of a phase-field functional and is controlled by a smoothening length scale $\ell$. The result of the phase-field minimization is a phase map $\phi(\mathbf{x})$ that assigns to each voxel of the cell a phase value between 0 (empty space) and 1 (lattice material). Similar to damage models \cite{Lemaitre1985}, the stress resulting from applying the constitutive equation of the pristine material is multiplied by the value of the phase map at that point.

Let $\omega(\mathbf{x})$ be a function which represents the exact geometry of the lattice, and which in this case corresponds to the indicator function of the lattice material, \rev{$\omega(\mathbf{x})=1-\chi_v(\mathbf{x})$}  (eq. \ref{eq:indicator}) where the value of $1$ is attributed to the points belonging to the material and $0$ to the empty space. Then, the value of $\phi(\mathbf{x})$ is obtained as the minimizer of the functional defined by eq. \eqref{eq:phase}.
\begin{equation}\label{eq:phase}
    E[\phi]=\int_\Omega \frac{1}{2}\ell^2 \| \nabla \phi \|^2 +\frac{\epsilon}{2}(\phi-\omega)^2 \mathrm{d}\Omega \text{ ,}
\end{equation}
where $\ell$ is the characteristic length, a parameter which defines the width of the smoothening region and $\epsilon$ is a weight which penalizes the difference between the exact geometry $\omega$ and its smoothened counterpart $\phi$. The functional assumes periodicity of all the fields. The first term of the functional penalizes the smoothened area and it is modulated by the length of the diffusion and the second term penalizes the difference between the initial field and the smoothened one. The result of this minimization corresponds to the solution of the partial differential equation described in \revv{eq.} \eqref{eq:phase2}
\begin{equation}\label{eq:phase2}
    \frac{\ell}{\epsilon}^2\nabla^2\phi-\phi = -\omega
\end{equation}
under periodic boundary conditions in $\phi$. If the problem is discretized in a regular grid and  $\omega(\mathbf{x})$ is replaced by its discrete counterpart, defined by the value of the indicator function at the center of each voxel of the grid, the equation can be explicitly \rev{solved on} the Fourier space as
\begin{equation}\label{eq:phase3}
    \widehat{\phi}\left(\boldsymbol{\xi}\right)=\frac{1}{1-\frac{\ell^2}{\epsilon}\boldsymbol{\xi}\cdot\boldsymbol{\xi}}\widehat{\omega}\left(\boldsymbol{\xi}\right) \text{ ,}
\end{equation}
where the frequency vector $\boldsymbol{\xi}$ is given in  eq. \eqref{eq:freq} and all the fields involved are periodic. 

The numerical implementation of this approach is done using two different discretization levels. \rev{In a pre-processing step a very fine grid is used with voxel size $d_{fine}$ to discretize the real geometry and solve the phase-field problem to obtain $\phi(\mathbf{x})$. During the simulation of the mechanical problem a coarser grid is used, $d_{coarse}$.} The ratio between $d_{coarse}/d_{fine}$ is between \rev{2} and 20. The fine discretization is used for having an accurate representation of the indicator function $\omega(\mathbf{x})$. This fine grid is used for solving eq. (\ref{eq:phase2}) using as characteristic length of diffusion $\ell$  the half of the length of the voxel in the coarse discretization ($\ell=0.5 d_{coarse}$) which \rev{encompasses} several voxels of the fine grid. Finally, the resulting field $\phi(\mathbf{x})$ in the fine grid is averaged for each voxel of the coarse discretization to define the phase map to be used during the mechanical simulations. Note that the values of the phase field below 5\% are taken as 0, to prevent the spurious presence of material detached \revv{from} the truss. In all the phase-field smoothening cases th e characteristic weight $\epsilon$ is taken equal to 1.

\subsection{Voigt approaches}\label{sec:lbs}
%
\rev{The third approach to smooth out the lattice surfaces is based on the Voigt rule, following \cite{GELEBART2015}. Under this approach, the stiffness of the voxels partially occupied by the lattice material is obtained using the Voigt homogenization approach. This rule \revv{establishes} that the effective elastic stiffness of that composite voxel is the volume average of the stiffness of the materials present in the voxel. Therefore, since the stiffness of the empty phase is zero, the effective stiffness corresponds to the product of the volume fraction of lattice material in the voxel, $\phi$, multiplied by the stiffness of the lattice material ($\mathbb{C}_m$) 
$$\mathbb{C} = \phi \ \mathbb{C}_m.$$The volume fraction $\phi \in [0,1]$ is then equivalent to a phase map, as the one generated using phase-field smoothening.
In \cite{GELEBART2015},  
to compute the volume fraction of each phase contained in the voxels partially occupied by different phases, it was proposed the use of a subgrid to count the number of points in the subgrid belonging to each phase. We have followed this approach first, using the same ratio for coarse and fine grids used in PFS, and have named this approach as Voigt fine grid (VFG). In parallel, since the surface of the lattice is known either by its mathematical expression or by an .stl file, we propose an alternative way to compute the phase map $\phi$ which does not require the use of a second grid and is just based on the distance of the voxel center to the lattice surface. This approach is named Voigt analytic (VA) smoothening. The method assigns  $\phi=1$ or $\phi=0$ to the voxels \revv{whose} centers are respectively inside or outside of the geometry considering an offset of $\ell/2$ with respect to the boundary of the struts $\Gamma$. For those voxels centers \revv{whose} distance to the boundary is smaller than $\ell/2$ , the value of $\phi$ is obtained as a linear function of the signed distance of that center to the boundary, $D$, following eq. \eqref{eq:phaselin},
\begin{equation}\label{eq:phaselin}
    \phi\left(\mathbf{x}\right) =1- \frac{D+\ell/2}{\ell} \quad \text{for} \quad D=\left\{\begin{array}{c} d(\mathbf{x},\Gamma) \ \text{if} \  \mathbf{x} \in \Omega_m  \\ -d(\mathbf{x},\Gamma) \ \text{if} \   \mathbf{x} \in \Omega_v\end{array} \right.  \text{ ,}
\end{equation}
where $d$ denotes the distance between a point and a surface.The characteristic length considered, $\ell$, is the length of one voxel. Note that this definition of $\phi$ corresponds exactly to the volume fraction of lattice material in the case of a planar interface. Since the strut curvature is normally small \revv{with reference to} the voxel length, the values obtained using eq. \eqref{eq:phaselin} are almost identical to the VFG in the case of fine discretizations. This equivalence has been assessed quantitatively and the phase map generated using the Voigt rule with a finer grid (the same used for the phase-field smoothening) was almost identical to the one obtained by VA (average difference below 0.2\%). For clarity, both methods are only considered for coarse grids while for finer discretization, where the results are almost identical, only the results of VA are represented. The practical benefit of the VA definition of $\phi$ with respect to \cite{GELEBART2015} is that the smoothening is obtained by an analytical expression using the exact geometry and not requiring the use of a smaller grid.}
\subsection{Combined smoothening}

This approach (CS) consists in applying the phase-field smoothening (Section \ref{sec:pfs}) to the coefficients map resulting from the \rev{Voigt analytic} method (Section \ref{sec:lbs}). The combination of these two approaches will result in a very smoothened phase map of weight coefficients that will be multiplied by the stresses within the different algorithms.

\begin{figure}[ht!]\centering
\begin{subfigure}{.24\linewidth}\centering\captionsetup{width=.8\linewidth}
\includegraphics[width=0.8\linewidth]{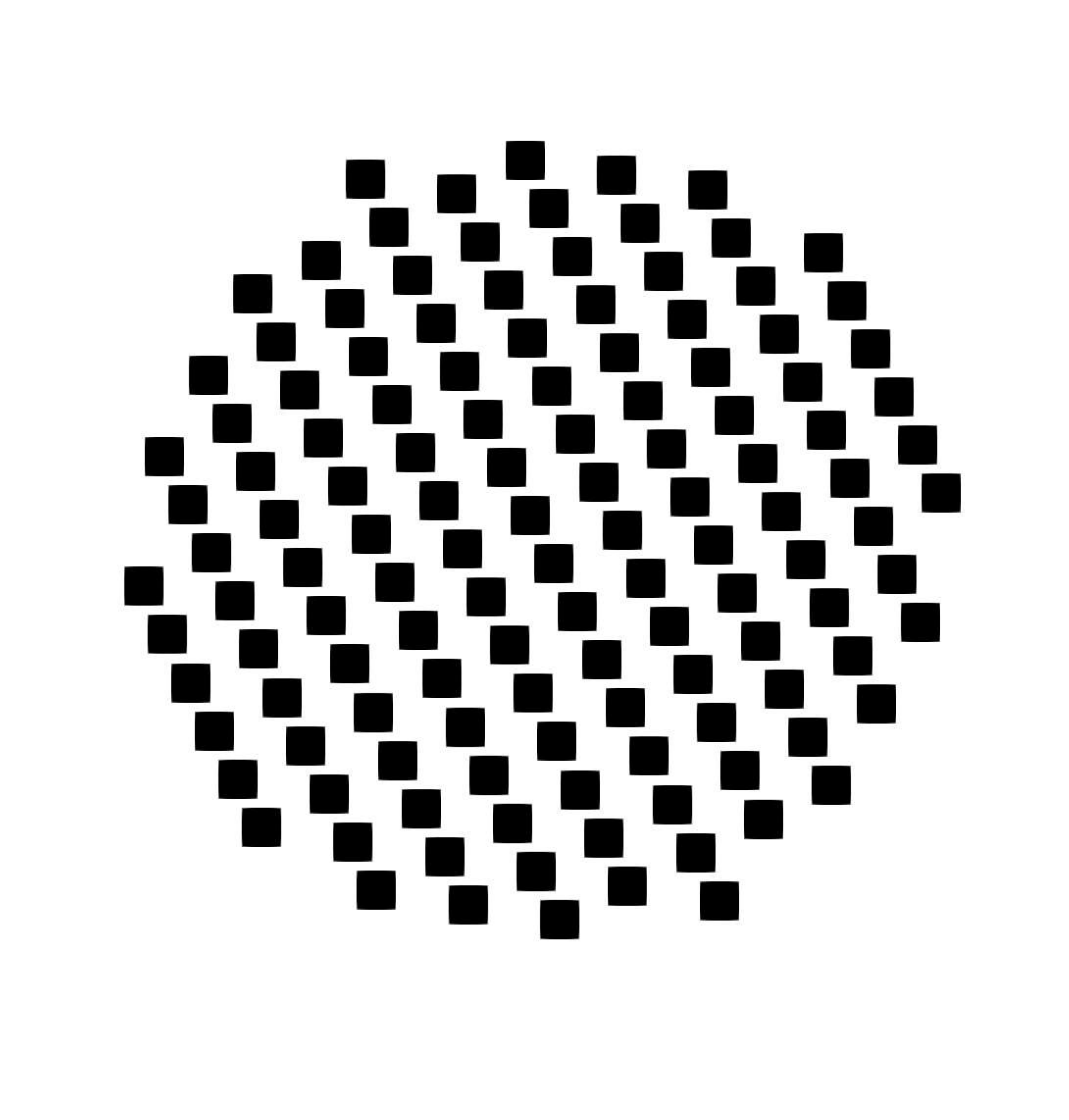}
\caption{Plain voxelized representation (PV).}\label{sfig:pure}
\end{subfigure}
\begin{subfigure}{.24\linewidth}\centering\captionsetup{width=.8\linewidth}
\includegraphics[width=0.8\linewidth]{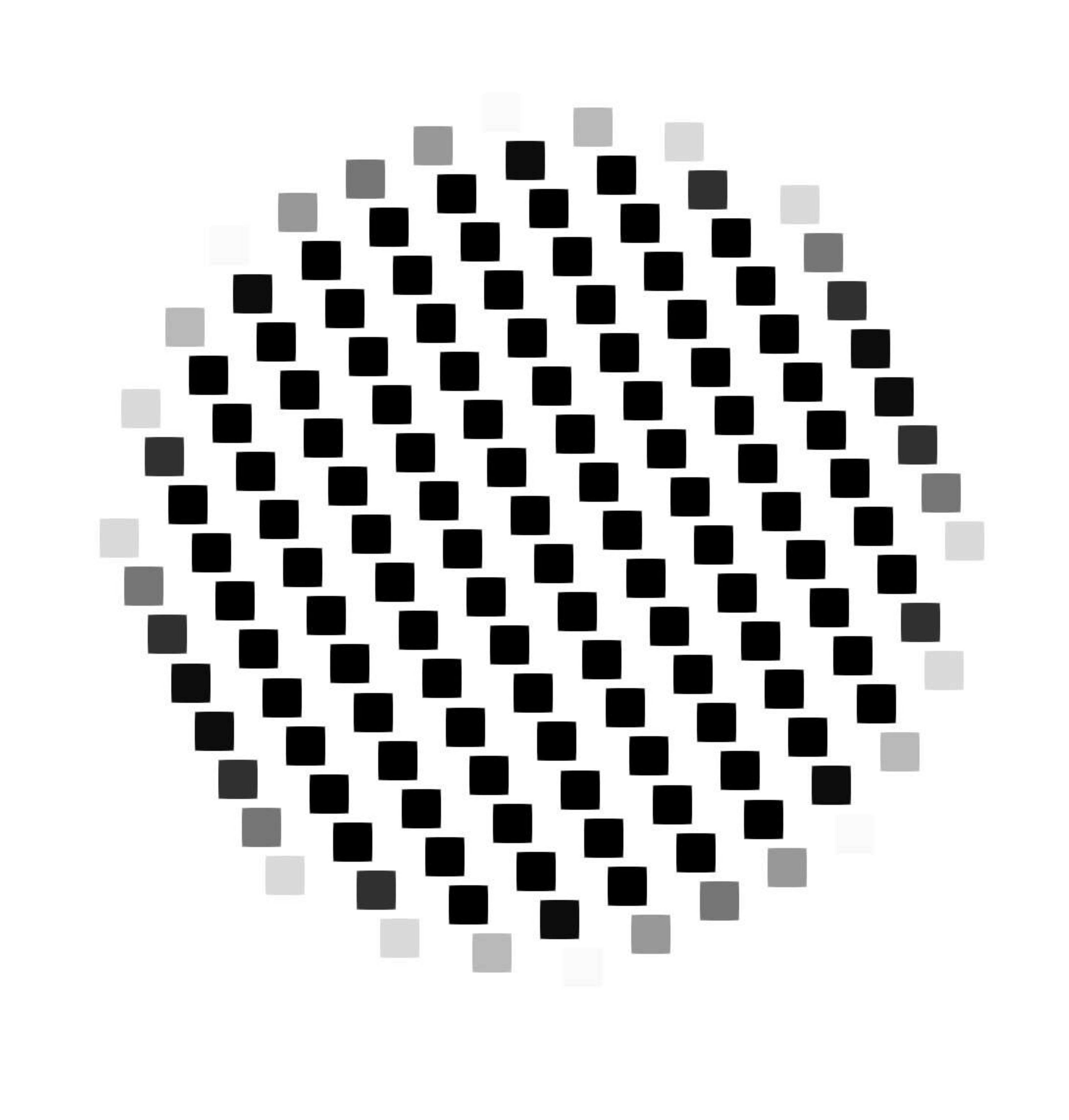}
\caption{Voigt approaches (VA and VFG).}\label{sfig:linear}
\end{subfigure}
\begin{subfigure}{.24\linewidth}\centering\captionsetup{width=.8\linewidth}
\includegraphics[width=0.8\linewidth]{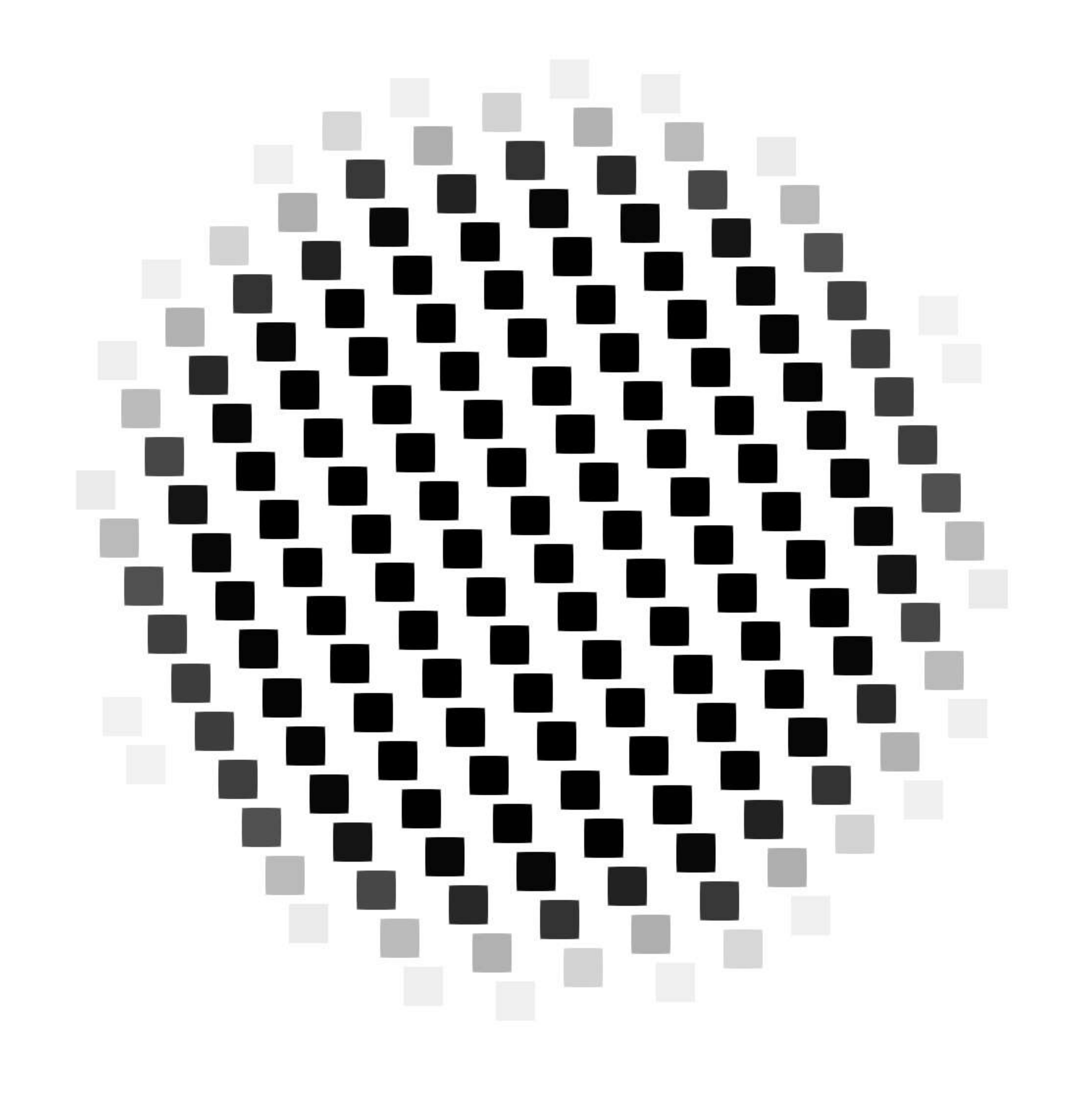}
\caption{Phase-field smoothening (PFS).}\label{sfig:phasefield}
\end{subfigure}
\begin{subfigure}{.24\linewidth}\centering\captionsetup{width=.8\linewidth}
\includegraphics[width=0.8\linewidth]{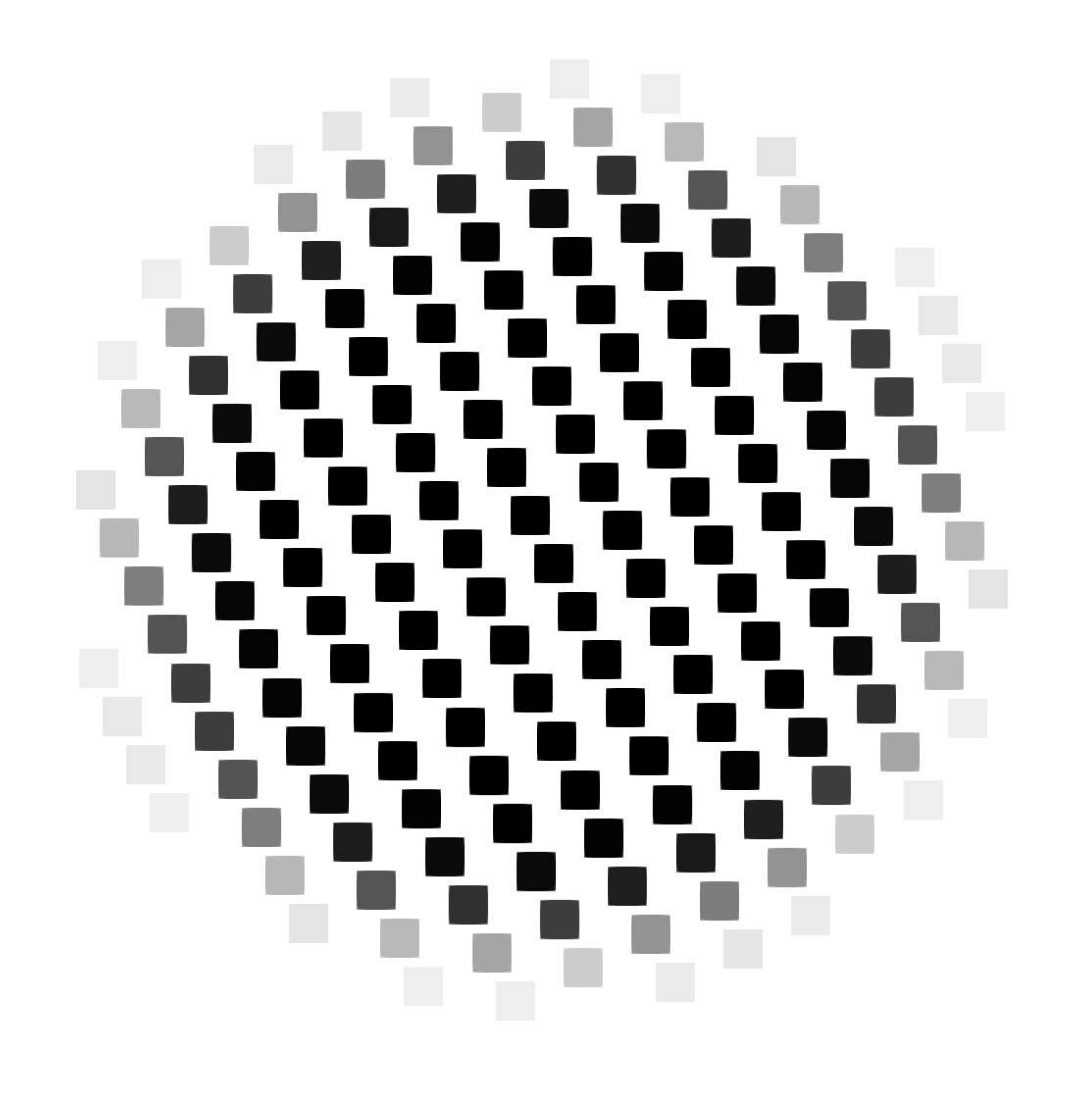}
\caption{Combined smoothening (CS).}\label{sfig:combined}
\end{subfigure}
\caption{Local densities (phase maps $\phi$) near the cross section of a strut for the different geometrical approaches. In the grey scale $\phi =1$ corresponds to black and 0 to white.}\label{fig:geomsmooth}
\end{figure}
\section{Validation for elastic materials}

The numerical methods proposed to homogenize the mechanical behavior of lattice based materials and their combination with the different geometrical representations are studied in this section for elastic materials. In order to evaluate the accuracy and efficiency of the two FFT methods, several numerical tests have been carried out, and both the macroscopic result and microscopic fields have been compared with  FEM simulations. \rev{For the microscopic solution, the} relative $L_2$ norm of the difference between the local fields of the solution in a given method, $f(\mathbf{x})$ compared to a reference solution ($f_{ref}(\mathbf{x})$) is used as metric of the error
\begin{equation}\label{eq:difff}
\text{Local\ diff.}\ \left[ \% \right]=\frac{\left\| f-f_{ref} \right\|_{L_2} }{\left\| f_{ref} \right\|_{L_2}}\text{ .}
\end{equation}


\subsection{Lattice geometry and simulation parameters}

The octet-truss lattice has been selected for the numerical studies. This structure is one of the most interesting lattice based materials since it presents both bending-dominated and stretching-dominated responses, depending on the strut thicknesses and the \rev{loading} conditions. Relative densities ranging from $0.5$\% to $30$\% are considered.

All the FFT simulations are performed using the FFTMAD code \cite{LUCARINI2019_CM}, to which the new algorithms have been added.  \rev{FEM simulations are done using the commercial code ABAQUS to serve} as reference solutions in order to assess the accuracy and efficiency of FFT approaches. The FEM model consists of a geometrically conforming mesh of quadratic tetrahedral elements (C3D10 in ABAQUS). The element size is controlled by the number of elements occupying the lattice strut diameter and is taken to be equal to the corresponding  FFT voxel size. An iterative solver (CG) has been selected to carry out a fair comparison with FFT approaches which are based on the same type of iterative solver. Periodic boundary conditions are used and introduced using multipoint linear constraints \cite{Segurado2002}.

An example of a FEM mesh  \rev{of the unit cell of an octet lattice with 10\% relative density is represented in Figure \ref{fig:geos} together with two FFT voxelized models of the same geometry with different discretization levels}
\begin{figure}[ht!]\centering
\includegraphics[width=0.32\linewidth]{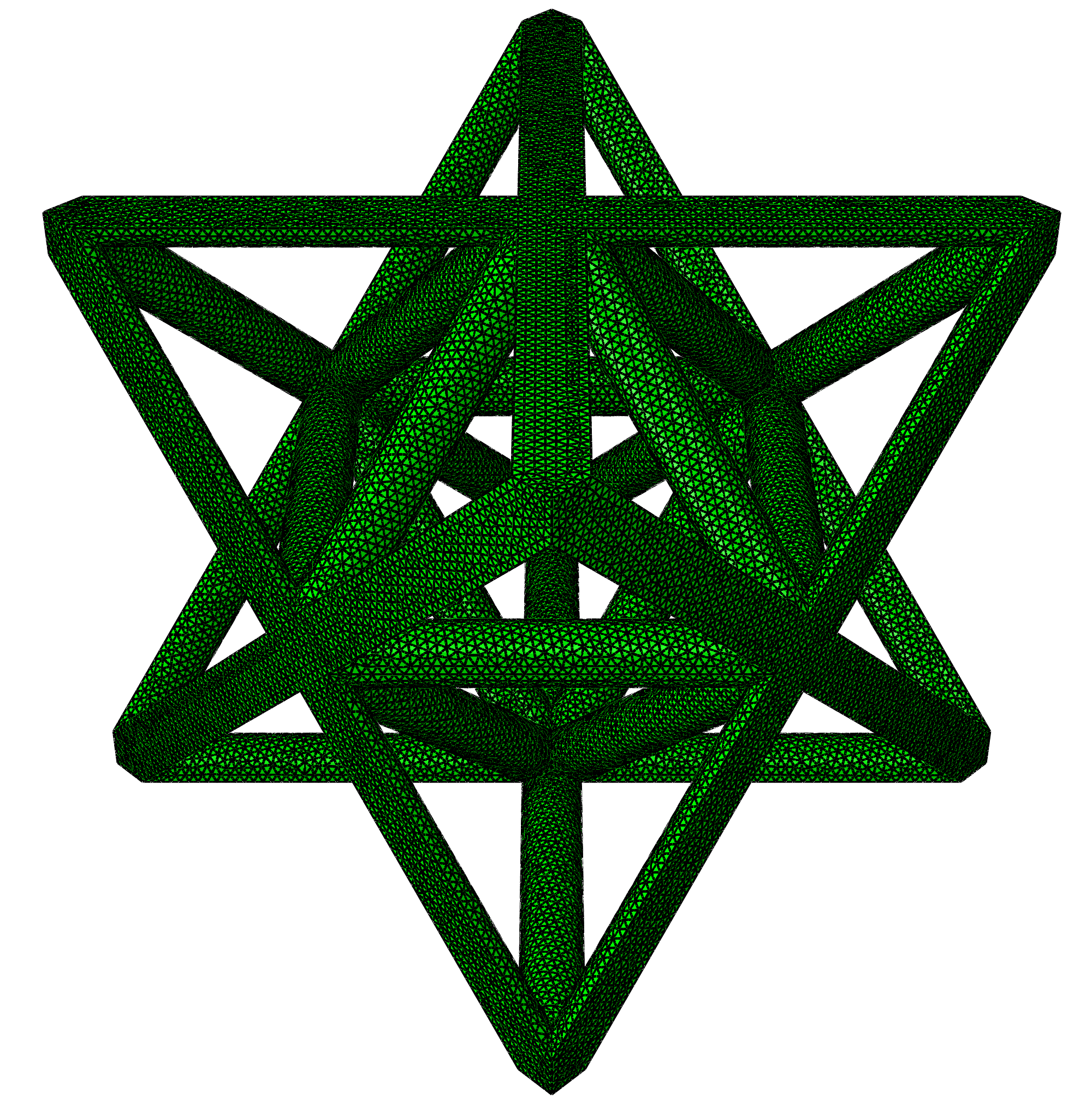}
\includegraphics[width=0.32\linewidth]{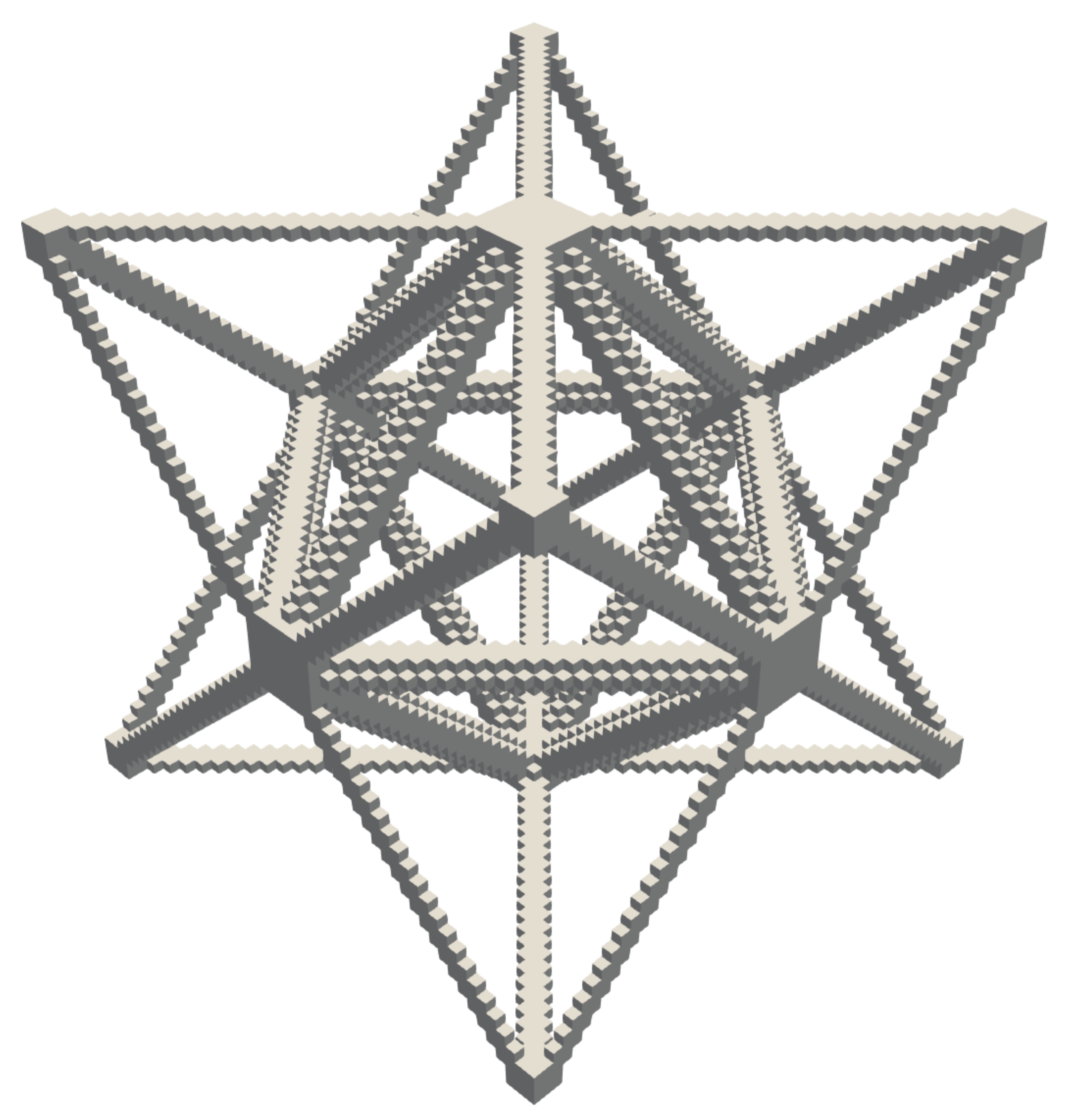}
\includegraphics[width=0.32\linewidth]{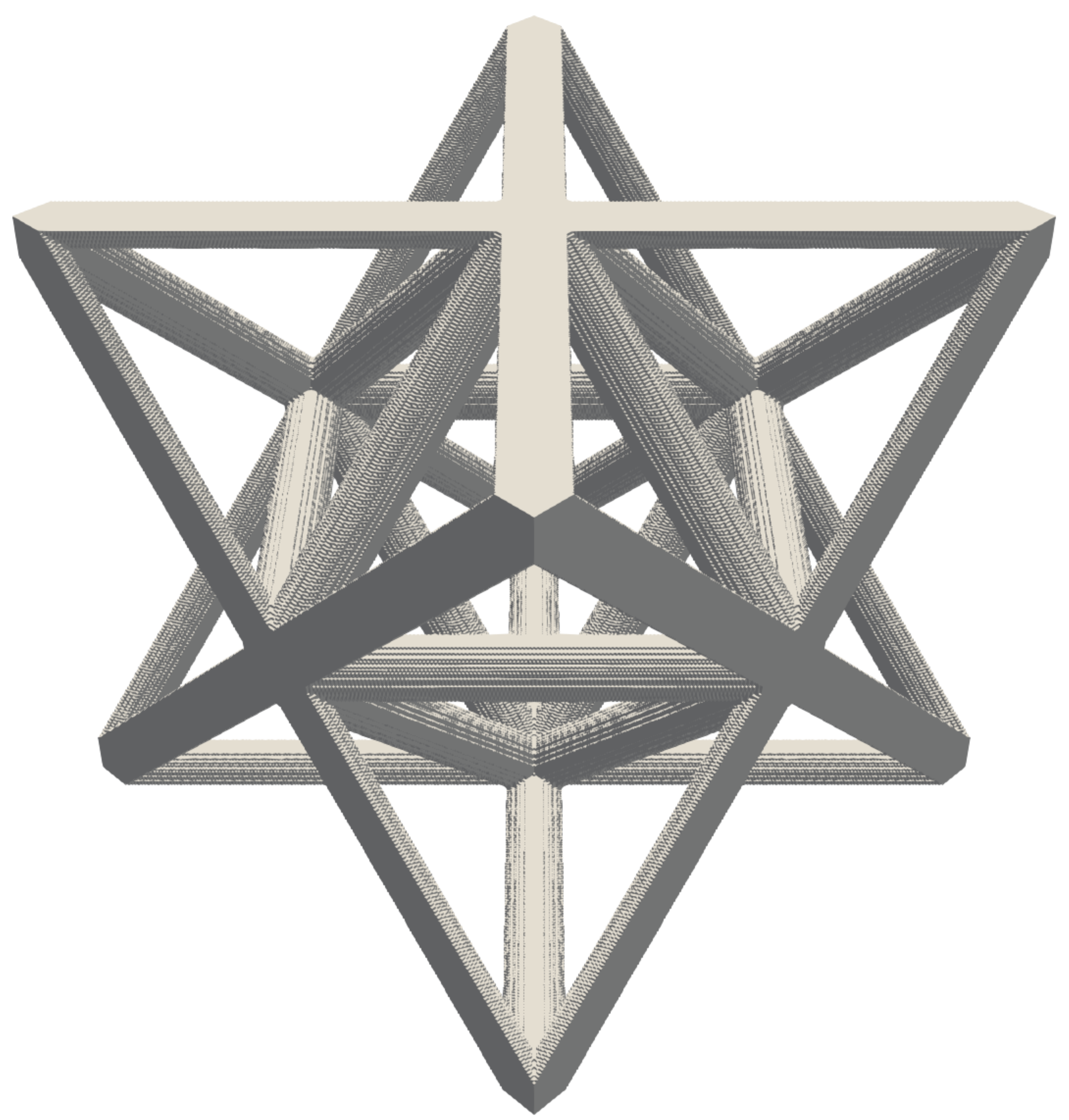}
\caption{Octet lattice with 10\% of relative density, FEM model with 15 elements per diameter and FFT discretizations with $54^3$ and $269^3$ voxels }\label{fig:geos}
\end{figure}
All simulations have been done in a single node workstation Dual 10 core Intel(R) Xeon(R) CPU E5-2630 v4 @ 2.20GHz with 64GB RAM memory. Both ABAQUS and FFTMAD use parallelization by threading (20 threads) both for the evaluation of the constitutive equations and for linear algebra operations.

Regarding the elastic properties of the material, the parameters correspond to a typical polyamide PA12 with isotropic linear elastic behavior with $E=1.7$GPa and $\nu=0.4$. The numerical tolerances for the relative errors in the linear iterative solvers in the Galerkin FFT, MoDBFFT and FEM have been set to $10^{-6}$. 

\subsection{Analysis of the numerical performance of FFT approaches}

In this section, the convergence rate of the different adaptations of FFT approaches for infinite phase contrast will be studied. To this aim, the evolution of the residual of the linear iterative solvers of eq. \eqref{eq:gallin} for the adapted Galerkin FFT (with modified frequencies and use of MINRES) and of eq. \eqref{eq:modbfftf} for the MoDBFFT (using standard frequencies and CG) will be compared.
 
Although the MoDBFFT method \rev{results in} a fully determined system of linear equations with a unique solution, the well-posedness of the resulting coefficient matrix depends on the numerical parameter  $\alpha$ . The system becomes ill-posed when $\alpha$ decreases and gets very low values with respect to the stiffness of the material domain. On the other hand, larger values of $\alpha$ induce an artificial stiffness in the BVP that affects the computed effective properties. \rev{In this case, although the null traction at the void interface is still considered explicitly by eqs. (\ref{eq:delta2}-\ref{eq:strongnew}), the overall results are slightly affected. This effect is only relevant for large values of $\alpha$ and is due to the components of the stress tensor which are not contained in the surface traction. Therefore, those components which are not directly cancelled by eqs. (\ref{eq:delta2}-\ref{eq:strongnew}) are indirectly affecting the overall response.}

In this study, a linear elastic test under uniaxial tension has been simulated for different alphas in a $\overline{\rho}=0.1$ relative density octet-truss lattice \rev{discretized using}  $215^3$ voxels (20 voxels/diameter) \revv{and PV representation}. Figure \ref{fig:alpha} represents the relative residual values for the different linear equilibrium equations and different $\alpha$ values, being $E$ is the Young's modulus of the \rev{lattice material}. 
\begin{figure}[ht!]\centering
\includegraphics[width=0.49\linewidth]{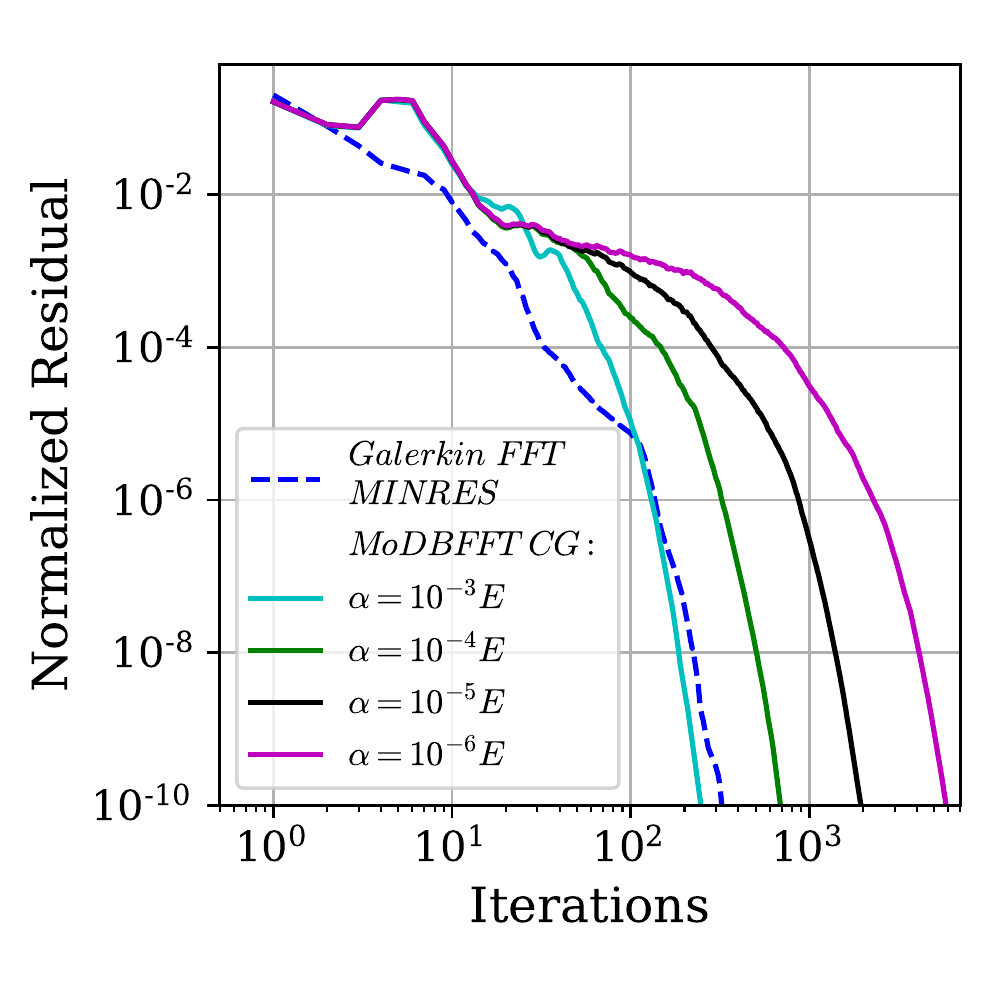}
\includegraphics[width=0.49\linewidth]{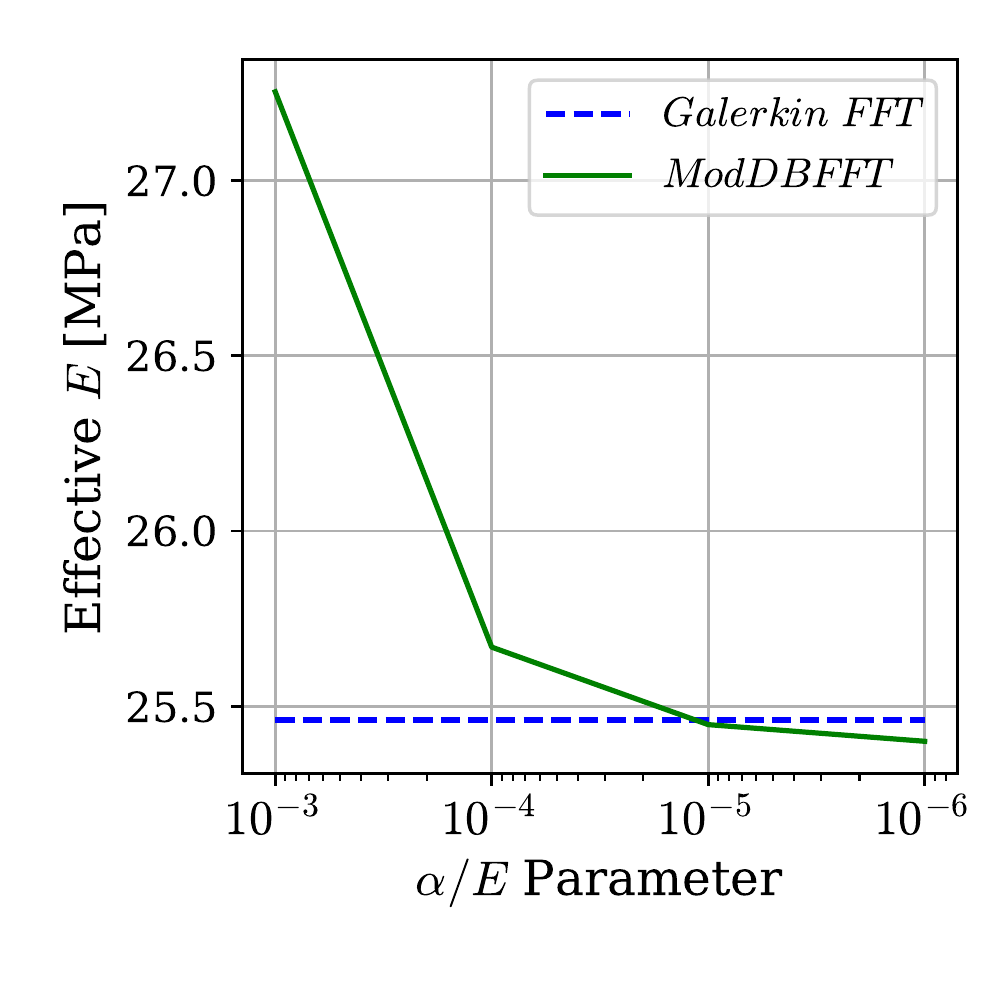}
\caption{\rev{Residual evolution and effective Young's modulus on an octet truss lattice for the different $\alpha$s considered}}\label{fig:alpha}
\end{figure}
\rev{Lower values of $\alpha$ in the MoDBFFT lead to worse convergence compared to} the adapted Galerkin FFT method. As a trade-off, large $\alpha$ values affect the macroscopic properties calculated inducing an artificial stiffness, and in Figure \ref{fig:alpha} it can be observed that the induced differences can go up to 10\% in terms of \rev{the} effective stiffness. In this work, $\alpha=10^{-4}E$ is selected as a compromise of convergence rate and accuracy of computed effective properties. \revv{It is interesting to note that the use of discrete frequencies is the most important ingredient for the success of the adapted Galerkin scheme. With discrete frequencies, the use of MINRES improves the performance of the CG version, but the reduction in computation time is only around 5\%. On the contrary, when standard Fourier discretization is used the difference in the performance between both solvers becomes substantial}. 

\subsection{Analysis of the surface smoothening approaches}

The regular discretization used in FFT can lead to actual densities slightly different from the target one, especially for coarse discretizations. The use of different geometrical representations (section \ref{sec:geoadapt}) to smooth out the surface also has an impact on the actual value of the relative density of the model. To quantify this geometrical misrepresentation, the actual relative density considered on the RVE has been calculated for each geometrical approach as the volume integral of the phase map, corresponding to
$$\overline{\rho}=\frac{1}{N_{vox}}\sum_{i \in N_{vox}} \phi_i.$$
A target relative density of $\overline{\rho}=0.1$ has been analyzed for the four geometrical representations, plain voxelized approach (PV), Voigt fine grid (VFG), Voigt analytic (VA), phase-field smoothening (PFS), and combined smoothening (CS). A range of discretizations from $54^3$ to $269^3$ voxels are considered, which corresponds approximately to a range from 5 to 25 voxels per diameter in this particular case. The resulting densities for different levels of the RVE discretization are represented in Figure \ref{fig:calcdens}. 
\begin{figure}[ht!]\centering
\includegraphics[width=0.5\linewidth]{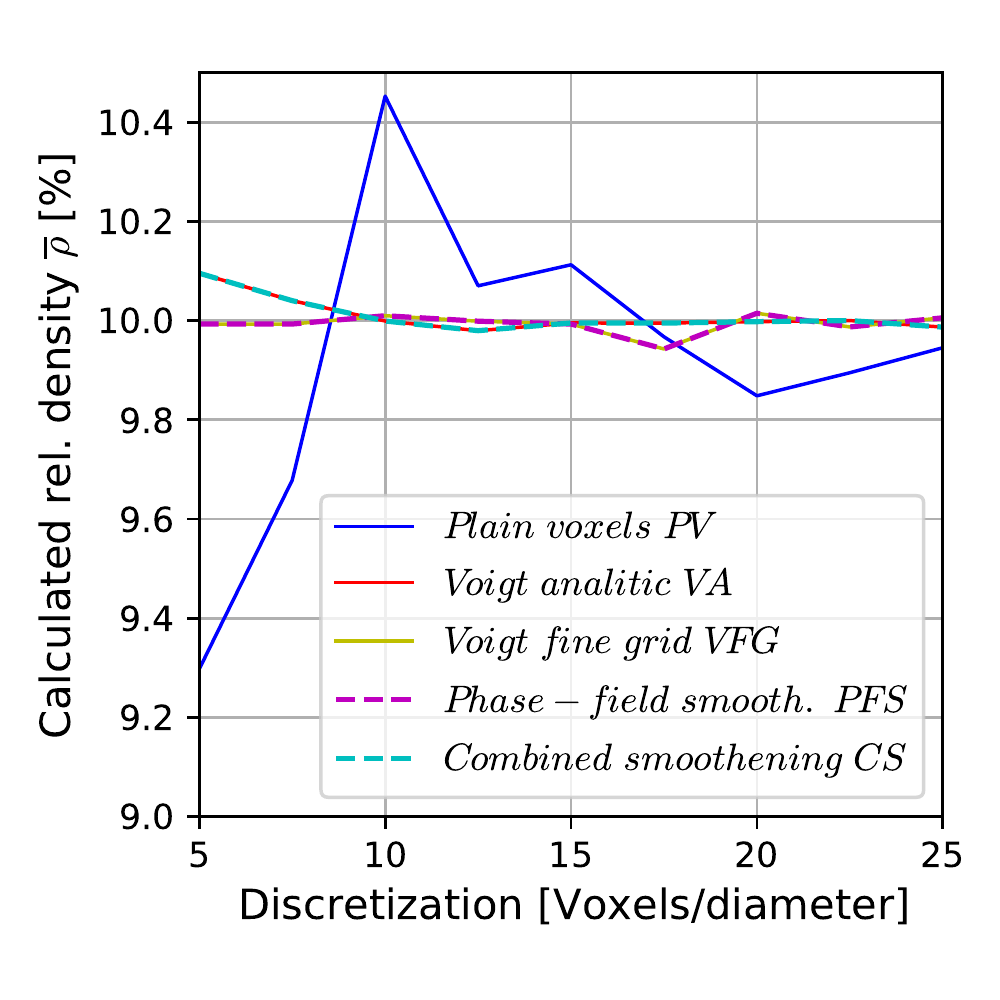}
\caption{Resulting RVE relative density for the different surface smoothening techniques with target relative density of $10$\%.}\label{fig:calcdens}
\end{figure}
From Figure \ref{fig:calcdens} it can be observed that the misrepresentation of the density is quite limited and the maximum error is below 0.7\% in all the cases. If the discretization is refined up to $162^3$ then the error is reduced below 0.1\% in the worst case. The maximum deviations occur for the plain voxelized representation. On the contrary,  the \rev{Voigt approaches} give a fairly good approximation of the density for all the mesh sizes considered, always below 0.1\%. \rev{In the case of VFG and PFS the small deviations from the target density are caused by the change on the ratio between the discretization used for the simulation and the finer one to compute the phase map. Finally, it can be observed that the relative densities obtained using phase-field smoothening (before thresholding) do not modify the relative density of the geometry function used as input. This mass conservation in the phase-field smoothening is due to the periodic boundary conditions which makes that volume integration of  eq. \eqref{eq:phase2} leads to $\overline{\phi}=\overline{\omega}$. Therefore PFS gives the same density as VFG and CS the same density as VA.}

\subsection{Accuracy of the methods}

The accuracy of the macroscopic and microscopic numerical results obtained with the FFT approaches combined with the different smoothening techniques will be assessed for different discretization levels, ranging from $54^3$ to $269^3$ voxels (Fig. \ref{fig:geos}). To this end, the uniaxial tensile deformation of an octet truss lattice with a relative density of $\overline{\rho}=0.1$ is simulated for  the combinations of FFT solvers and surface smoothening. The RVE is deformed along one of its edges and stress free conditions \rev{are imposed} in the perpendicular directions. To assess the result of the simulations, the FFT results are compared with FEM results with an equivalent discretization level in terms of number of elements per truss diameter. 

The macroscopic strain and stress tensors are extracted from the simulation results to obtain the effective properties. The effective Young's modulus and Poisson's ration \rev{have} been represented in Figure \ref{fig:etresol} together with the corresponding FEM results. \rev{First, it is observed that the Voigt fine grid (VFG) approach shows almost identical behavior than the Voigt analytic (VA) and therefore, for the shake of clarity, only VA will be considered for the rest of the discussion.}

Figure \ref{fig:etresol} shows that the convergence of FFT results with the discretization is slower than FEM results, except the combination of the Galerkin approach with the VA that provides a solution almost independent on the grid for model sizes \revv{greater than 10} elements per diameter ($108^3$ voxels). The largest oscillations of the effective response with the discretization are obtained for the plain voxelized models (PV) and are a direct result of the variations in the actual relative density of the cells.  The results of the phase-field smothening (PFS) and combined smoothening (CS) converge better than the plain voxelized representations but worse than the VFG and VA approaches separately. Therefore, it can be concluded that using phase-field smoothening has a non-negligible effect \revv{on} the resulting stiffness for the same average density. In the case of plain voxel approaches, the MoDBFFT method provides the same tendency as the Galerkin approach with a small offset, being the first one slightly stiffer.  In the case of Poisson's ratio, which is less dependent on the cell density, the convergence with the discretization is faster and smoother. The maximum difference between FEM and FFT results is below 0.6\% for every discretization. In all the cases, \rev{the} effective FFT response converges to the same value that coincides with the FEM results.

\begin{figure}[ht!]\centering
\includegraphics[width=0.49\linewidth]{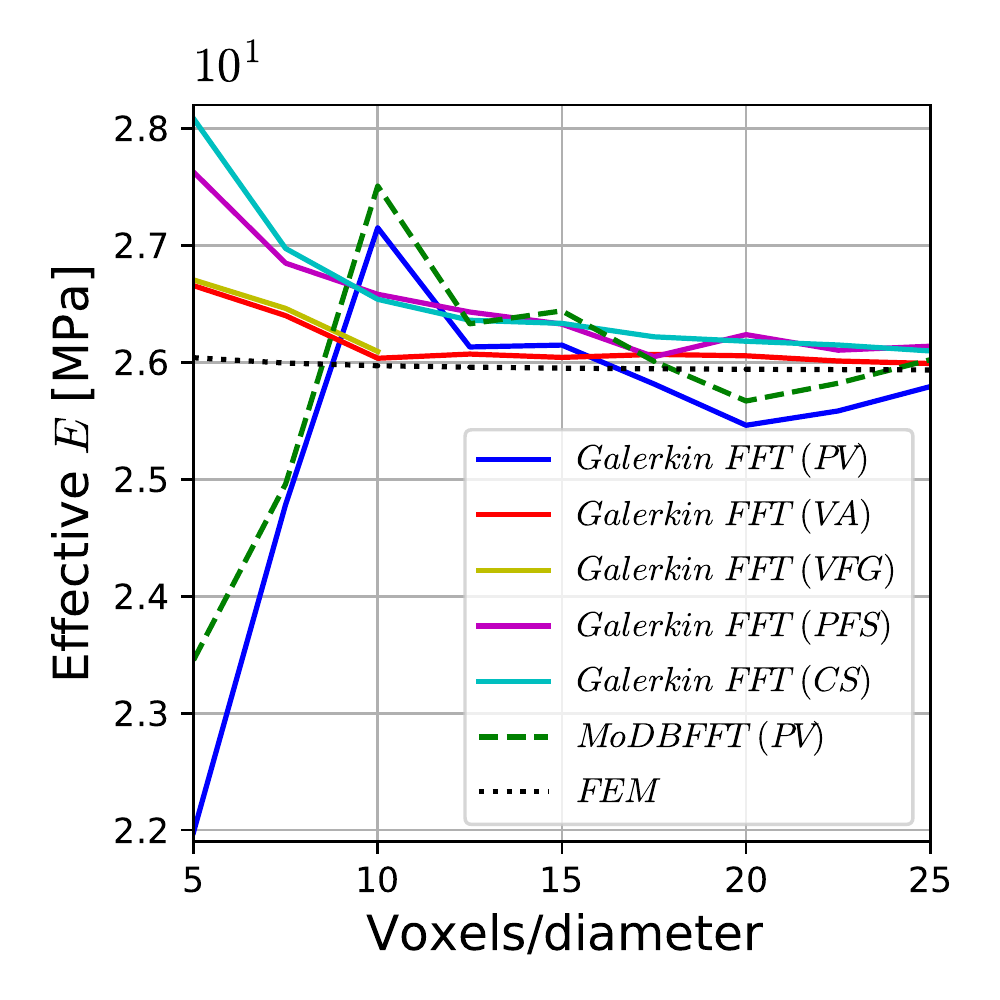}
\includegraphics[width=0.49\linewidth]{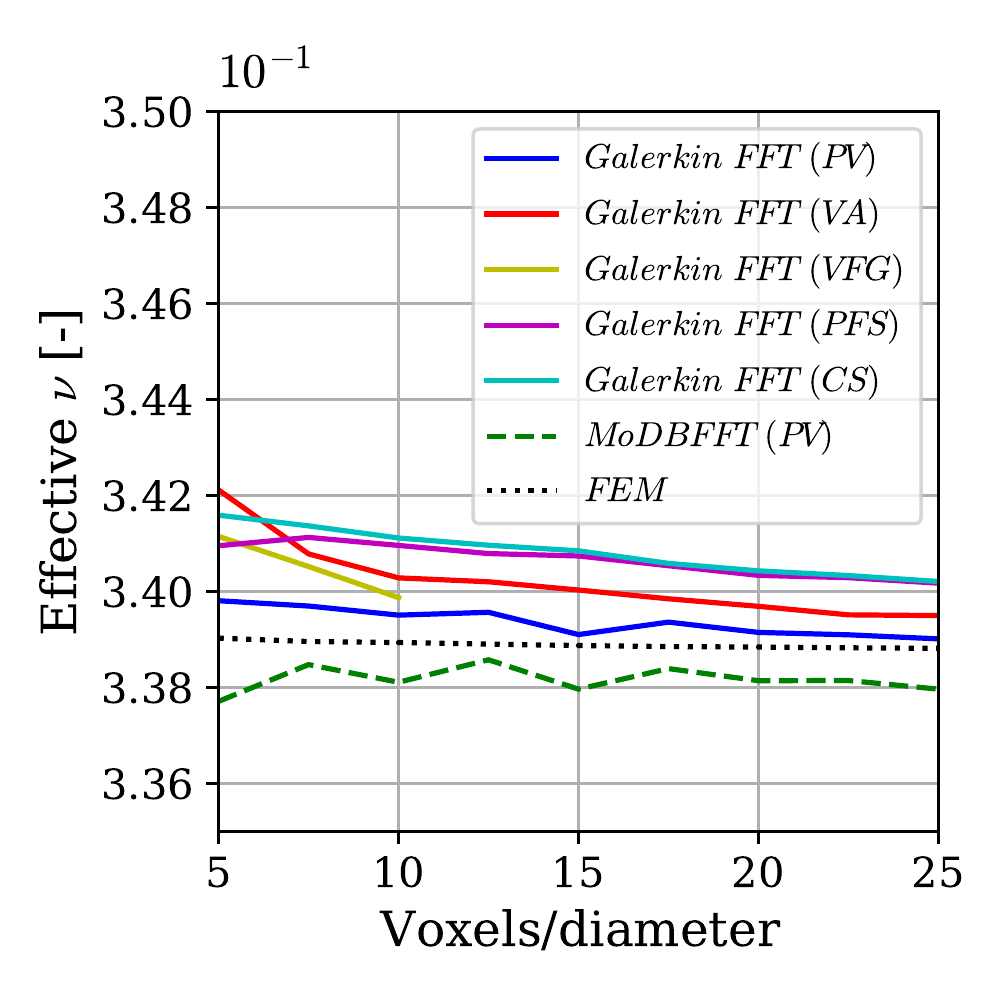}
\caption{Effective Young's modulus and Poisson's ratio for different discretizations.}\label{fig:etresol}
\end{figure}

The accuracy of the microscopic fields has also been analyzed and compared with FEM results. In Figure \ref{fig:fieldsresol}, the microscopic stress in the loading direction has been superposed to the deformed geometry,  magnified by a factor $\times 20$,  for the two most representative FFT approaches (Galerkin with \rev{Voigt analytic smoothening} and MoDBFFT with a plain voxel approach). Qualitatively it can be observed that the deformed shapes are almost identical and the concentrations of stress fields are very similar both in location and intensity. 

\begin{figure}[ht!]\centering
\includegraphics[width=0.3\linewidth]{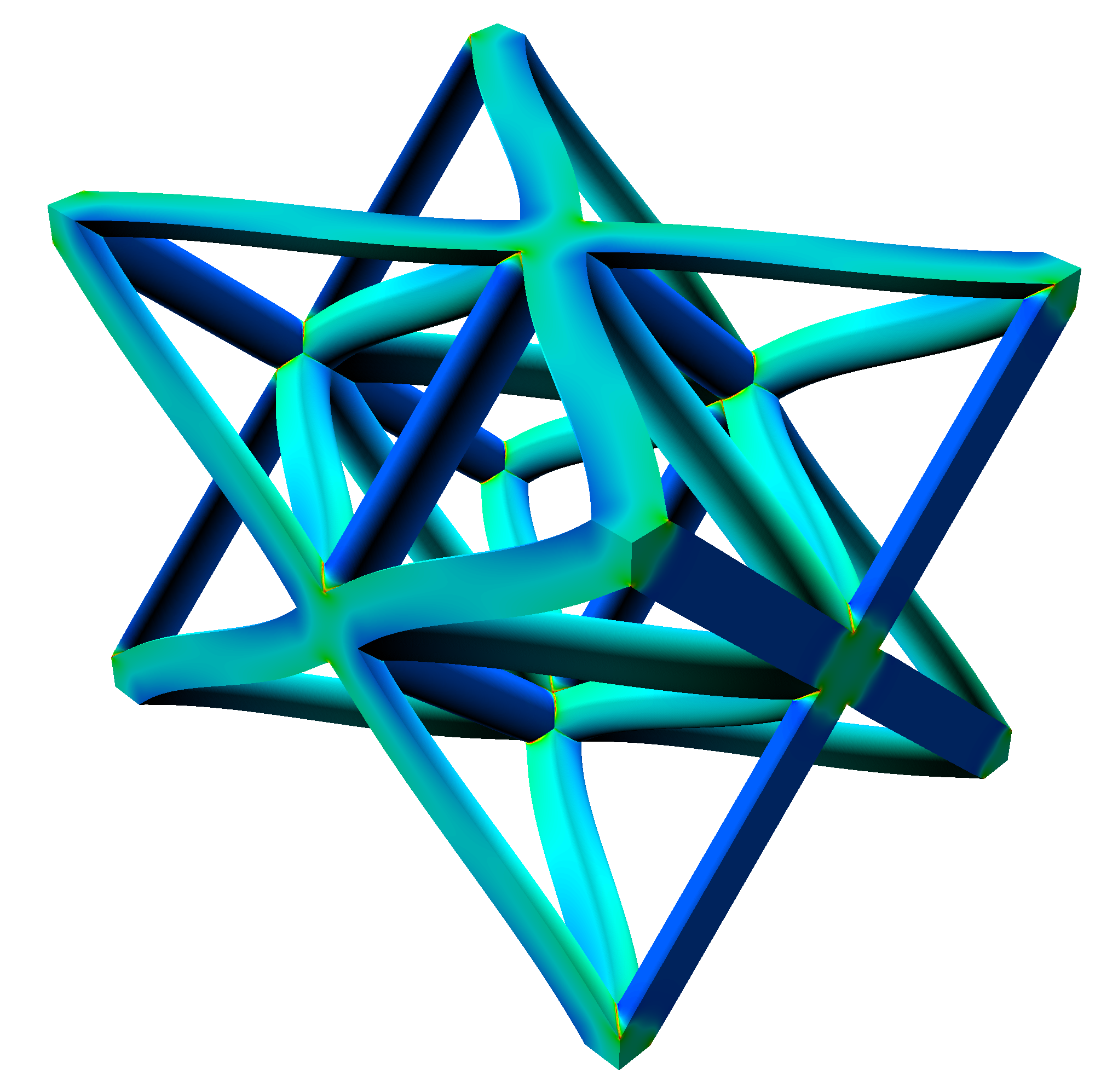}
\includegraphics[width=0.3\linewidth]{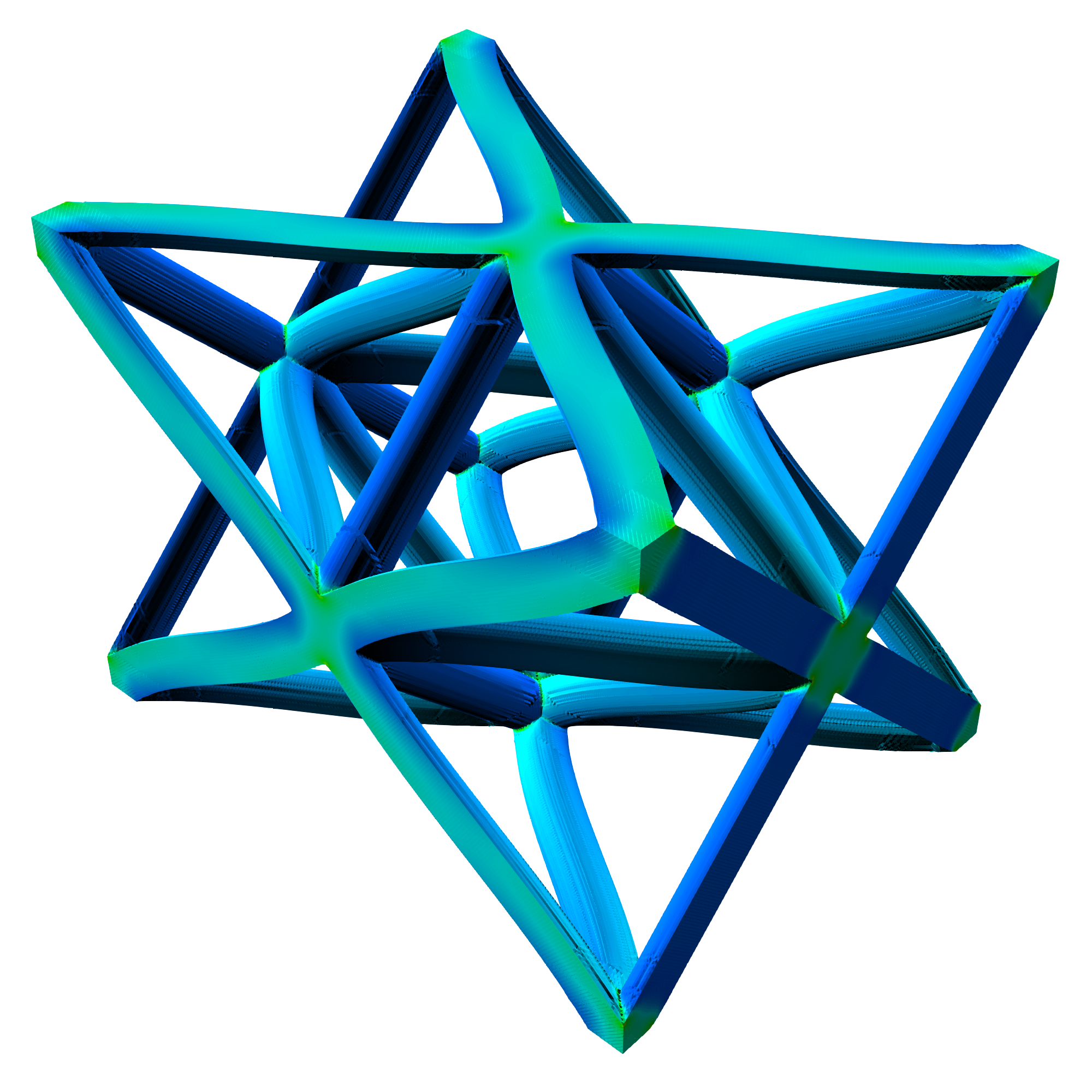}
\includegraphics[width=0.3\linewidth]{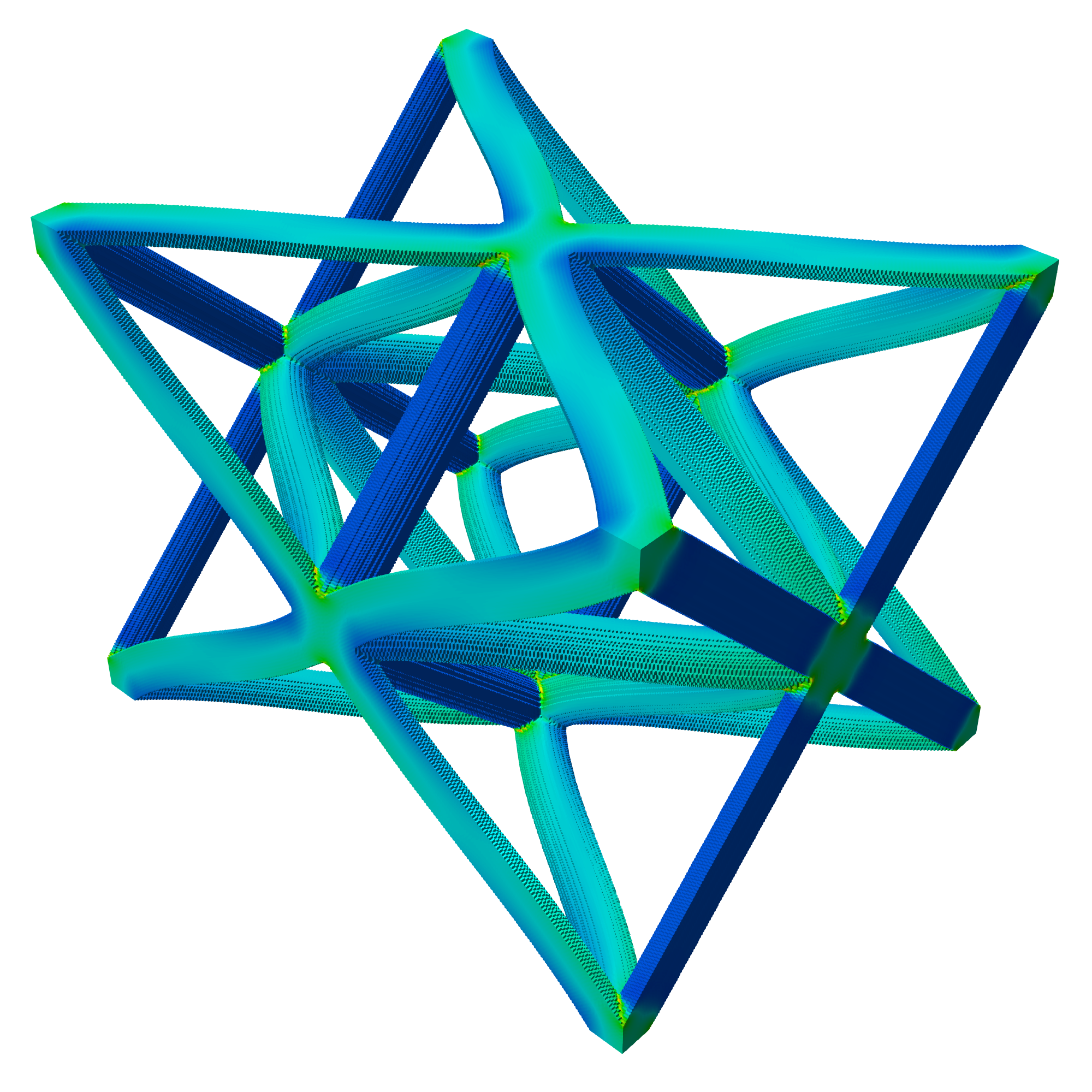}
\includegraphics[width=0.07\linewidth]{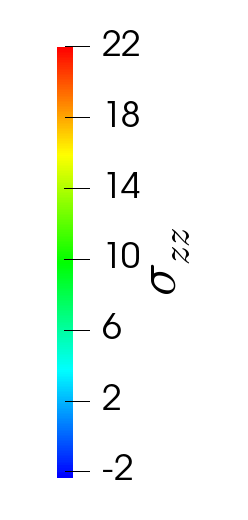}
\caption{Local stress fields in loading direction ($\sigma_{zz}$) on the deformed configuration (x$20$) for FEM, Galerkin FFT (VA) and MoDBFFT (PV).}\label{fig:fieldsresol}
\end{figure}

To quantify this difference, the $L_2$ norm of the difference in the stress in the loading direction $\sigma_{zz}$ (eq. \ref{eq:difff}) \revv{is} computed with respect to the local fields of the FEM method and the result has been represented in Fig. \ref{fig:locresol}. It can be observed that, as it happened with the effective response, FFT results converge to FEM solutions. It is also remarkable that for discretizations \rev{finer} than $15$ voxels/diameter, the differences are always below $10$\% except \rev{for} those methods where the phase-field smoothening technique is used.

\begin{figure}[ht!]\centering
\includegraphics[width=0.49\linewidth]{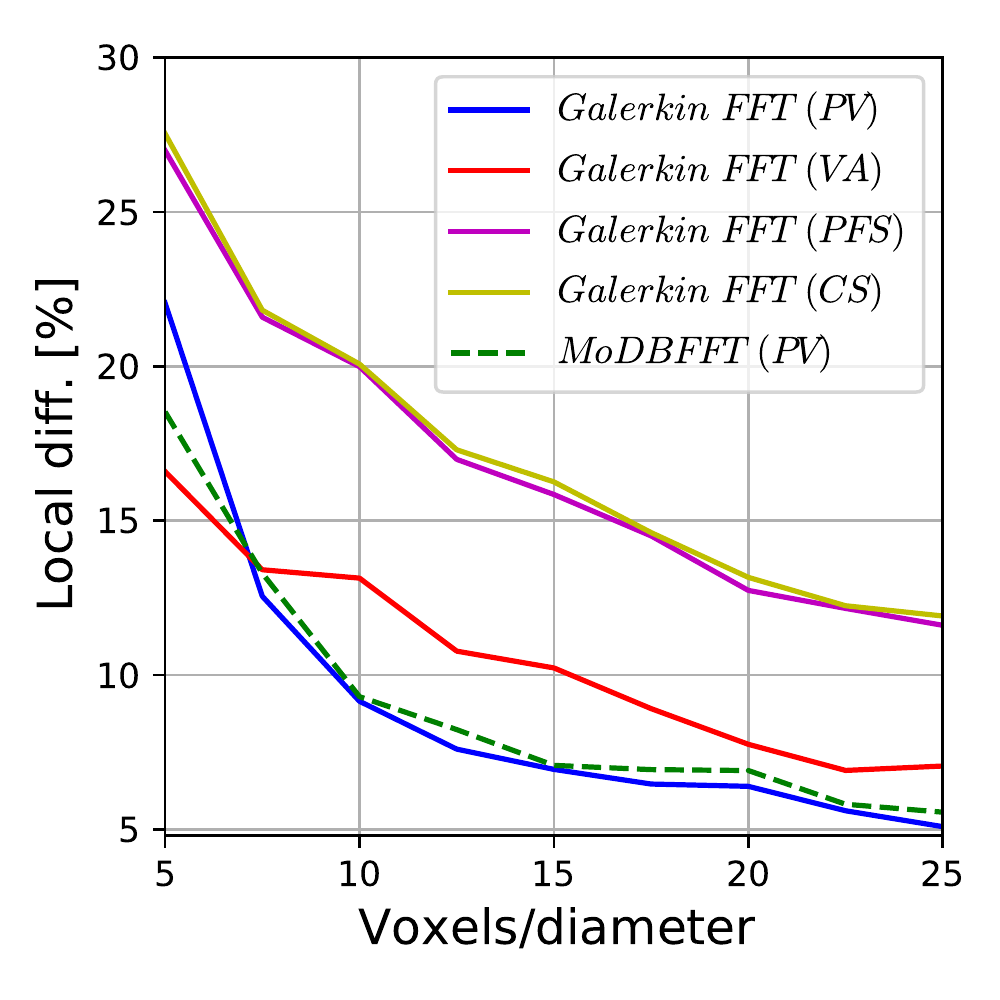}
\caption{Local stress differences in the loading direction ($\sigma_{zz}$) as function of the discretization level.}\label{fig:locresol}
\end{figure}

The solutions using the plain voxels geometric approach show a slightly better microscopic response but need a larger number of voxels to accurately predict the overall behavior due to the density variations.  It is interesting to note that, for a PV geometrical representation, both modified Galerkin (which uses a rotated scheme) and MoDBFFT ---which has a standard discretization--- provide very similar microfields. This result indicates that the terms included in the MoDBFFT to approach the free surface condition have a similar effect in smoothening the response to the use of discrete frequencies in the modified Galerkin,  as it can be observed in Fig. \ref{fig:locresol}. Phase-field smoothening \revv{alleviates the noise efficiently} but, for the value of $\ell$ considered here, induces non-negligible changes in both macroscopic and microscopic responses.  As a summary, Galerkin method combined with Voigt analytic smoothening shows the best combination of   accuracy in the microfields and effective response.

\subsection{Effect of  the relative density and numerical efficiency}

The effect of the relative density on the accuracy and efficiency of the different FFT methods is studied in this section. Relative densities $\overline{\rho}$ ranging from $0.5$\% to $30$\% are considered for the octet truss lattice. A discretization of $15$ voxels(elements)/diameter is used for every volume fraction, leading to models with different total number of voxels. The loading case applied is \rev{uniaxial stress}, which is accounted \rev{for} using macroscopic mixed boundary conditions. For comparison purposes, finite element simulations with the same conditions are also performed for every cell, using in this case 15 elements per diameter.

\rev{The macroscopic specific stiffness, $E/\bar{\rho}$, and Poisson' s ratio, $\nu$, obtained using the different approaches are represented in Figure \ref{fig:etdens} \revv{as functions} of the cell relative density}.  It can be observed that, in most of the cases, the specific Young's modulus ($E/\bar{\rho}$) is very close to the FEM results. The maximum relative difference \revv{with respect to} FEM is $10$\% for the elastic modulus in the case of the phase-field smoothening method, showing again that although near-to-surface local fields are smoothened with this approach, \rev{the} macroscopic response is slightly altered. On the other \rev{hand}, the difference of the response obtained using the Galerkin approach with discrete frequencies and \rev{Voigt analytic smoothening (VA) with respect to} FEM is always below 2\%. The prediction of the Poisson's ratio was very accurate for all the methods and densities considered, with maximum differences below 1.5\%.

\begin{figure}[ht!]\centering
\includegraphics[width=0.49\linewidth]{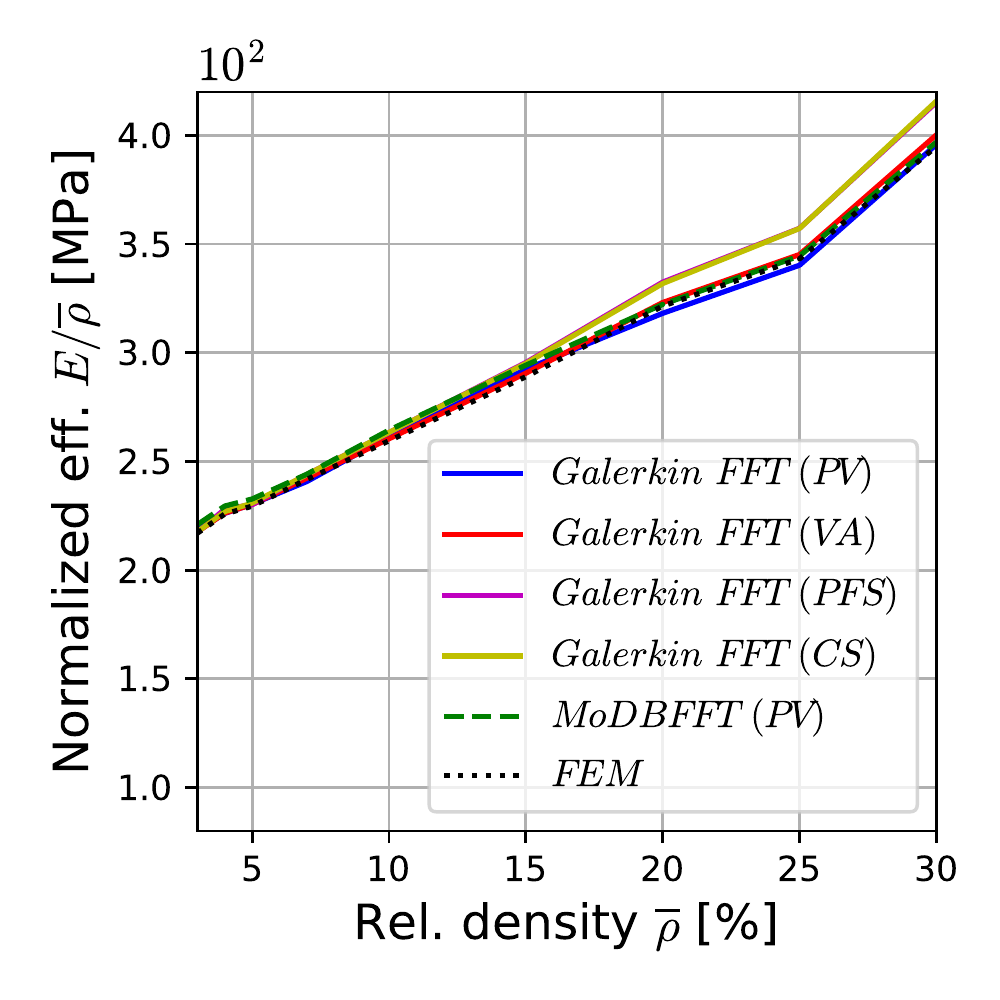}
\includegraphics[width=0.49\linewidth]{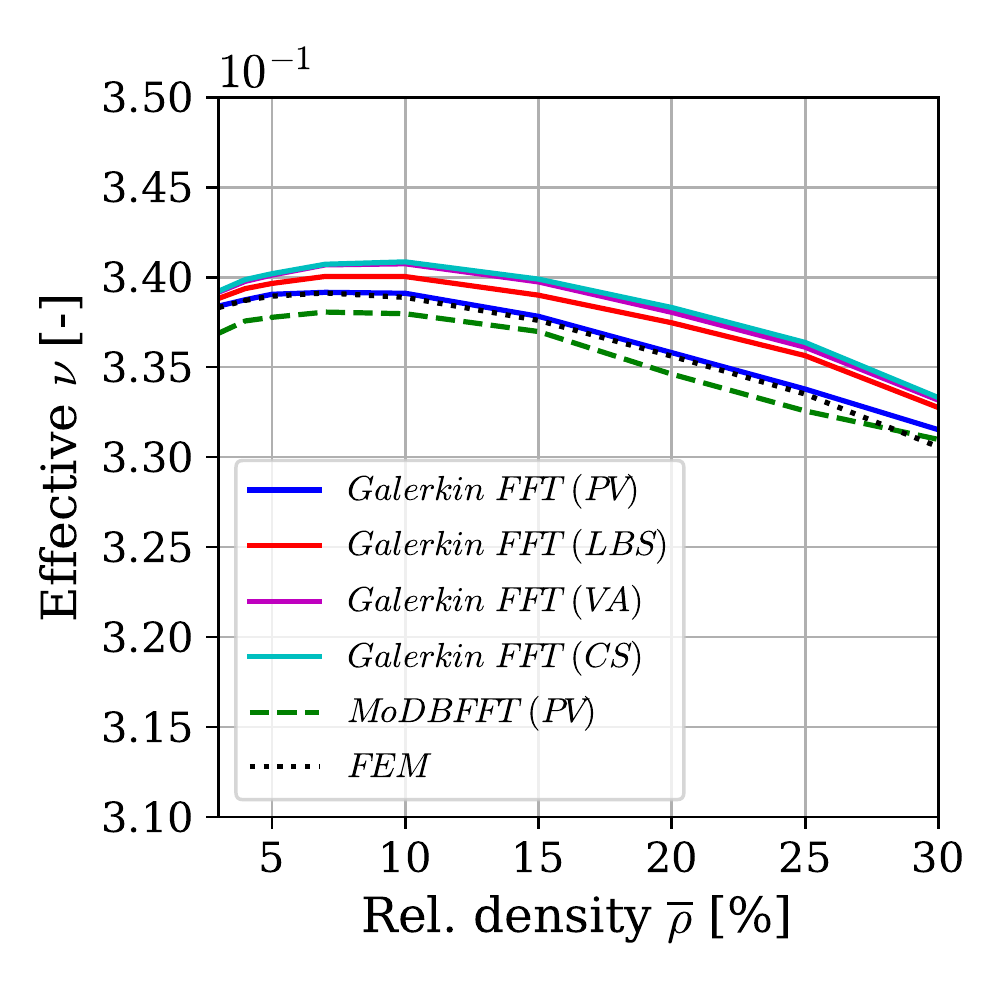}
\caption{Effective Young's modulus and Poisson's ratio time for different relative densities.}\label{fig:etdens}
\end{figure}

The microscopic stresses obtained are also analyzed and compared with the FEM counterparts. The localization of stresses and their intensity are strongly dependent on the relative density due to the change from a stretch to a bending dominated \rev{behavior}. In all the cases the response of all the FFT approaches considered was very similar to the FEM results. This different behavior is also reflected in the deformation modes, which were also accurately captured for the FFT approaches for all the densities. As an illustration, the diagonal stress component in the loading direction, $\sigma_{zz}$, is represented in Figure \ref{fig:feildsdens} for the cell with $\bar{\rho}=30\%$ superposed to the deformation of the cell with a magnification of $\times20$. \rev{Qualitatively, it can be observed how both stresses and deformed shape of the cell are very similar. }
\begin{figure}[ht!]\centering
\includegraphics[width=0.3\linewidth]{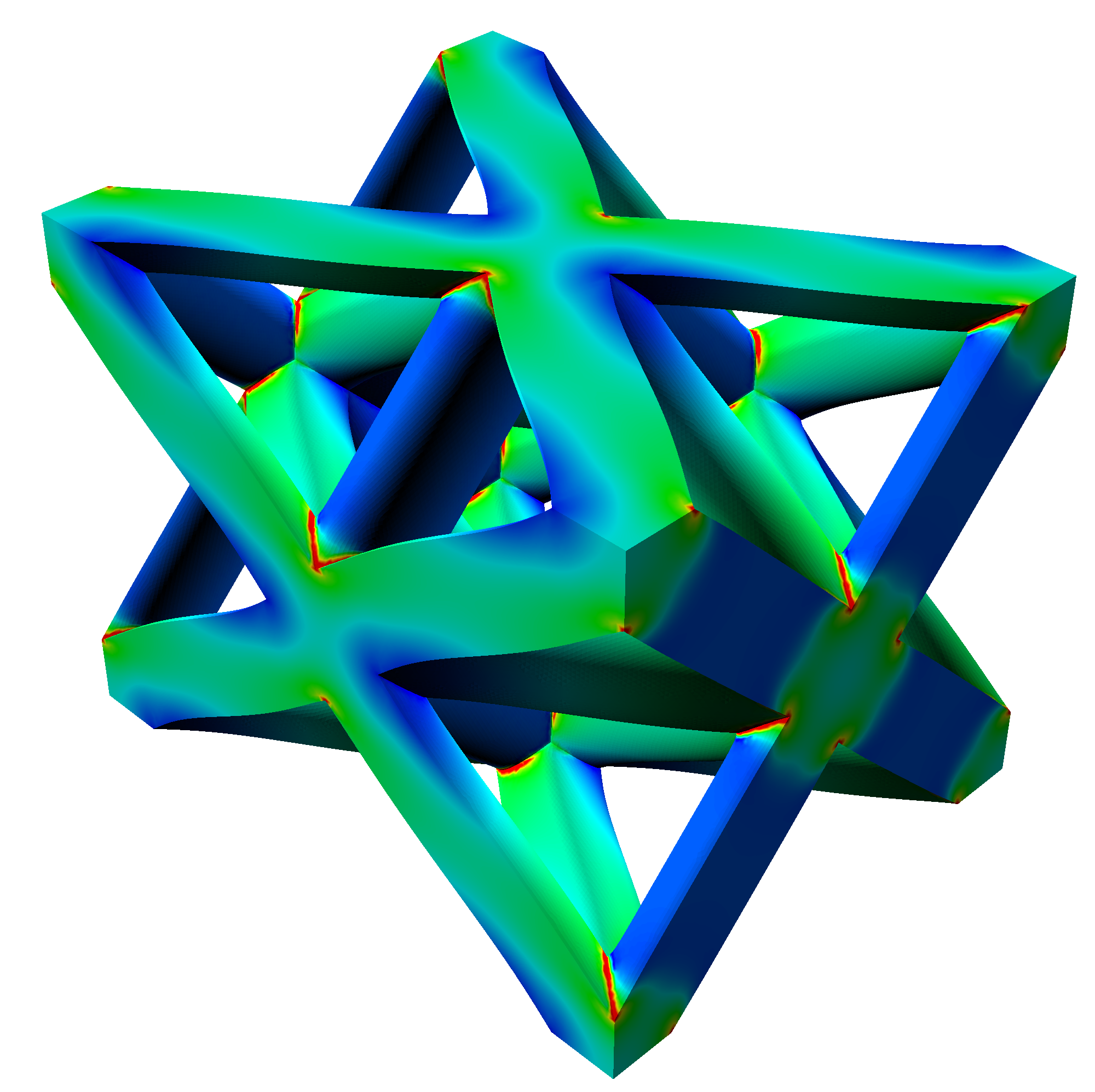}
\includegraphics[width=0.3\linewidth]{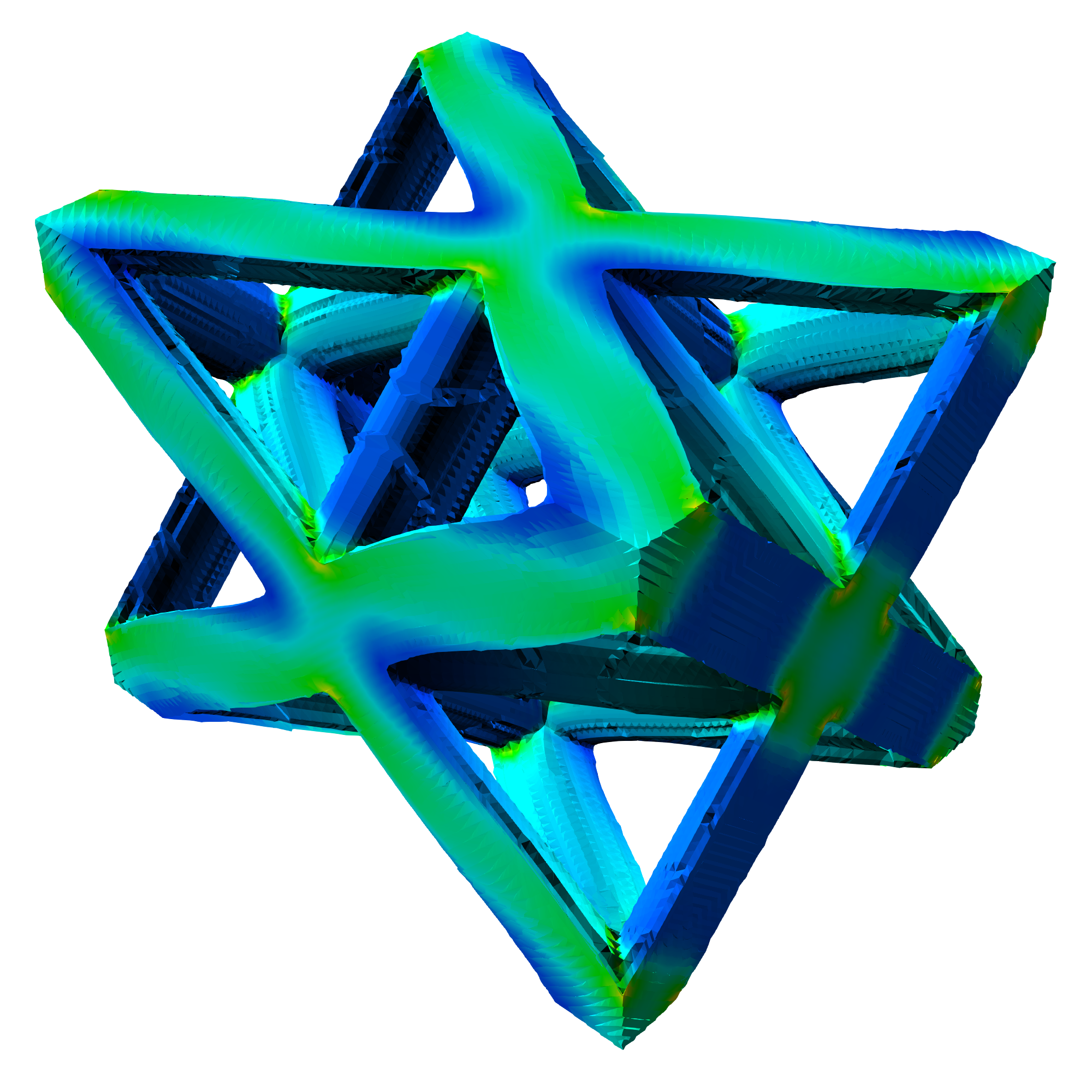}
\includegraphics[width=0.3\linewidth]{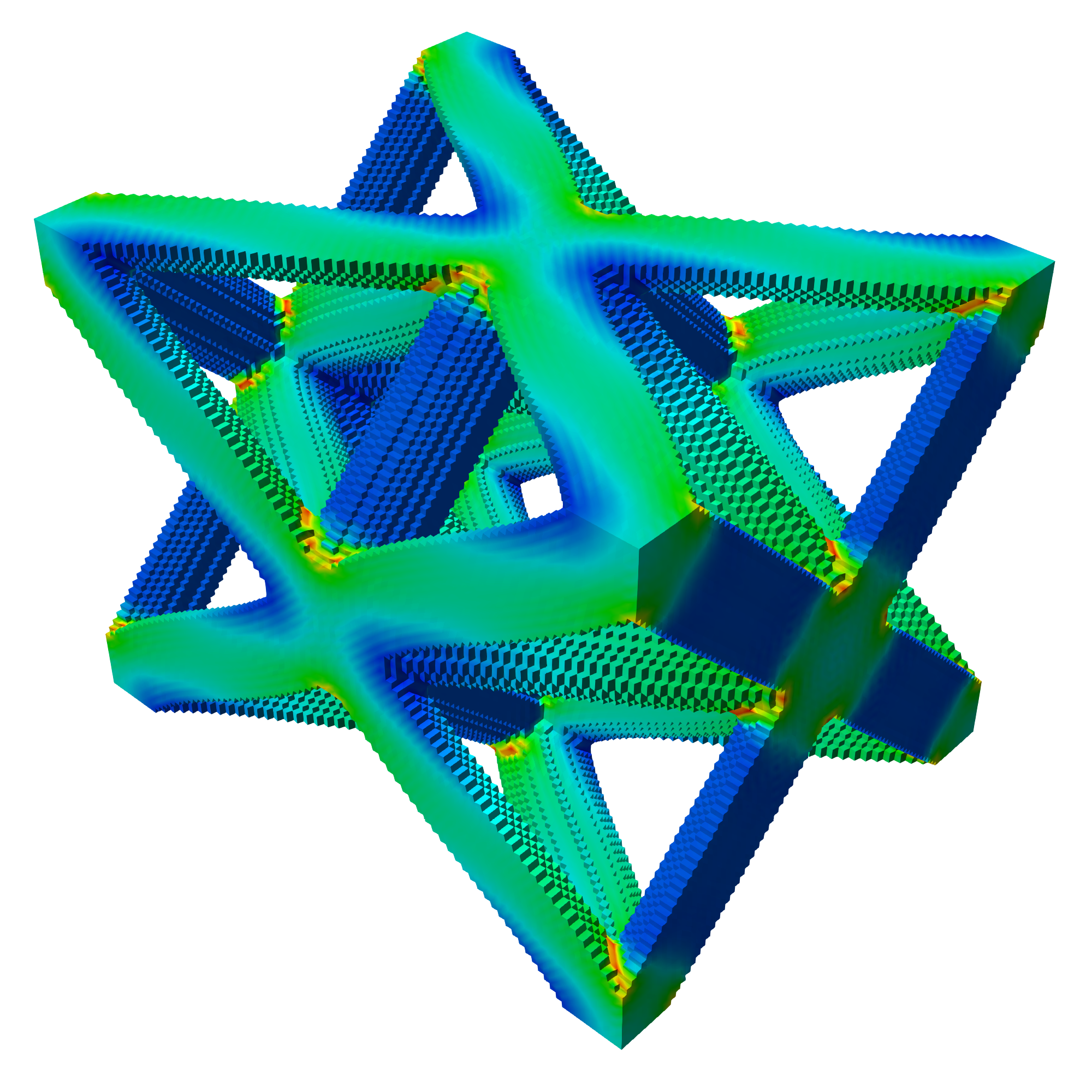}
\includegraphics[width=0.07\linewidth]{leg1.png}
\caption{Local stress fields in loading direction ($\sigma_{zz}$) on the deformed configuration (x$20$) for FEM, Galerkin FFT (VA) and MoDBFFT (PV).}\label{fig:feildsdens}
\end{figure}

From a quantitative viewpoint, the $L_2$ norm of the difference  (eq. \ref{eq:difff}) between the FFT microscopic stress component $\sigma_{zz}$ and the FEM value was computed and represented in Fig. \ref{fig:locdens}. The norm of the difference was around 15\% in most of the cases. Again, the microscopic response of PV approaches are more near to the FEM results than the other smoothening approaches.


\begin{figure}[ht!]\centering
\includegraphics[width=0.49\linewidth]{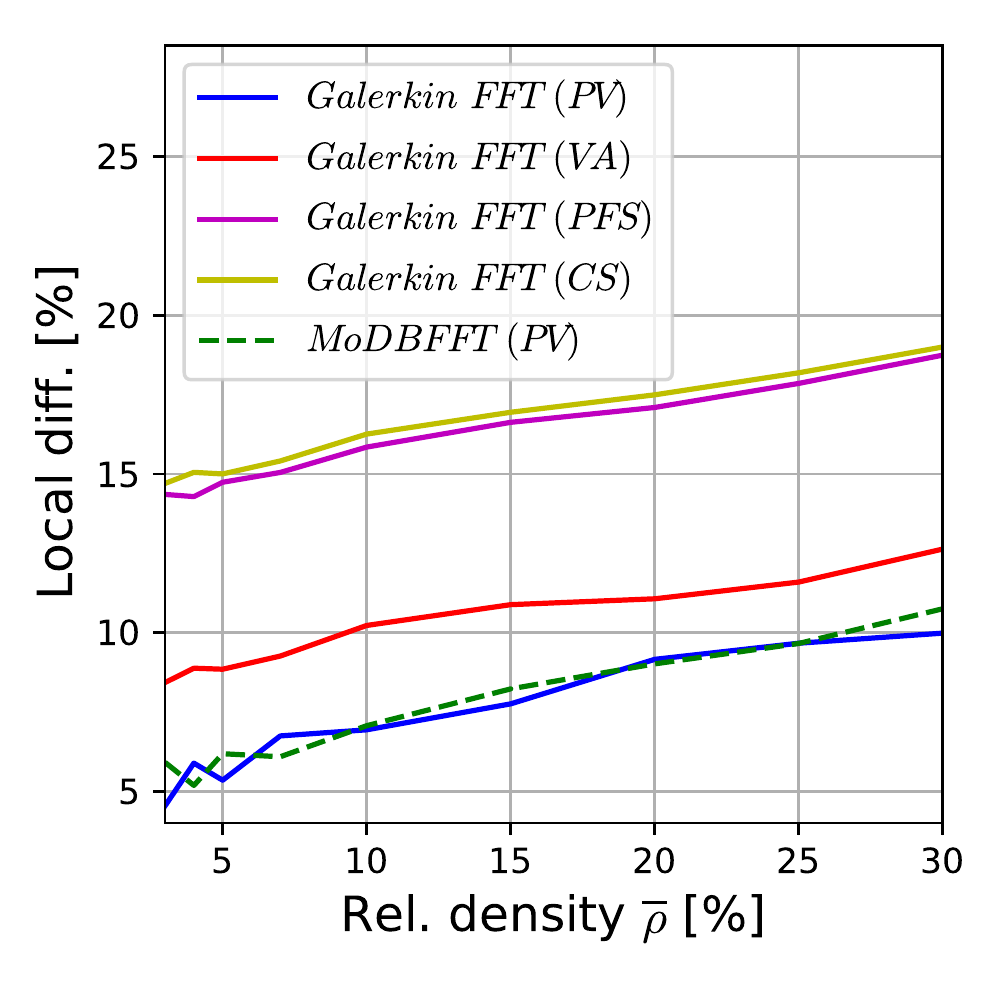}
\caption{Local stress differences in the loading direction ($\sigma_{zz}$) for different relative densities.}\label{fig:locdens}
\end{figure}
As stated in the study of the cell with $\bar{\rho}=0.1$ \rev{for} different discretization levels, the Galerkin FFT combined with the \rev{Voigt} smoothening shows the best compromise between the accuracy of effective and the local properties.

Finally, the numerical performance of the FFT approaches is analyzed for the different relative densities. The time spent \rev{on} the simulations was obtained for different  FFT approaches and FEM and was represented in Figure \ref{fig:tdens}. It can be observed first that all FFT approaches were more efficient than FEM method for relative densities greater than $\approx 5$\%. Note that this comparison is made for a particular choice of 15 elements/voxels per diameter and, if this number were increased, better performance of the FFT solver would be expected. Second, curves in Fig. \ref{fig:tdens} show that if the number of elements/voxel per strut diameter is kept constant, the simulation times decrease with the density for both FEM and FFT due to the reduction in the total number of elements/voxels in the lattice. Nevertheless, the time reduction grows faster in FFT than in FEM, and the FFT simulations for a relative density of $\bar{\rho}=0.3$ were 4 to 8 times faster than FEM ones. The improvement of the performance ratio FFT/FEM with the density can be easily explained by the number of voxels of the full RVE that belong to the interior of the lattice. For low densities most of the voxels of the RVE  belong to the empty space, not contributing to the cell response but having to be considered for FFT operations. Therefore, the use of FFT for densities below $\bar{\rho}<7\%$ is not competitive \rev{with respect to} FEM. On the contrary, it is remarkable that even with this strong disadvantage, FFT becomes clearly more efficient for relative densities \rev{exceeding} 10\% making the approaches here \rev{proposed} very competitive for foams and porous materials.  \rev{As a conclusion, among the different FFT approaches and smoothening techniques proposed, the Galerkin FFT combined with the Voigt analytical smoothening is the most interesting one since it combines very accurate results with the best numerical performance. The displacement approach developed, the MoDBFFT, can provide smooth results which are as accurate as the modified Galerkin}, but is not competitive in terms of efficiency since for obtaining such accurate results a small parameter $\alpha$ is required (i.e. $\alpha=10^{-4} \ E$) and, for this value, the number of iterations is larger than the adapted Galerkin approach.

\begin{figure}[ht!]\centering
\includegraphics[width=0.49\linewidth]{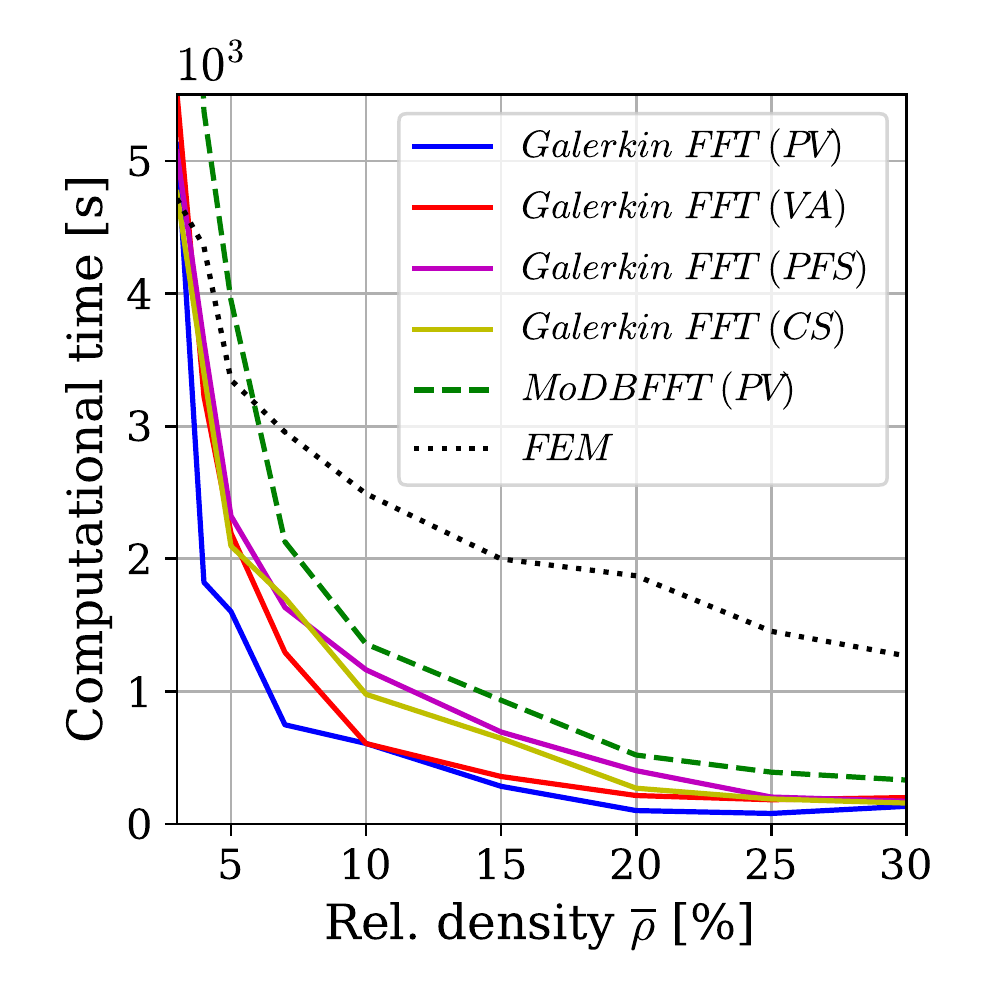}
\caption{Simulation time for different relative densities, \rev{using models with around 15 elements/voxels per diameter.}}\label{fig:tdens}
\end{figure}

\section{Validation for non linear material response}

To assess the non linear extension of the methods proposed, simulations have been made using a Von Mises J2 plasticity model as lattice material behavior. The elastic constants are the same as in the previous section and perfectly plastic behavior is considered (no strain hardening), being the yield stress $\sigma_y=70$MPa. Uniaxial compression tests have been carried out with a maximum strain of $10$\%. The strain is applied using a ramp divided in $20$ regular strain increments. The tolerance used for the Newton-Raphson method is $5 \cdot 10^{-3}$. The approaches selected for this study are the Galerkin with \rev{Voigt analytic smoothening} (VA) and the MoDBFFT with a plain voxelized (PV) representation, considered as the most representative methods from previous results. Octet truss lattices with relative densities of 10\%, 20\% and 30\% are studied for a fixed discretization of 15 voxels(elements)/diameter. FEM simulations with equivalent discretization are performed for comparison purposes.

The resulting macroscopic stress-strain curves are represented in Figure \ref{fig:j2ss}. The simulations predict a large elastic region followed by a plastic regime with a very small hardening rate. The elastic-plastic transition is smooth and the strength reached increases with the relative density. The results of all the simulations are very similar, being the maximum difference between FEM and FFT results smaller than $5$\% for all the densities considered. 

\begin{figure}[ht!]\centering
\includegraphics[width=0.49\linewidth]{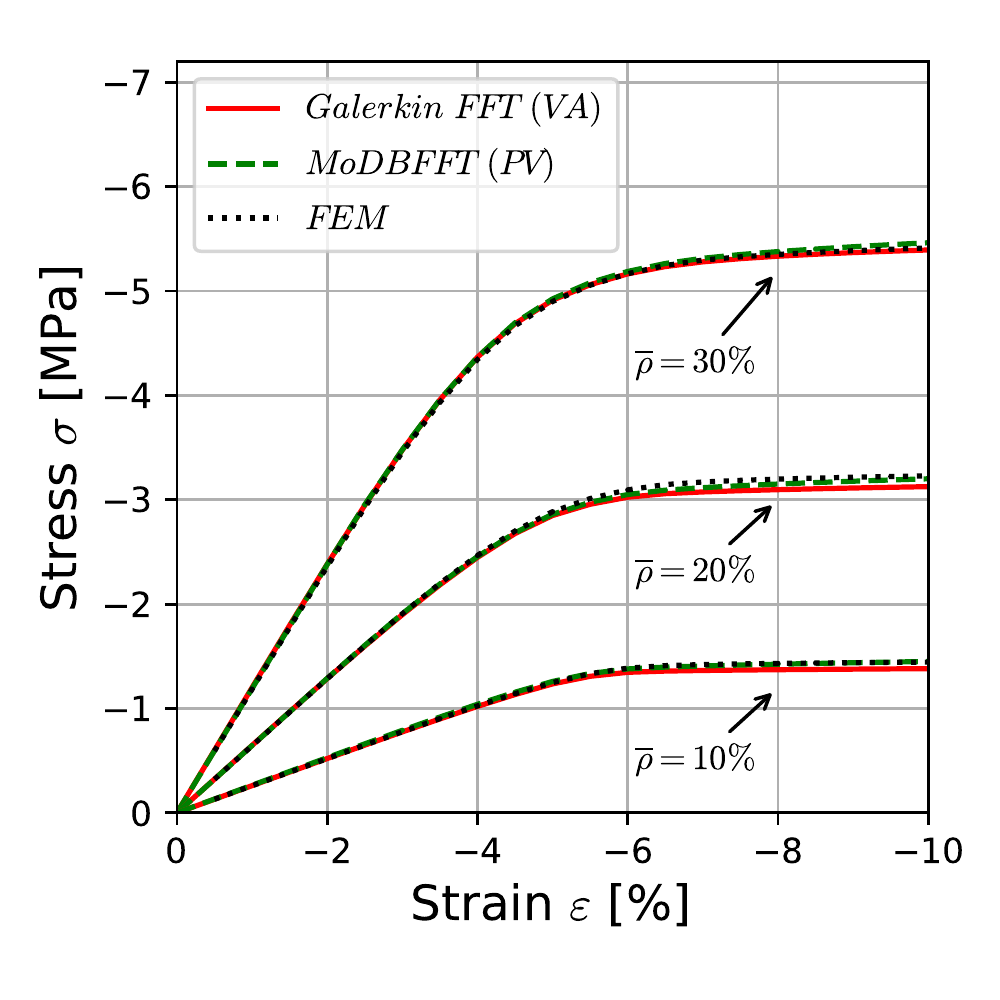}
\caption{Stress-strain curves with non linear material}\label{fig:j2ss}
\end{figure}

The microscopic accumulated equivalent plastic strain field $\varepsilon_{p}$, defined as \rev{
$$\varepsilon_{p} =\int_t \left ( \dot{\boldsymbol{\varepsilon}}^P: \dot{\boldsymbol{\varepsilon}}^P \right )^{1/2} \mathrm{d}t
$$ with $\dot{\boldsymbol{\varepsilon}}^P$ } the plastic strain rate tensor, has been represented in Figure \ref{fig:feildsj2} for both FFT and FEM approaches \rev{and time corresponding to a total compressive strain of -10\%}. The  iso-plots are represented in the deformed cell (x2) to observe the deformed cell shape. During the simulations, it was observed that the plastification started in the strut joints and after plastifying these joints behave as ball joints \rev{resulting in} an almost uniaxial stretching of the struts, as it can be observed in Figure \ref{fig:feildsj2}. The deformed configurations and stress distributions predicted by both FFT approaches are very similar to the FEM results.
\begin{figure}[ht!]\centering
\includegraphics[width=0.3\linewidth]{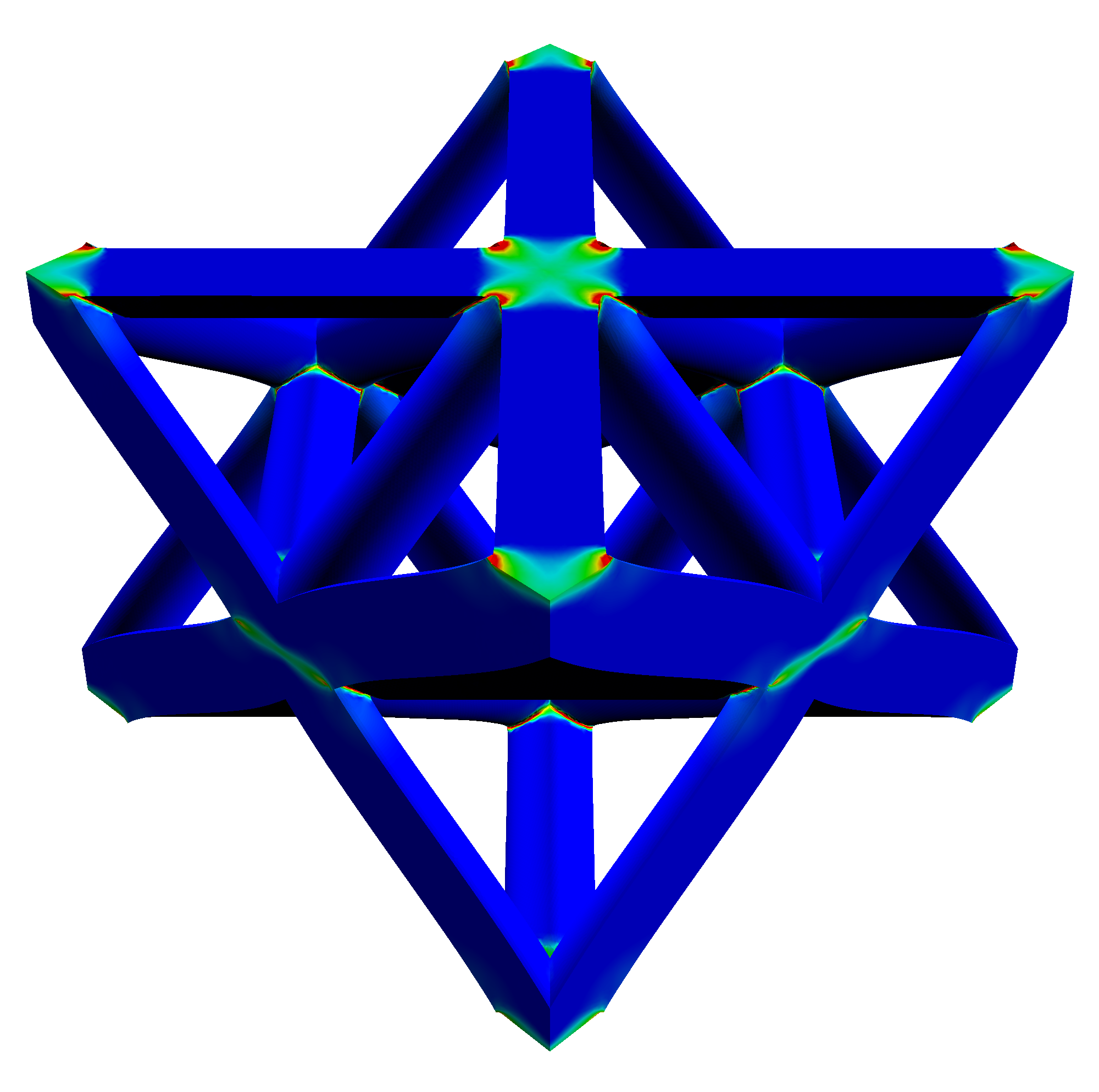}
\includegraphics[width=0.3\linewidth]{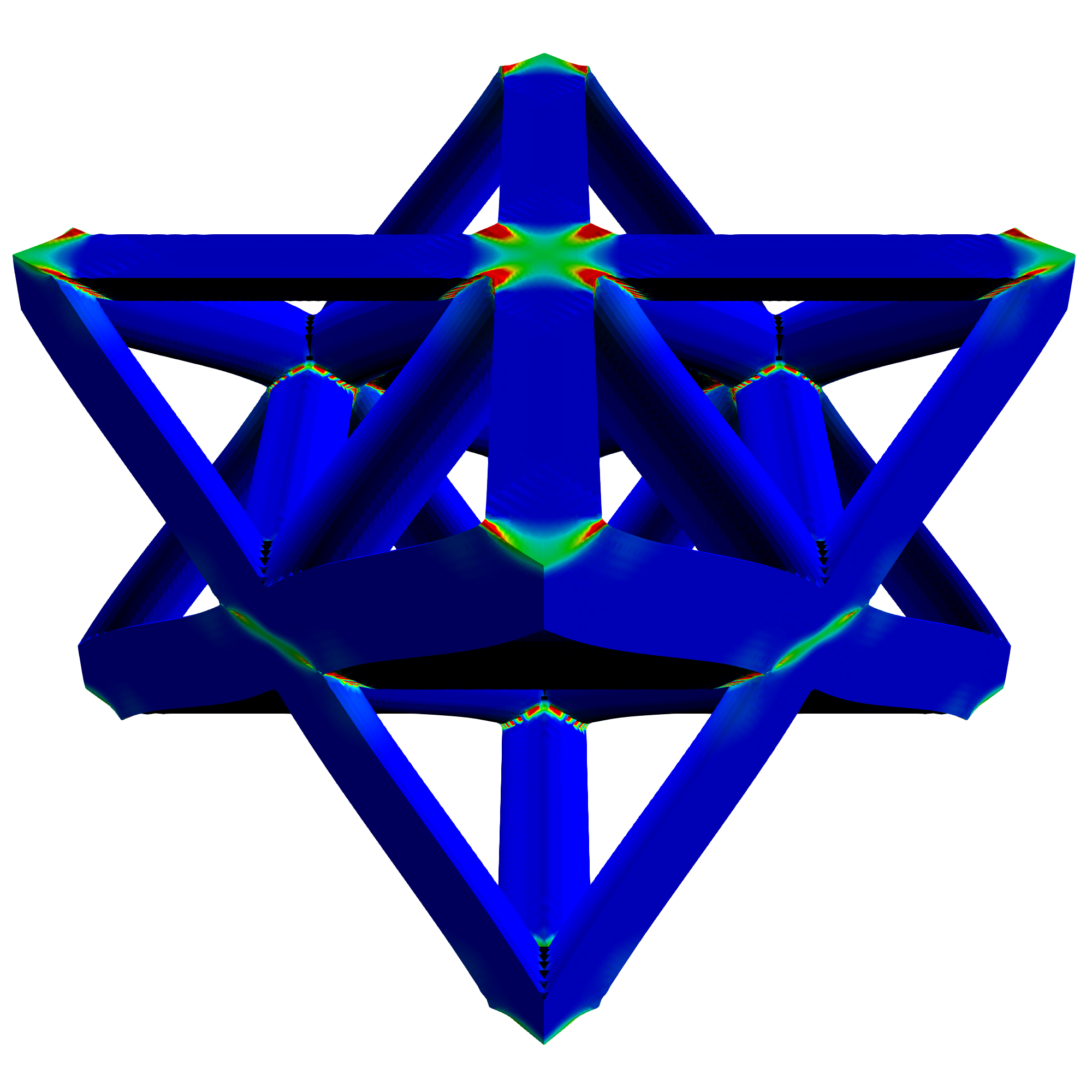}
\includegraphics[width=0.3\linewidth]{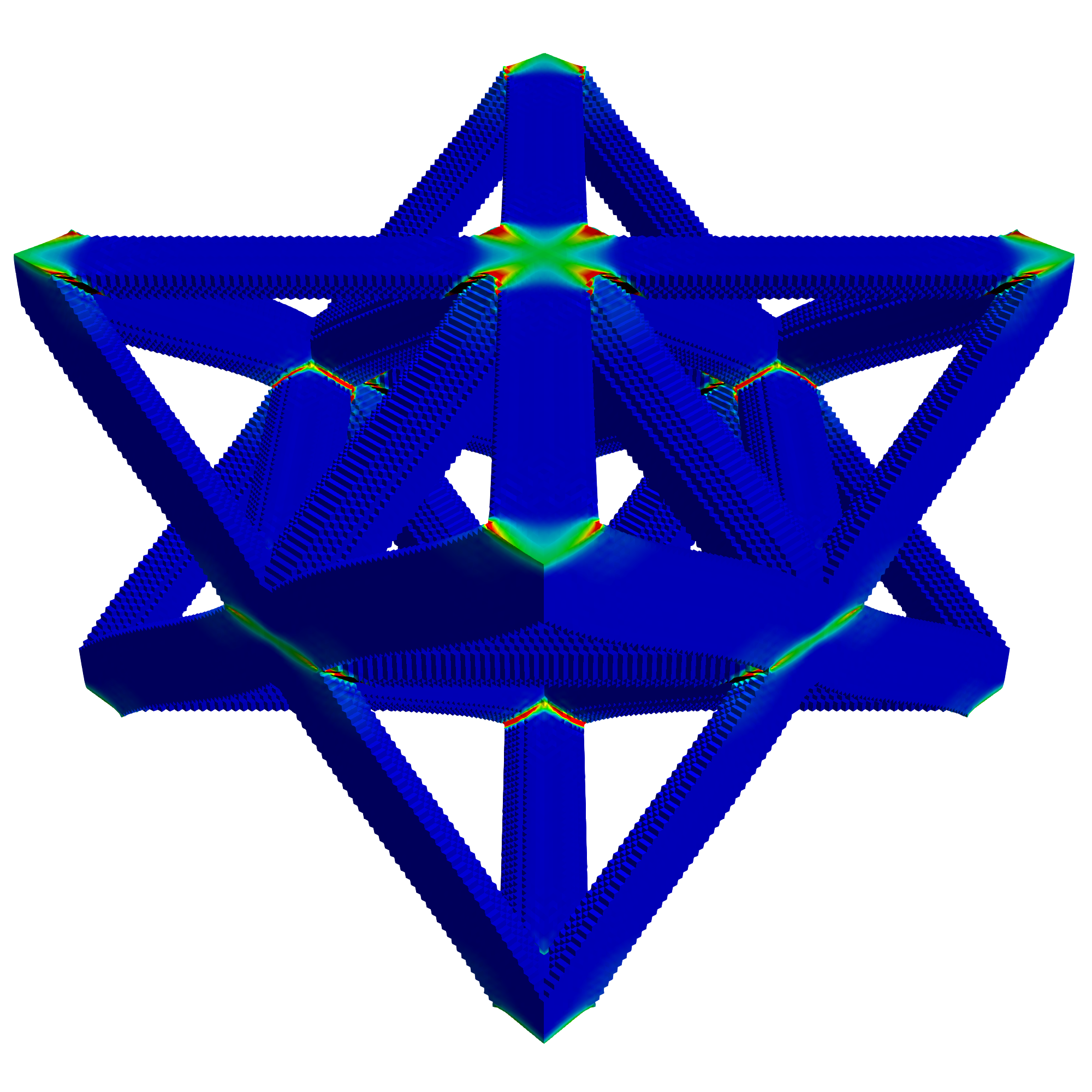}
\includegraphics[width=0.07\linewidth]{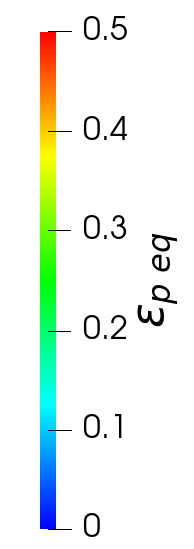}
\caption{Equivalent plastic strain for FEM, Galerkin FFT with \rev{Voigt analytic} smoothening and the MoDBFFT with plain voxelized approach.}\label{fig:feildsj2}
\end{figure}

For a quantitative measure of the local difference between FFT and FEM microscopic results, the $L_2$ norm of the differences in the stress component in the loading direction is represented in Fig. \ref{fig:j2norms}. In all the cases the agreement is good also from a microscopic viewpoint and the differences were always below 15\%.  These differences vary with the volume fraction and also with the geometrical representation, since plastic strain is very localized and a smoothened surface representation might affect the intensity of the localization in those regions.

\begin{figure}[ht!]\centering
\includegraphics[width=0.49\linewidth]{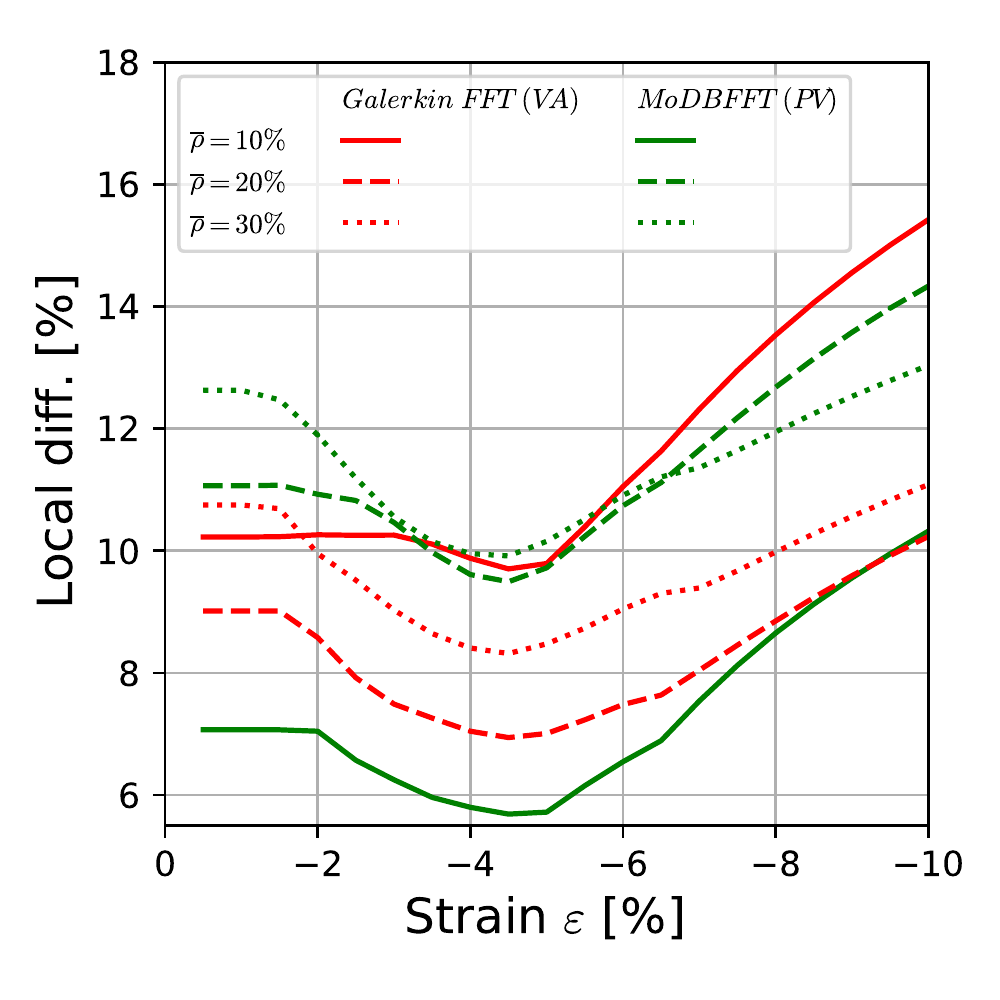}
\caption{Local stress differences in the loading direction ($\sigma_{zz}$) for the different relative densities with non linear material.}\label{fig:j2norms}
\end{figure}

\section{Application \rev{to} a real 3D tomography}

The principal application of the FFT framework adaptation for lattices is the ability to analyze directly the actual topology obtained by 3D tomography or a similar approach. In this section, the potential of the FFT framework proposed will be shown by simulating and comparing the responses of an ideal lattice and its real counterpart considering the fabrication defects resulting from the additive manufacturing process. \rev{This example illustrates that full-field simulation of the cell microstructure obtained by tomography is an extremely powerful technique to quantify the changes in the cell response due to fabrication defects.}  Under this framework, the real porosity can be considered explicitly inside the RVE, without performing any post-processing to obtain averaged porosities and without the need of using homogenization models to account for the effect of the average porosity \cite{AMANI2018395}.

The unit cell selected is a cubic-diagonal lattice with a designed relative density of \rev{14.2\%} and manufactured in PA12 by Selective Laser Melting powder deposition. The cell edge length is 6.2mm and the trust nominal diameter is 0.9mm. The cell was fabricated by CIRP \verb|https://www.cirp.de/comp/comp\_EN.php5| following standard fabrication parameters. 

The cell microstructure was analyzed by 3D tomography using a GE (Phoenix) Nanotom 160 kV with a Hamamatsu 7942-25SK detector and nanofocus X-ray source. The resolution of each voxel was $\approx 4\mu$m and the tomographic data included $1551\times 1557 \times 1581$ voxels. An image of the full resolution 3D tomography is shown in Figure \ref{fig:tomo}. The analysis of the 3D tomography data of the actual cell microstructure shows a volume fraction of porosity of around \rev{3.6\%}. The presence of that porosity inside the struts might affect the properties of the cell and the actual behavior can differ from the one expected for the design geometry. 

\begin{figure}[ht!]\centering
\includegraphics[width=0.49\linewidth]{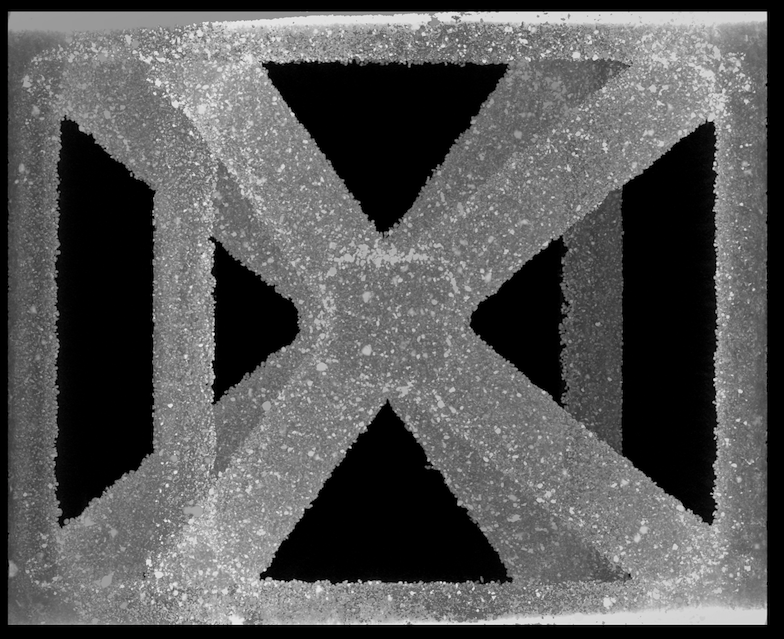}
\includegraphics[width=0.48\linewidth]{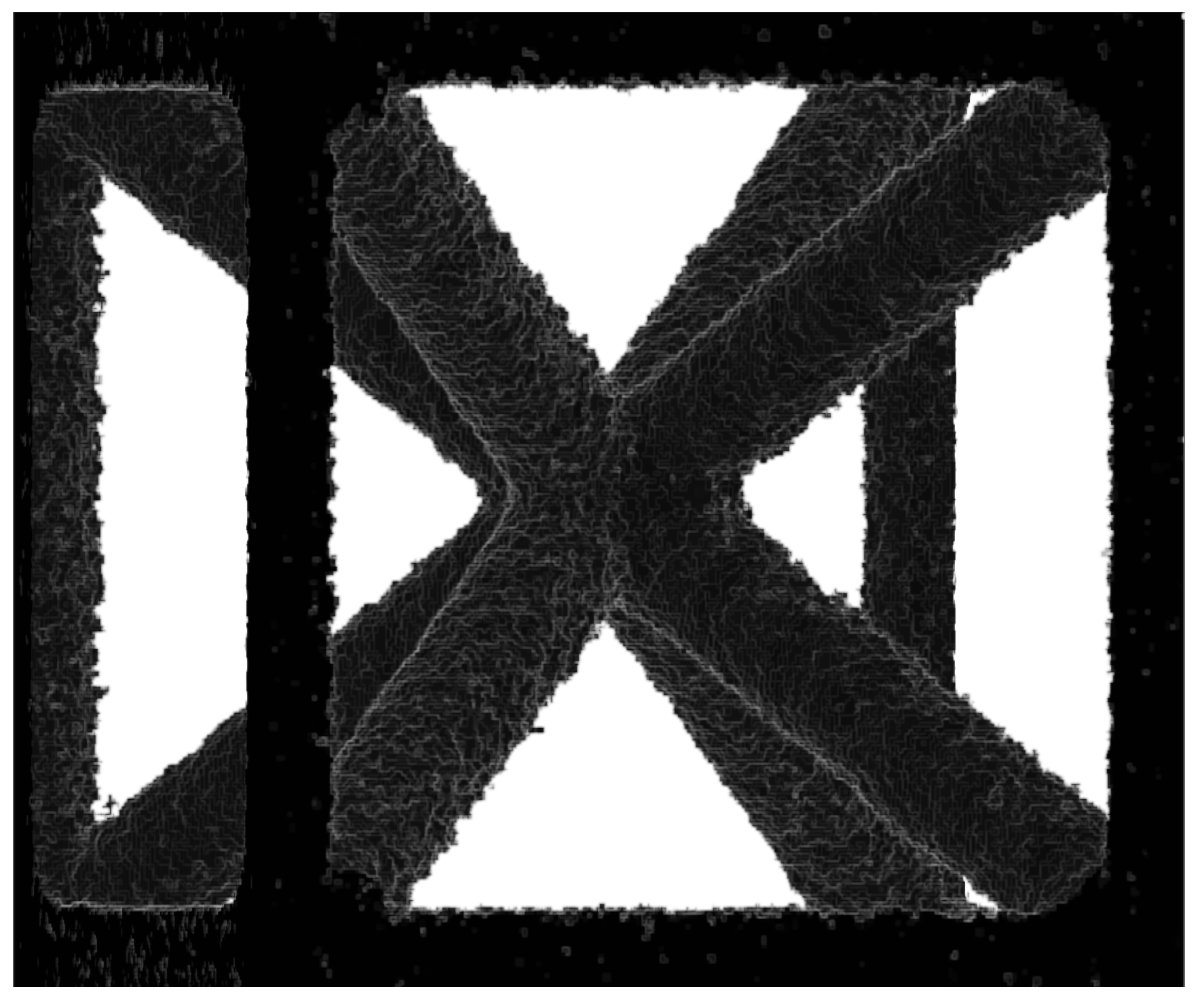}
\caption{Left: Full resolution 3D tomography of the cell, Right: FFT model from tomographic data.}\label{fig:tomo}
\end{figure}

To quantify the effect of the porosity in the elastic response, the design and actual geometries have been subjected to uniaxial test of 1\% of deformation using the two different FFT approaches. For the design geometry, a discretization of $25$ voxels per diameter was used ($216^3$ voxel RVEs). In the case of the real tomography, the original $1500^3$ pixel 3D tomography image has been compressed to $256^3$ voxelized model by averaging the densities obtained from the tomography (Fig. \ref{fig:tomo}, right figure). In the case of the Galerkin FFT, the smooth map of averaged densities has been directly introduced as the phase-map $\phi$ into the simulation. \rev{In the case of the MoDBFFT, thresholding of the densities \revv{has} been performed, distinguishing as a material point all the local densities above a value that enforces an average relative density equal to the measured one, and as empty region all the rest of the points.}

The effective response obtained are given in Table \ref{tab:tomo}. \revv{The FFT simulations predict a decay in the effective Young's modulus of around 10\%, a relatively large reduction considering the low porosity volume fraction measured (3.6\%). }The prediction of the overall stiffness reduction is a very interesting characteristic of the FFT framework because it cannot be accurately obtained using a mean field approach since the location of the porosity within the struts influences the macroscopic response of the cell.
\begin{table}[ht!]\centering
\begin{tabular}{c|c|c|c|c|}
\cline{2-5}
 & \multicolumn{2}{c|}{Galerkin FFT (density map)} & \multicolumn{2}{c|}{MoDBFFT (threshold map)} \\ \cline{2-5} 
 & $E\ [MPa]$ & $\nu$ & $E\ [MPa]$ & $\nu$ \\ \hline
\multicolumn{1}{|c|}{Design geometry} & 45.76 & 0.314 & 46.25 & 0.313 \\ \hline
\multicolumn{1}{|c|}{3D Tomography} & 41.06 & 0.292 & 43.00 & 0.291 \\ \hline
\end{tabular}
\caption{Effective properties extracted from uniaxial tests on cubic-diagonal lattice design and actual geometries.}\label{tab:tomo}
\end{table}

In addition to the changes of the macroscopic response, thanks to the resolution of the local fields, the FFT analysis can be used to estimate the microscopic fields and hot spots of the structure and, using damage indicators, the reduction of the lattice strength due to the presence of defects. As an example of these microscopic fields, the stress component in the loading direction $\sigma_{zz}$ obtained in FFT simulations has been represented in Fig. \ref{fig:tomoloc}. \rev{It must be noted that the differences in local fields using Galerkin FFT (using a density map as phase map) and MoDBFFT (using a pure voxelized approach) are below 10\%.} In the cell with a perfect microstucture, stress concentrates in the trust joints and varies smoothly through the geometry of the bars. On the contrary, on the simulations with the actual microstructure, large stress concentrations localized near bigger pores are observed superposed to the concentrations near the joints. The maximum stresses found with the real microstructures are approximately 50\% larger than the ones obtained with the designed cell.  Due to these stress concentrations, if the maximum local stress were taken as a rough estimation of the fracture initiation, the real structure would fail at stress level 50\% lower than the design geometry.

\begin{figure}[ht!]\centering
\includegraphics[width=0.45\linewidth]{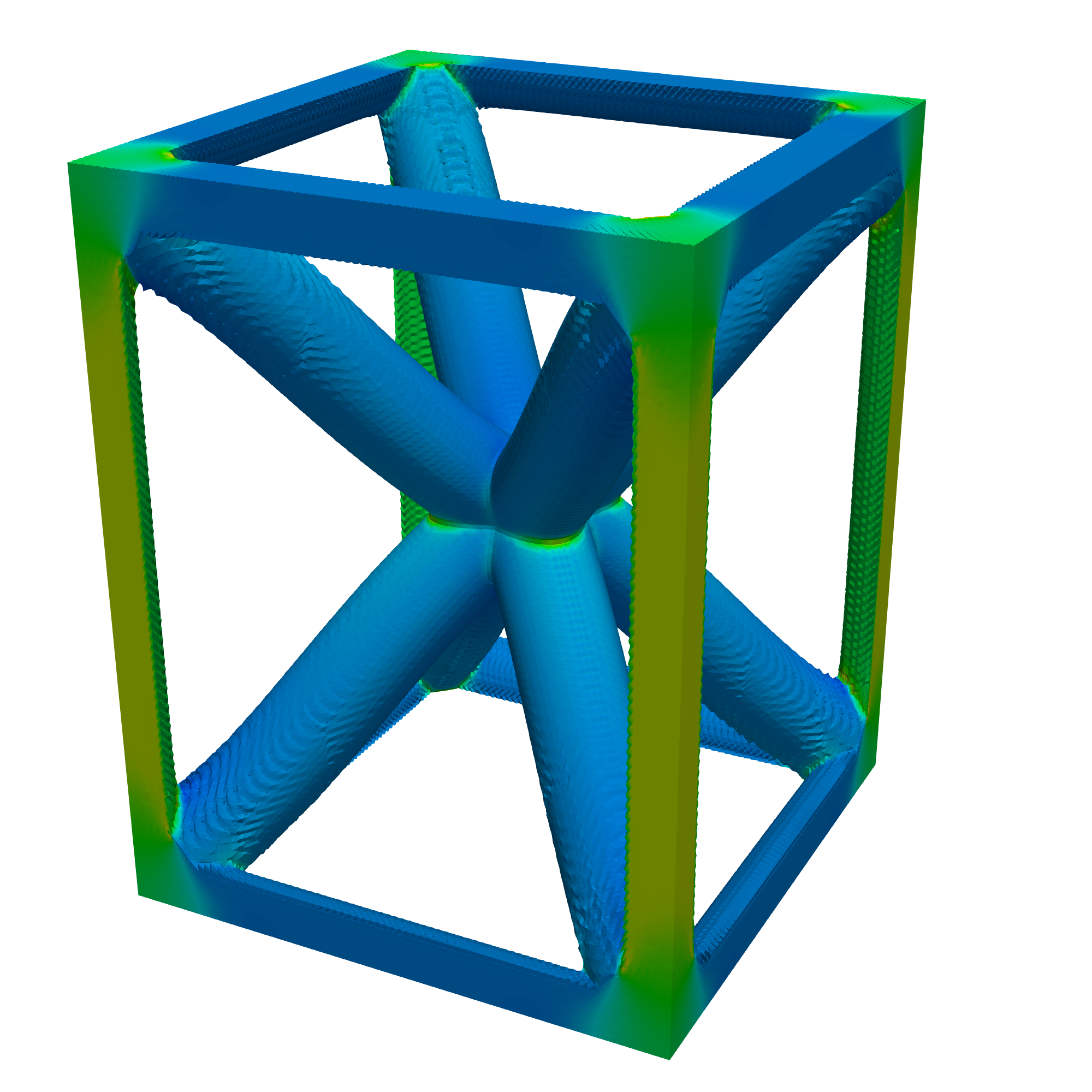}
\includegraphics[width=0.45\linewidth]{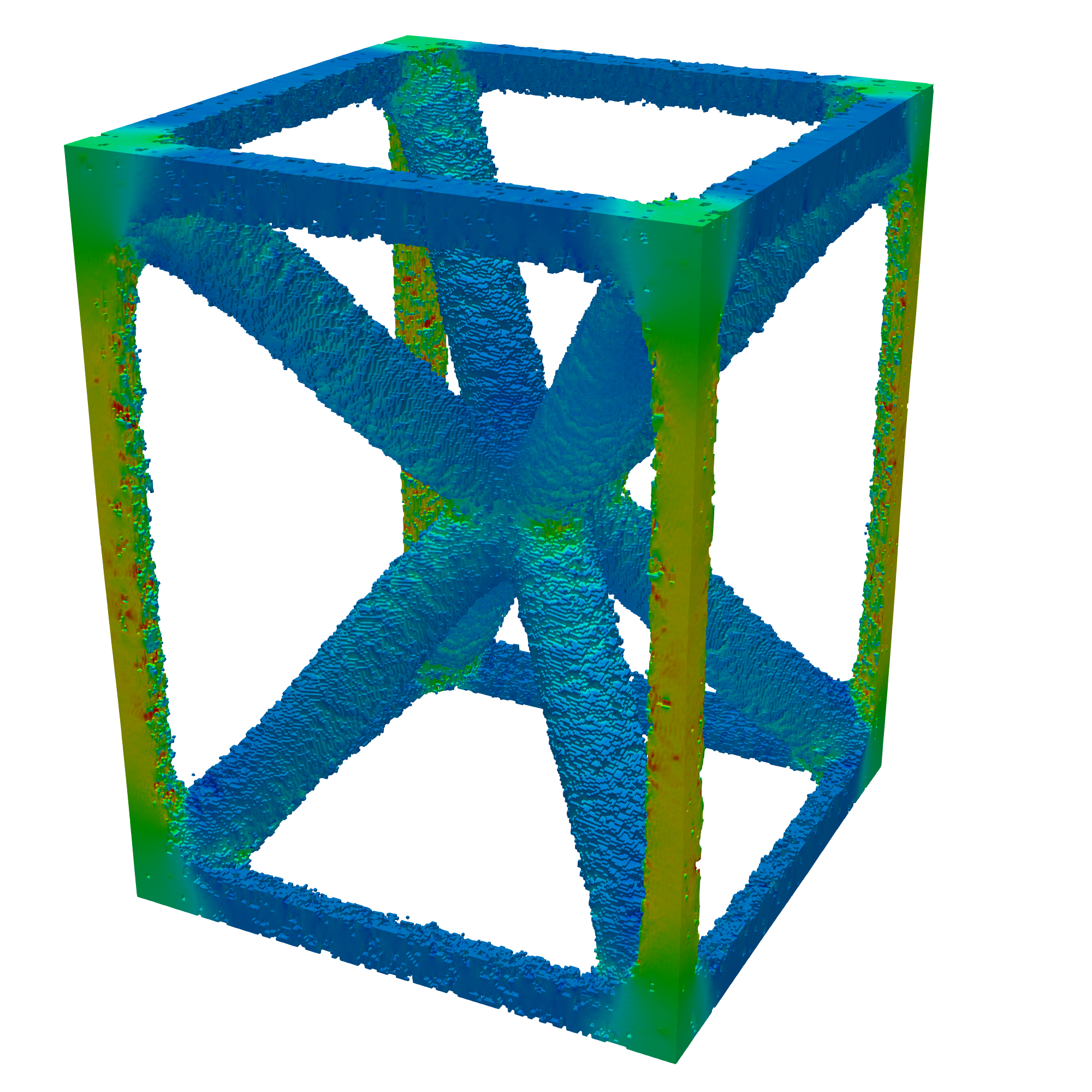}
\hspace{0.07\linewidth}

\includegraphics[width=0.45\linewidth]{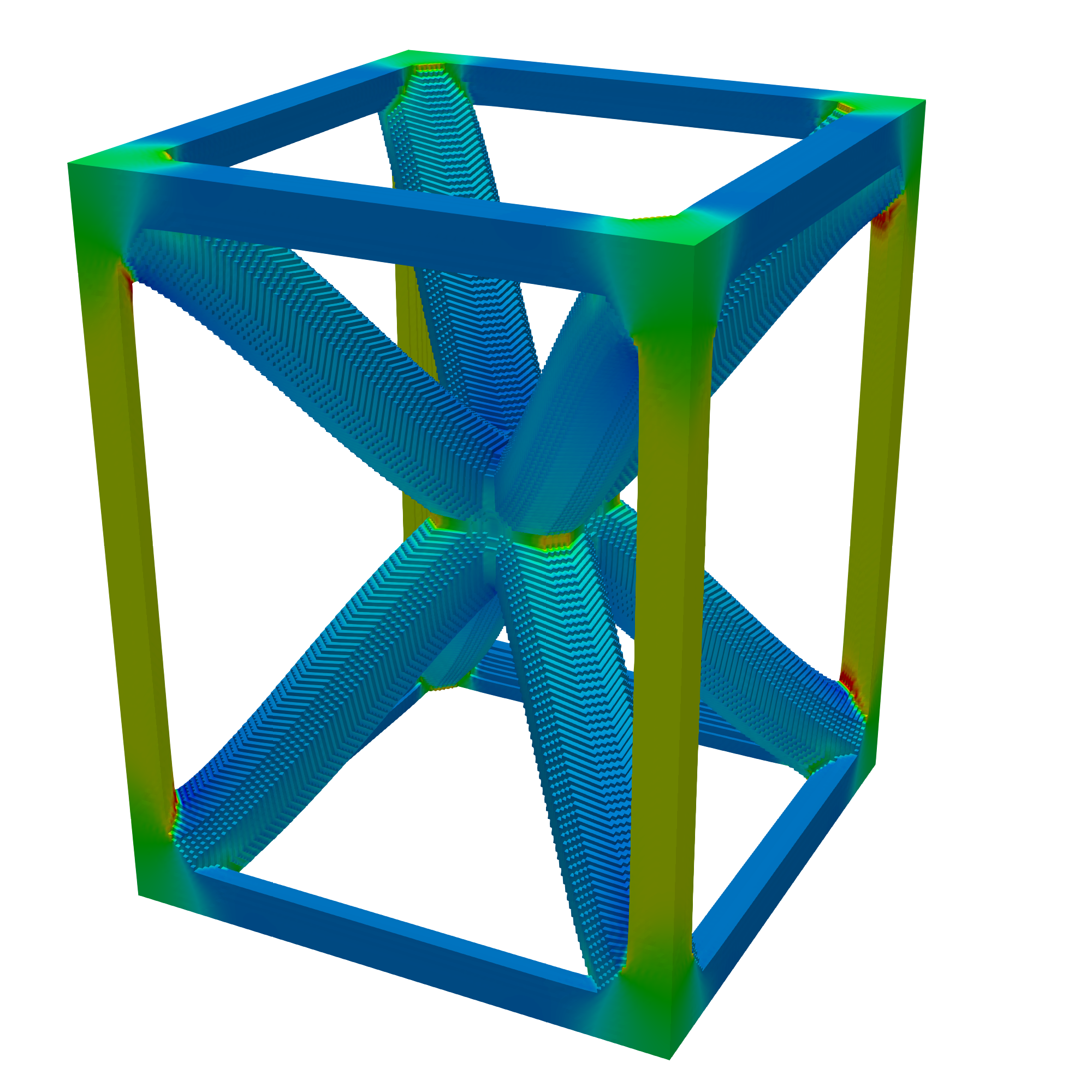}
\includegraphics[width=0.45\linewidth]{tomomodgood.png}
\includegraphics[width=0.07\linewidth]{leg1.png}
\caption{Local stress fields in loading direction ($\sigma_{zz}$) on the deformed configuration (x$20$) for Galerkin FFT (top) and MoDBFFT (bottom), design and real geometries.}\label{fig:tomoloc}
\end{figure}

\section{Conclusions}

In this paper, an optimal FFT framework for the homogenization of lattice materials has been searched and validated. The \rev{challenge} was finding an FFT approach that preserves the accuracy, good numerical performance and ability to use images/tomographies as direct input, in the case of domains with large regions of empty space. To this aim, two different FFT approaches able to solve problems containing phases with zero stiffness were combined with several approaches to smooth out the lattice surface in order to improve its geometrical representation and reduce the noise in the microscopic solution. 

\rev{Regarding} FFT solvers, after a first analysis, two algorithms have been \rev{selected} as the best options for infinite phase contrast. The first one is an adaptation of the Galerkin FFT approach using MINRES as linear solver and modified Fourier frequencies to consider a discrete differentiation scheme, the rotated forward approach. The second one, the MoDBFFT, \rev{is a method based on the displacement FFT approach in \cite{Lucarini2019b} which eliminates the indeterminacy} of strains in the empty regions leading to a fully determined system of equilibrium equations which allows the use of \rev{standard Fourier discretization and differentiation}. The accuracy of the two FFT solvers considered has been validated by comparison with FEM simulations of an octet cell for several volume fractions and discretization levels. The homogenized response of both FFT approaches was almost identical to FEM macroscopic response for linear elastic and elastoplastic materials, and differences in microfields were below 20\%.

Regarding the surface smoothening, several approaches have been considered based on modifying the actual stiffness of the voxels not fully embedded in the lattice or empty space. The impact of these geometrical representations in the effective response and local fields has been analyzed. The \rev{Voigt analytic} smoothening technique, which interpolates the stiffness of the interfacial voxels with the distance to the real lattice surface, was the best option since it allowed to represent exactly the relative density of the cell allowing to use coarser grids with very accurate macroscopic response. 

In terms of numerical efficiency, both FFT solvers succeed in converging in a relatively small number of iterations considering actual zero stiffness for the empty regions. Nevertheless, the adaptation of the Galerkin framework convergence rate was slightly superior in all the cases. When comparing the efficiency \rev{with respect to} FEM with the same discretization in the interior of the cell, FFT became competitive for relative densities greater than 7\%. For relative densities of 30\% (70\% of porosity), the simulation of this last FFT approach was 4 to 8 times faster than FEM. As a conclusion, the modified Galerkin approach combined with \rev{Voigt analytic} smoothening was the best FFT framework considering accuracy, numerical efficiency, and best h-convergence. 
 
Finally, to show the real potential of the approaches presented, both FFT frameworks are used to simulate the behavior of an actual printed lattice by using direct 3D tomographic data as input. The simulation volume element explicitly included the actual surface roughness and internal porosity (around \rev{3.6}\%) resulting from the fabrication process. The macroscopic elastic response was around \rev{10}\% more compliant than the ideal designed geometry, and local stress concentrations of 50\% were found near large pores. As a summary, it is shown that this technology can help to optimize the lattice fabrication parameters as well as accurately determine the actual lattice response taking into account the real fabrication defects.

\section*{Acknowledgements}

The authors gratefully acknowledge the support provided by the Luxembourg National Research Fund (FNR), Reference No. 12737941 and the European Union’s Horizon 2020 research and innovation programme for the project “Multi-scale Optimisation for Additive Manufacturing of fatigue resistant shock-absorbing MetaMaterials (MOAMMM)”, grant agreement No. 862015, of the H2020- EU.1.2.1. - FET Open Programme.

\bibliographystyle{unsrt}      
\bibliography{mybibfile}   

\end{document}